\pdfoutput=1
\documentclass[11pt,a4paper]{article}
\usepackage{jheppub}

\RequirePackage[utf8]{inputenc}

\usepackage{amsfonts}
\usepackage{amssymb}
\usepackage{comment}
\usepackage{graphicx}
\usepackage{amsmath}
\usepackage{tabu}
\usepackage{dsfont}
\usepackage{float}
\usepackage{amsthm}
\usepackage{slashed}
\usepackage{setspace}
\usepackage[labelformat=simple]{subcaption}

\usepackage[utf8]{inputenc}
\usepackage{amsmath, amsthm, amssymb, amsfonts}
\usepackage[font=small, labelfont=bf]{caption}
\usepackage{amsmath, amsthm, amssymb, amsfonts}
\usepackage{multirow}
\usepackage{graphicx}
\usepackage{float}
\usepackage[numbers,sort&compress]{natbib}
\usepackage{jheppub}
\usepackage{enumerate}

\newcommand{\nn}{\nonumber}

\newcommand{\p}{\partial}

\newcommand{\cY}{\mathcal{Y}}
\newcommand{\cF}{\mathcal{F}}
\newcommand{\cN}{\mathcal{N}}

\newcommand{\cV}{\mathcal{V}}
\newcommand{\cG}{\mathcal{G}}
\newcommand{\cO}{\mathcal{O}}

\newcommand{\ee}{\mathrm{e}}
\newcommand{\kom}{\, ,\quad }

\newcommand{\cK}{\mathcal{K}}
\newcommand{\bK}{\mathbb{K}}

\newcommand{\nc}{\newcommand}
\nc{\beq}{\begin{equation}}
\nc{\eeq}{\end{equation}}
\nc{\bea}{\begin{eqnarray}}
\nc{\eea}{\end{eqnarray}}

\def\ov{\overline}

\allowdisplaybreaks[2]
\numberwithin{equation}{section}

\usepackage[framemethod=TikZ]{mdframed}
\mdfsetup{skipabove=\topskip, skipbelow=\topskip}

\newcounter{equ}[section]
\newenvironment{equ}[1][]{%
\stepcounter{equ}%
\ifstrempty{#1}%
{\mdfsetup{%
frametitle={%
\tikz[baseline=(current bounding box.east),outer sep=0pt]
\node[anchor=east,rectangle,fill=blue!20]
{\strut };}}
}%
{\mdfsetup{%
frametitle={%
\tikz[baseline=(current bounding box.east),outer sep=0pt]
\node[anchor=east,rectangle,fill=blue!20]
{\strut ~#1};}}%
}%
\mdfsetup{innertopmargin=10pt,linecolor=blue!20,%
middlelinewidth=2pt,topline=true,
frametitleaboveskip=\dimexpr-\ht\strutbox\relax,}
\begin{mdframed}[]\relax%
}{\end{mdframed}}

\newcounter{Boxequ}[section]

\title{Systematics of type IIB moduli stabilisation\\ with odd axions}
\author[a,b]{Michele Cicoli,}
\author[c]{Andreas Schachner,}
\author[d]{Pramod Shukla}

\affiliation[a]{\small Dipartimento di Fisica e Astronomia, Universit\`a di Bologna, \\ via Irnerio 46, 40126 Bologna, Italy}
\affiliation[b]{\small INFN, Sezione di Bologna, viale Berti Pichat 6/2, 40127 Bologna, Italy}
\affiliation[c]{DAMTP, Centre for Mathematical Sciences, Wilberforce Road, Cambridge, CB3 0WA, UK.}
\affiliation[d]{\small Abdus Salam ICTP, Strada Costiera 11, Trieste 34151, Italy}
\emailAdd{michele.cicoli@unibo.it}
\emailAdd{as2673@maths.cam.ac.uk}
\emailAdd{pramodmaths@gmail.com}

\abstract{Moduli stabilisation in superstring compactifications on Calabi-Yau orientifolds remains a key challenge in the search for realistic string vacua. In particular, odd moduli arising from the reduction of 2-forms $(B_2,C_2)$ in type IIB are largely unexplored despite their relevance for inflationary model building. This article provides novel insights into the general structure of 4D $\mathcal{N}=1$ $F$-term scalar potentials at higher orders in the $\alpha^{\prime}$ and $g_{s}$ expansion for arbitrary Hodge numbers. We systematically examine superpotential contributions with distinct moduli dependences which are induced by fluxes or non-perturbative effects. Initially, we prove the existence of a no-scale structure for odd moduli in the presence of $(\alpha^\prime)^{3}$ corrections to the Kähler potential. By studying a partially $\mathrm{SL}(2,\mathbb{Z})$-completed form of the Kähler potential, we derive the exact no-scale breaking effects at the closed string $1$-loop and non-perturbative D-instanton level. These observations allow us to present rigorous expressions for the $F$-term scalar potential applicable to arbitrary numbers of moduli in type IIB Calabi-Yau orientifold compactifications. Finally, we compute the Hessian for odd moduli and discuss potential phenomenological implications.}

\keywords{Moduli stabilisation, Calabi-Yau orientifolds, Flux compactification}

\usepackage{todonotes}

\begin{document}

\maketitle

\bigskip

%\listoftodos

\section{Introduction}
\label{sec_intro}

The landscape of string vacua encompasses an immensely rich and diverse structure of Effective Field Theories (EFTs) arising from compactifications of critical superstring theory.
It is the result of a plethora of degeneracies involved in choosing compact geometries and suitable backgrounds of generalised electromagnetic fluxes, D-branes and O-planes.
In recent years,
investigations into the properties of EFTs from string compactifications have brought enormous advances in our capacity to engineer realistic string models.
However, it is widely known that achieving arbitrary good control over quantum corrections in string theory is impossible due to the absence of freely tunable parameters.
This is the so-called Dine-Seiberg problem \cite{Dine:1985he} which complicates the construction of fully trustable string vacua.
Although it is an unavoidable string theory condition,
computational control is fortunately attainable through the presence of extra parameters such as
the plenty of integer fluxes and ranks of condensing gauge groups.
Notwithstanding, a major challenge continues to be a unifying framework for stabilising moduli in well-controlled de Sitter minima.\footnote{See also \cite{Vafa:2005ui,Palti:2019pca} for recent swampland conjectures.}

In this context, we provide innovative methods to computing $F$-term scalar potentials
for type IIB orientifold compactifications with $ h^{1,1}_-\neq 0$ signalling the presence of odd moduli $G^{a}$.
To this day, they remain largely unexplored as compared to the simpler set-ups with $h^{1,1}_- = 0$ such as KKLT \cite{Kachru:2003aw}, LVS \cite{Balasubramanian:2005zx} and their plethora of variants.
Notwithstanding,
these moduli are ubiquitous in the string landscape and play a prominent role in the context of axion monodromy \cite{Silverstein:2008sg,McAllister:2008hb}.
Over the years, aspects of moduli stabilisation and inflationary model building have for instance been discussed in \cite{Lust:2006zg,Lust:2006zh, Grimm:2007hs, McAllister:2008hb, Hristov:2008if, Flauger:2009ab, Gao:2013rra, Long:2014dta, Gao:2014uha, Ben-Dayan:2014lca, McDonough:2018xzh, Carta:2020oci}.
Similarly, explicit constructions including orientifold-odd 2-cycles in concrete Calabi-Yau (CY) threefolds are available in the literature \cite{Blumenhagen:2008zz,Cicoli:2012vw, Cicoli:2013mpa, Cicoli:2013zha, Gao:2013pra, Carta:2020ohw}.
In addition,
odd moduli appear in the tree-level superpotential from generalised flux in (non-)geometric set-ups as analysed in \cite{Benmachiche:2006df,Robbins:2007yv, Shukla:2015rua, Shukla:2015bca, Blumenhagen:2015kja, Blumenhagen:2015lta, Shukla:2015hpa, Shukla:2016hyy, Shukla:2016xdy, Plauschinn:2018wbo, Shukla:2019wfo}.

This paper concerns a systematic treatment of the $(\alpha^{\prime})^{3}$-corrected $F$-term scalar potential in $4$D $\cN=1$ supergravity (SUGRA) theories obtained from general type IIB CY orientifold compactifications.
%Our results provide a unified framework for addressing open challenges in moduli stabilisation and inflationary model building in string theory.
%We derive elegant formulae for $F$-term scalar potentials that make a numerical implementation straight forward.
We begin our endeavours by computing exact expressions for derivatives of the Kähler potential
for complex structure moduli $U^{i}$,
axio-dilaton $S$,
Kähler (or even) moduli $T_{\alpha}$
and odd moduli $G^{a}=c^{a}+Sb^{a}$ in terms of the NS-NS $B_{2}$-axions $b^{a}$ and R-R $C_{2}$-axions $c^{a}$.
In particular,
we prove explicitly that the no-scale identity\footnote{This identity was observed at tree level in \cite{Grimm:2004uq} for $A,B\in\lbrace S,T_{\alpha},G^{a}\rbrace$, but already in \cite{Giddings:2001yu,Becker:2002nn,DeWolfe:2002nn,DAuria:2004pqc,DAuria:2004qsi} for models without odd moduli.}
\begin{equation}\label{eq:WeakNoScale} 
{K}_A\, {K}^{{A} \ov {B}}\, K_{\ov{B}}=4\kom A,B\in\lbrace S,T_{\alpha},G^{a}\rbrace
\end{equation}
holds for the $(\alpha^{\prime})^{3}$-corrected Kähler potential derived by BBHL \cite{Becker:2002nn}.
This can be anticipated from homogeneity arguments which hold not only at tree level \cite{Grimm:2005fa},
%page 50
but also in the presence of tree level $(\alpha^{\prime})^{3}$ corrections,
while being broken at higher order in the loop expansion.
Indeed, we derive the precise coefficient from the breaking of the above no-scale result from a partially $\mathrm{SL}(2,\mathbb{Z})$-completed Kähler potential depending on the non-holomorphic Eisenstein series of weight $3/2$.
In this way,
we are able to write down the exact $F$-term scalar potential including closed string loop and non-perturbative D-instanton corrections.

Our results lead us to the general master formula for the $(\alpha^{\prime})^{3}$-corrected $\mathcal{N}=1$ $F$-term scalar potential
\begin{equation}
V = V_{\text{cs}} + \ee^{{\cal K}} \,\biggl(|W|^2 + W_A\, K^{{A} \ov {B}} \, \ov W_{\ov B} +\sum_{A \in \{{S}, G^a, T_\alpha\}} (A - \ov A) \, \, (W \, \ov W_{\ov A} - \ov W\, W_A) \biggr)
\end{equation}
where $A,B \in \{{S}, G^a, T_\alpha\}$ and where all quantities are known exactly without resorting to any approximation in a large volume or small coupling expansion in 4D.
This form of the scalar potential is well suited for investigating moduli stabilisation aspects in (non-)SUSY (A)dS$_{4}$ minima in the presence of odd moduli.
We illustrate its usefulness by analysing a variety of superpotentials.

We begin with the simplest models with superpotentials $W=W(U^{i},S)$ induced by $3$-form fluxes \cite{Gukov:1999ya,Giddings:2001yu} for arbitrary numbers of ($U^i, T_\alpha, G^a$) moduli.
Here, we show explicitly that the shift symmetry of NS-NS axions $b^{a}$ remains intact even after including the BBHL correction, due to an exact cancellation of terms at sub-leading order in the volume.
This is clearly expected given that shift symmetries are protected against perturbative corrections which makes them sensitive to non-perturbative effects only.
However,
whenever the supersymmetric stabilisation of the axio-dilaton is enforced by hand,
terms of the form $(t^{\alpha}b^{a}b^{b})^{2}/\cV^{4}$ remain as traces in the effective scalar potential as observed e.g. in \cite{Hristov:2008if,Ben-Dayan:2014lca}.
Although such pieces usually get nullified at the minimum $\langle b^a \rangle = 0$, they are potentially misleading when studying fluctuations around the minimum such as in inflationary model building.
We argue that, in the presence of odd moduli, the stabilisation of the axio-dilaton and the complex structure moduli $U^{i}$ should be treated together with the $T_{\alpha}$ and $G^{a}$, especially since odd moduli may have masses heavier than the overall volume modulus \cite{Gao:2013rra}.

As another direct application of our results,
we examine superpotentials of the form $W=W(U^{i},S,T_{\alpha})$
which are commonly studied in the context of even-sector moduli stabilisation with $h^{1,1}_{-}=0$.
Due to the way the chiral fields $T_{\alpha}$ and $G^{a}$ are coupled in the Kähler potential,
the $B_{2}$-axions $b^{a}$ receive a potential,
while the $C_{2}$-axions $c^{a}$ remain flat.
In this context,
we explicitly compute the exact Hessian for $b^{a}$ axions and general superpotentials $W(U^{i},S,T_{\alpha})$.
At SUSY AdS$_{4}$ minima,
$h^{1,1}_{-}$ tachyons arise due to the unfixed axionic superpartners $c^{a}$,
thereby reproducing the results of \cite{Conlon:2006tq}.
Further,
we derive the Hessian for non-SUSY vacua from non-perturbative D3/D7-brane superpotentials where the presence and number of tachyons can be identified directly from properties of the model dependent triple intersection numbers.

Lastly,
we discuss superpotentials with explicit $G^{a}$ dependences which induce a potential for the R-R axions $c^{a}$.
We provide explicit formulae for the scalar potential from D5-brane gaugino condensation \cite{Grimm:2007xm,Grimm:2011dj} as well as geometric fluxes \cite{Benmachiche:2006df,Robbins:2007yv, Shukla:2015rua, Shukla:2015bca, Blumenhagen:2015lta,Blumenhagen:2015kja}.
These results will lead to novel insights into full moduli stabilisation in $\cN=1$ CY orientifold compactifications.
Indeed,
previous investigations for set-ups with $h^{1,1}_{-}=0$ in \cite{AbdusSalam:2020ywo} already obtained hybrid vacua sharing certain characteristics with KKLT and LVS type solutions.
Given that cases with $h^{1,1}_{-}\neq 0$ introduce additional axionic modulations to the potential,
our systematic framework gives rise to a much richer structure of string vacua which are highly attractive for phenomenological model building.

This paper is organised as follows.
After defining our conventions in Sect.~\ref{sec_basics},
we analyse the Kähler potential and its derivatives in Sect.~\ref{sec_scalarpotential}.
The main results of this paper are the three master formulae for the $F$-term scalar potential which are derived in Sect.~\ref{sec:GeneralScalarPotExpressions}.
The remainder of the paper is devoted to successively including more and more moduli couplings in the superpotential and applying the aforementioned formulae for advancing our ability to stabilise moduli.
We begin with the simple scenario of superpotentials induced by 3-form flux in Sect.~\ref{sec_Loopholes}.
Subsequently,
Sect.~\ref{sec:SPAnalysisWT} provides an extensive analysis for $T_{\alpha}$-dependent superpotentials where the scalar potential develops a dependence on the NS-NS axions $b^{a}$.
To generate a scalar potential for R-R axions $c^{a}$,
we study superpotentials depending on odd moduli $G^{a}$ in Sect.~\ref{sec:GDepSup}.
Lastly,
we briefly comment on phenomenological implications in Sect.~\ref{sec:PhenoImplications} before summarising our results in Sect.~\ref{sec_conclusions}.

\section{Preliminaries}
\label{sec_basics}

We focus on type IIB string theory compactified on CY threefolds $X_{3}$ which gives rise to $\cN=2$ SUGRA theories in 4D.
Including D-branes and O-planes reduces the supersymmetry of the EFTs to $\cN=1$, see e.g. \cite{Grimm:2005fa} for details.
The massless states in the 4D effective theory are in one-to-one correspondence with harmonic forms which are either even or odd under the action of an isometric, holomorphic involution $\sigma$ acting on the internal CY threefold $X_3$, thereby generating the equivariant cohomology groups $H^{p,q}_\pm (X_3)$. 

We denote the bases of even/odd 2-forms as $(\mu_\alpha, \, \nu_a)$ and of 4-forms as $(\tilde{\mu}_\alpha, \, \tilde{\nu}_a)$ where $\alpha\in h^{1,1}_+(X_3)$ and $a\in h^{1,1}_-(X_3)$. In addition, the bases for the even/odd cohomologies of 3-forms $H^3_\pm(X_3)$ are denoted as the symplectic pairs $(a_K, b^J)$ and $({\cal A}_\Lambda, {\cal B}^\Delta)$ respectively. Using the conventions of \cite{Robbins:2007yv}, we fix the normalisation in the various cohomology bases as
\begin{align}
\label{eq:intersection}
\int_{X_3} \, \mu_\alpha \wedge \mu_\beta \wedge \mu_\gamma &= k_{\alpha \beta \gamma} \kom \int_{X_3} \, \mu_\alpha \wedge \nu_a \wedge \nu_b = \hat{k}_{\alpha a b}\, , \nn\\
\int_{X_3} \, \mu_\alpha \wedge \tilde{\mu}^\beta &= {\delta}_\alpha^{\, \, \, \beta}  \kom\int_{X_3} \, \nu_a \wedge \tilde{\nu}^b = {\delta}_a^{\, \, \,b}\, , \quad\\
\int_{X_3} \, a_K \wedge b^J &= \delta_K{}^J  \kom\int_{X_3} {\cal A}_\Lambda \wedge {\cal B}^\Delta = \delta_\Lambda{}^\Delta \, . \nonumber
\end{align}
Here, depending on the orientifold choice, we have two possibilities:
\begin{itemize}
\item $O3/O7$-planes: $K\in \{1, ..., h^{2,1}_+(X_3)\}$ and $\Lambda\in \{0, ..., h^{2,1}_-(X_3)\}$,
\item $O5/O9$-planes: $K\in \{0, ..., h^{2,1}_+(X_3)\}$ and $\Lambda\in \{1, ..., h^{2,1}_-(X_3)\}$. 
\end{itemize}

Now, the various $p$-form fields can be expanded in appropriate bases of the equivariant cohomologies.
Specifically, the K\"{a}hler form $J$, the
2-forms\footnote{Note that the even component of the Kalb-Ramond field $B_{+} = b^\alpha \, \mu_\alpha$, though not a continuous modulus, can take the two  discrete values $b^\alpha\in\{0, 1/2\}$. } $B_2$,  $C_2$ and the R-R 4-form $C_4$ can be expanded as \cite{Grimm:2004uq}
\begin{align}
\label{eq:fieldExpansions}
J &= t^\alpha\, \mu_\alpha \kom  B_2= b^a\, \nu_a \kom C_2 =c^a\, \nu_a\, , \nonumber\\[0.6em]
C_4 &= {\rho}_{\alpha} \, \tilde\mu^\alpha\, + V^{K}\wedge a_K - V_{K}\wedge b^K\, + D_2^{\alpha}\wedge \mu_\alpha\, ,
\end{align}
where $t^\alpha$ denotes the 2-cycle volume moduli and $b^a, \, c^a,\rho_\alpha$ are various axions which inherit their shift symmetry from 10D $p$-form gauge symmetries. Further, ($V^K$, $V_K$) forms a dual pair of space-time 1-forms
and $D_2^{\alpha}$ are space-time 2-forms dual to the scalars $\rho_\alpha$. Due to the self-duality of $C_{4}$, half of the degrees of freedom of $C_4$ are removed.

Further, since $\sigma^*$ reflects the holomorphic three-form $\Omega_3$, there are $h^{2,1}_-(X_{3})$ complex structure deformations parametrised by complex moduli $U^{i}$, $i=1,\ldots ,h^{2,1}_{-}$.
In fact, the three-form $\Omega_3$ can be written as
\bea
\label{eq:Omega3}
& &  \Omega_3\, \equiv  {\cal X}^\Lambda \, {\cal A}_\Lambda - \, {\cal F}_{\Lambda} \, {\cal B}^\Lambda 
\eea
where the periods $\mathcal{X}^{\lambda}$, $\mathcal{F}_{\lambda}$ are obtained from
\begin{equation}\label{eq:PeriodsInt} 
{\cal X}^\Lambda=\int_{X_{3}}\Omega_{3}\wedge {\cal B}^\Lambda\kom {\cal F}_{\Lambda}=\int_{X_{3}}\Omega_{3}\wedge {\cal A}_\Lambda\, .
\end{equation}
Since the complex structure moduli space is equipped with a special Kähler structure,
we can compute Kähler and superpotential from a pre-potential ${\cal F} = ({\cal X}^0)^2 \, \, f( {\cal X}^{i})$, cf.~Sect.~\ref{sec:InverseMetric}.

Apart from the complex structure moduli, the spectrum of the 4D $N = 1$ effective theory is encoded in the chiral variables ($U^i, {S}, G^a, T_\alpha$) defined as \cite{Benmachiche:2006df},
\begin{align}\label{eq:N1coords}
U^i &= v^i + i\, u^i\kom {S} \, = C_0 + \, i \, \ee^{-\phi} = c_0 + i\, s\, , \nonumber\\[0.6em]
G^a &= c^a + {S} \, b^a \kom T_\alpha = \bigl({\rho}_\alpha +  \hat{{k}}_{\alpha a b} c^a b^b + \frac{1}{2} \, {S} \, \hat{{k}}_{\alpha a b} b^a \, b^b \bigr)  -\frac{i}{2} \, {k}_{\alpha\beta\gamma} t^\beta t^\gamma\,.
\end{align}
It will be convenient to also define
\begin{align}\label{eq:ConventionModAx} 
\hat{c}^{a}&= c^a + c_{0}\, b^a\, ,\nn\\
\tilde{\rho}_{\alpha}&={\rho}_\alpha +  \hat{{k}}_{\alpha a b} c^a b^b + \frac{1}{2} \, c_{0} \, \hat{{k}}_{\alpha a b} b^a \, b^b\kom\sigma_{\alpha}=-\frac{1}{2} \, {s} \, \hat{{k}}_{\alpha a b} b^a \, b^b  +\frac{1}{2} \, {k}_{\alpha\beta\gamma} t^\beta t^\gamma
\end{align}
so that
\begin{equation}
G^{a}=\hat{c}^{a}+isb^{a}\kom T_{\alpha}=\tilde{\rho}_{\alpha}-i\sigma_{\alpha}\, .
\end{equation}
The sign in front of $\sigma_{\alpha}$ arises from the way the chiral variables $T_{\alpha}$ are defined in Eq.~\eqref{eq:N1coords} which follows from consistently fixing the signs in various coordinates in a manifestly $T$-dual manner \cite{Shukla:2019wfo}. %{\bf Such field redefinitions help in working with effectively three independent classes of RR axions, namely $c_0, \hat{c}^a$ and $\tilde{\rho}_{\alpha}$. Also, let us mention that the axion combination $\tilde{\rho}_{\alpha}$ is a $S$-dual invariant and has its own relevance in different models \cite{Shukla:2015rua, Shukla:2016hyy}.}

The ${\cN}=1$ $F$-term scalar potential is determined from the K\"ahler potential $\cK$ and superpotential $W$ via
\begin{equation}
\label{eq:Vtot}
V=e^{\cal K}\Big({\cal K}^{{\cal A}\bar {\cal B}} D_{\cal A} W\, \ov{D}_{\ov{\cal B}} \ov W-3\, |W|^2\Big)\kom D_{\cal A}W=\p_{\cal A}W+\cK_{\cal A}\, W  \,,\,
\end{equation}
where the covariant derivatives are defined with respect to all the chiral variables $\mathcal{A}\in\lbrace U^{i},S,T_{\alpha},G^{a}\rbrace$.
At the perturbative level, the K\"ahler potential receives corrections from the $\alpha^\prime$ and string-loop ($g_s$) expansion. Using appropriate chiral variables and the leading order $\alpha^\prime$ corrections of \cite{Becker:2002nn}, the K\"{a}hler potential splits in two pieces from their underlying ${\cN}=2$ special K\"ahler and quaternionic structure such that
\begin{equation}
\label{eq:K}
{\cal K} = \mathbb{K}(U^i, \ov{U}^i) + K({S}, G^a, T_\alpha; \ov{S}, \ov{G}^a, \ov{T}_\alpha)
\end{equation}
where
\begin{align}
\label{eq:Kgennn}
\mathbb{K}(U^i, \ov{U}^i) &= -\ln\left(i\int_{X}\Omega_3\wedge{\ov\Omega_3}\right) = -\ln\biggl[i\, \left(\ov {\cal X}^\Lambda \, {\cal F}_\Lambda - {\cal X}^\Lambda \, \ov {\cal F}_\Lambda \right)\biggr]\, , \\[0.6em]
\label{eq:KgenKGS} K({S}, G^a, T_\alpha; \ov{S}, \ov{G}^a, \ov{T}_\alpha) &= - \ln\left(-i({S}-\ov{S})\right) -2\ln{\cal Y}\,. 
\end{align}
Here, ${\cal Y}$ denotes the $\alpha^\prime$-corrected volume of the CY threefold \cite{Becker:2002nn},
\begin{align}\label{eq:DefYTree} 
{\cal Y}({S}, G^a, T_\alpha; \ov{S}, \ov{G}^a, \ov{T}_\alpha) &\equiv {\cal V} + \frac{\xi}{2}\, \left(\frac{{S}-\ov{S}}{2\,i}\right)^{3/2}\\
&= \frac{1}{6} \,{k_{\alpha \beta \gamma} \, t^\alpha\, t^\beta \, t^{\gamma}} + \frac{\xi}{2}\, \left(\frac{{S}-\ov{S}}{2\,i}\right)^{3/2}\kom \xi = -\frac{\zeta(3)\, \chi(X_{3})}{2\,(2\pi)^3}\,, \nonumber
\end{align}
where the 2-cycle moduli $t^\alpha$ are implicitly functions of all the complexified chiral variables ${S}, G^a$ and $T_\alpha$ and their complex conjugates.
In the absence of any open-string moduli, the additive structure within the K\"ahler potential \eqref{eq:K} results in the block-diagonal nature of the K\"ahler metric, and its inverse.

%the superpotential $W$ is highly background dependent.
Throughout this paper,
we investigate scalar potentials for a great variety of superpotential couplings.
A prominent example for non-trivial $3$-form flux backgrounds
is the tree level Gukov-Vafa-Witten (GVW) flux superpotential \cite{Gukov:1999ya,Giddings:2001yu}
%\cite{Aldazabal:2006up,Aldazabal:2008zza, Guarino:2008ik,Blumenhagen:2015kja},
\begin{equation}
\label{eq:WGVW}
W_0(U^{i},S) =\int_{X} \left({F}_{3} +{S} \, {H}_{3} \right) \wedge \Omega_3 = e_\Lambda \, {\cal X}^\Lambda + m^\Lambda \, {\cal F}_\Lambda\,.
\end{equation}
Here the periods are defined in \eqref{eq:PeriodsInt} and the components of the symplectic vectors $e_\Lambda$ and $m^\Lambda$ are
\begin{equation}
\label{eq:eANDm}
e_\Lambda = \left({F}_\Lambda + {S} \, H_\Lambda \right)\kom  m^\Lambda = \left({F}^\Lambda + {S} \, {H}^\Lambda \right)\, ,
\end{equation}
where the fluxes are obtained from integrals over 3-cycles, i.e., ${F}_\Lambda = \int_{X_{3}}\,  F_{3}\wedge \mathcal{A}_{\Lambda} $ etc.
Beyond 3-form fluxes, 
$W$ is protected against perturbative corrections by non-renormalisation theorems \cite{Wen:1985jz,Grisaru:1979wc,Dine:1986vd,Burgess:2005jx}.
However,
there can be other terms in the superpotential
induced by non-perturbative effects \cite{Witten:1996bn} or (non-)geometric fluxes \cite{Shelton:2005cf, Shelton:2006fd, Aldazabal:2006up, Benmachiche:2006df, Robbins:2007yv, Guarino:2008ik, Blumenhagen:2013hva, Shukla:2015rua, Shukla:2015bca, Blumenhagen:2015kja, Blumenhagen:2015lta, Shukla:2015hpa, Shukla:2016hyy, Shukla:2016xdy, Plauschinn:2018wbo, Shukla:2019wfo}.
For example, non-perturbative contributions can arise from Euclidean D3-instantons \cite{Witten:1996bn} or gaugino condensations effects on stacks of D7-branes \cite{Novikov:1983ek,Novikov:1983ee,Ferrara:1982qs} wrapping suitable 4-cycles.
Introducing fluxes on top of these corrections gives rise to ``fluxed-instantons" or magnetised gaugino condensation effects with a more intricate moduli dependence \cite{Grimm:2011dj}.
We are going to study these choices of superpotentials in more detail throughout the paper.

%\section{No-scale structure, odd moduli and $\alpha^\prime$ corrections}
\section{No-scale properties, odd moduli and $\alpha^\prime$ corrections}
\label{sec_scalarpotential}

In this section,
we derive exact expressions for the Kähler metric in the presence of $(\alpha^{\prime})^{3}$ corrections \cite{Becker:2002nn}.
We show that certain tree level identities hold even at this order in $\alpha^{\prime}$.
Beyond that, we argue that they are broken once $1$-loop and D-instanton effects are taken into account.
These results will be applied in the subsequent section to deriving master formulae for $(\alpha^{\prime})^{3}$-corrected $F$-term scalar potentials.

\subsection{Exact inverse K\"ahler metrics at order $(\alpha^{\prime})^{3}$}\label{sec:InverseMetric} 

Given that the full K\"ahler potential ${\cal K}$ in \eqref{eq:K} contains two decoupled pieces with $\mathbb{K}$ depending only on the complex-structure moduli $U^i$, and $K$ on all other moduli ${S}, G^a$ and $T_\alpha$,
one arrives at a block-diagonal structure for the inverse K\"ahler metric, 
\bea\label{eq:InverseKMProdModSpace} 
& & {\cal K}_{{\cal A} \ov{\cal B}} = \left[
\begin{array}{cc}
\bK_{i \ov{j}} \qquad & {\cal O} \\
& \\
%\hline
{\cal O} & \qquad K_{A \ov B}
\end{array}
\right], \qquad {\cal K}^{{\cal A} \ov{\cal B}} = \left[
\begin{array}{cc}
\bK^{i \ov{j}} \qquad & {\cal O} \\
& \\
%\hline
{\cal O} & \qquad K^{A \ov B}
\end{array}
\right]\,.
\eea
In the remainder of this subsection,
we compute exact expressions for the inverse metric in set-ups with arbitrary Hodge numbers $h^{p,q}_{\pm}$.
They are subsequently utilised to compute identities for ${K}_A\, {K}^{{A} \ov{B}}$ and ${K}_A\, {K}^{{A} \ov{B}}K_{\ov B}$ that dramatically facilitate the computation of $F$-term scalar potentials in Sect.~\ref{sec:MasterFormulas}.

\subsection*{Complex structure moduli sector}

Let us first consider the complex structure moduli sector where the periods \eqref{eq:PeriodsInt} can be computed from solving Picard-Fuchs equations \cite{Hosono:1993qy,Hosono:1994ax,Cox:2000vi} or using asymptotic Hodge theory \cite{Bastian:2021eom}.
Throughout this paper, we restrict to the large complex structure regime where
the pre-potential is given by \cite{Morrison:1991cd,Hosono:1994av,Hosono:1994ax},
\bea
\label{eq:prepotentialNew}
& & {\cal F} = \frac{l_{ijk} \, {\cal X}^i\, {\cal X}^j \, {\cal X}^k}{6\, {\cal X}^0} +  \frac{1}{2} \,{p_{ij} \, {\cal X}^i\, {\cal X}^j} +  \,{p_{i} \,{\cal X}^0\,  {\cal X}^i} +  \frac{i}{2} \, \,{\tilde \xi}\, ({\cal X}^0)^2 +  ({\cal X}^0)^2\, {\cal F}_{inst}\,.
\eea
This is a homogeneous function of degree two in the symplectic coordinates ${\cal X}^\Lambda$, and one can use ${\cal X}^0 = 1$ and ${\cal X}^i = U^i$ to write it in terms of the non-homogeneous complex variables $U^i$. Further, the parameters $l_{ijk}$ are the triple intersection numbers on the mirror threefold $\tilde{X}_3$ which, along with the other real parameters, are defined as \cite{Mayr:2000as,Grimm:2009ef}
\begin{align}
l_{ijk} &= \int_{\tilde{X_3}} \, J_i \wedge J_j \wedge J_k\kom p_{ij} = -\frac{1}{2}\int_{\tilde{X_3}} \, J_i \wedge J_j \wedge J_j\,\text{mod}\,\mathbb{Z}\; ,\quad \nonumber\\
p_j &= \frac{1}{4\, 3!}\int_{\tilde{X_3}} \,c_2(\tilde{X_3}) \wedge J_j\kom \tilde{\xi} = -\, \frac{\zeta(3)\, \chi(\tilde{X_3})}{(2\, \pi)^3}\,.
\end{align}
Let us note that $\tilde{\xi}= - 2\xi$ as the Euler characteristics satisfy $\chi(\tilde{X}_{3})=-\chi(X_{3})$.
Finally,
the string worldsheet corrections on the mirror dual side give rise to \cite{Hosono:1994av,Hosono:1994ax}
\begin{equation}\label{eq:InstCorrections} 
\mathcal{F}_{\text{inst}}(U^{i})=\sum_{\beta\in H_{2}^{-}(\tilde{X}_{3},\mathbb{Z})\setminus\lbrace0\rbrace}\, n_{\beta}\, \text{Li}_{3}(q^{\beta})\kom \text{Li}_{3}(x)=\sum_{m=1}^{\infty}\, \dfrac{x^{m}}{m^{3}}\kom q^{\beta}=\ee^{2\pi i d_{i}U^{i}}
\end{equation}
in terms of Gopakumar-Vafa invariants $n_{\beta}$ \cite{Gopakumar:1998ii,Gopakumar:1998jq}
which naively count the number of rational (oriented) curves $\Sigma_{g}$ of genus $g$ and of class $\beta=d_{i}\beta^{i}$ that can be holomorphically mapped into $\tilde{X}_{3}$.
Now, the first derivatives of the prepotential ${\cal F}$ are given by
\bea
\label{eq:Prepder1}
& & {\cal F}_0 = -\, \frac{1}{6}\, l_{ijk}\, U^i \, U^j\, U^k + p_i \, U^i + i\, \tilde\xi +\left(2\,{\cal F}_{inst} - U^i\, \partial_{i} {\cal F}_{inst} \right), \\
& & \hskip0cm {\cal F}_i = \frac{1}{2}\, l_{ijk} \, U^j\, U^k + p_{ij}\, U^j + p_i + \left(\partial_{i} {\cal F}_{inst} \right) \nonumber
\eea
For simplicity,
we ignore $\mathcal{F}_{\text{inst}}$ in the subsequent analysis which we will explore in the future.

Now using (\ref{eq:Kgennn}), the complex structure moduli dependent part of the K\"ahler potential is simplified as,
\bea
& & \hskip-1cm {\mathbb K} = -\ln \biggl[-\frac{i}{6}\,l_{ijk} \, \left(U^i - \ov{U}^i\right)\, \left(U^j - \ov{U}^j\right)\, \left(U^k - \ov{U}^k\right)  - 2 \, \tilde\xi \biggr]\,.
\eea
Further,
we define $U^i= v^i + i\, u^i$ so that
\bea
\label{eq:derKcsSimp}
& & {\mathbb K}_i = \frac{3\, i\, l_i}{2\, l+ 3\,\tilde{\xi}}, \qquad {\mathbb K}_{i \ov j} = \frac{9}{(2 \, l + 3 \, \tilde\xi)^2} \left(l_i \,l_j  -\frac{2 \, l + 3 \, \tilde\xi}{3}\,l_{ij} \right)
\eea
in terms of the short hand notations
\begin{equation}
l= l_{ijk}\,u^i\, u^j\, u^k\kom l_{i}= l_{ijk}\, u^j\, u^k\kom l_{ij}= l_{ijk}\, u^k\kom l^{ik} \, l_k= u^i\, .
\end{equation}
The inverse K\"ahler metric is thus given by
\bea
\label{eq:invKcsSimp}
& & {\mathbb K}^{i\ov{j}} = \frac{2 \, l + 3 \, \tilde\xi}{l -3\, \tilde{\xi}} \left(u^i\, u^j - \frac{l -3\, \tilde{\xi}}{3} \, l^{ij} \right).
\eea
Note that the setting $\tilde\xi = 0$ leads to the following expressions arising from the classical triple intersection number $l_{ijk}$ on the mirror CY threefold,
\bea
& & {\mathbb K}_i = \frac{3\, i\, l_i}{2\, l}, \qquad {\mathbb K}_{i \ov j} = \frac{9}{4 \, l^2} \left(l_i \,l_j  -\frac{2 \, l}{3}\,l_{ij} \right), \qquad {\mathbb K}^{i\ov{j}} = 2\, u^i\, u^j - \frac{2\,l}{3} \, l^{ij},
\eea
which matches with the standard result for the CY threefold where $l \to 6 {\cal V}$ on the mirror K\"ahler sector. Finally, one obtains the following useful identities
\begin{align}
\label{eq:id-KcsSimp}
{\mathbb K}_{i} {\mathbb K}^{i\ov{j}}  & = i\, \frac{2\, l + 3\, \tilde{\xi}}{l - 3\, \tilde{\xi}} \, u^j = -\, {\mathbb K}^{j\ov{i}} {\mathbb K}_{\ov{i}}\kom {\mathbb K}_{i} {\mathbb K}^{i\ov{j}} {\mathbb K}_{\ov{j}} = 3 + \frac{9\, \tilde\xi}{l - 3\, \tilde{\xi}}\, .
\end{align}
This shows that in the absence of perturbative effects on the mirror-side, these identities reduce to the following simpler forms,
\bea
& & {\mathbb K}_{i} {\mathbb K}^{i\ov{j}}  =  2\, i\, u^j\, = (U - \ov{U})^j = -\,{\mathbb K}^{j\ov{i}} {\mathbb K}_{\ov{i}} \kom {\mathbb K}_{i} {\mathbb K}^{i\ov{j}} {\mathbb K}_{\ov{j}} = 3\, .
\eea
The breaking of the no-scale identity ${\mathbb K}_{i} {\mathbb K}^{i\ov{j}} {\mathbb K}_{\ov{j}} = 3$ (on the mirror dual side) through $\tilde{\xi}$ in \eqref{eq:id-KcsSimp} is of course expected from simple homogeneity arguments.

\subsection*{K\"ahler moduli sector}

Next, we compute the inverse K\"ahler metric for the ${S}, G^a$ and $T_\alpha$ moduli
by considering the K\"ahler potential \eqref{eq:KgenKGS} with $\cY$ defined in \eqref{eq:DefYTree}.
To this end, we have to rewrite the overall volume ${\cal V}$ in terms of the chiral variables ${S}, G^a, T_\alpha$ and their complex conjugates.
The definition of the chiral variable $T_\alpha$ in Eq.~(\ref{eq:N1coords}) allows us to write
\bea
& & {k}_\alpha \equiv {k}_{\alpha\beta\gamma} t^\beta\,t^\gamma = i\left(T_\alpha - \ov{T}_\alpha \right) - \frac{i\, \hat{{k}}_{\alpha ab}\, (G^a - \ov{G}^a)\, (G^b - \ov{G}^b)}{2\, ({S} - \ov{S})}
\eea
and hence we have
\bea\label{eq:YExplicit} 
& & \hskip-0.5cm {\cal V} \equiv \frac{1}{6}\, \, \, k_{\alpha} \, t^\alpha = \frac{t^\alpha}{6} \biggl[i\left(T_\alpha - \ov{T}_\alpha \right) - \frac{i\, \hat{{k}}_{\alpha ab}\, (G^a - \ov{G}^a)\, (G^b - \ov{G}^b)}{2\, ({S} - \ov{S})}\biggr] \,.
\eea
Here $t^\alpha$ is an implicit function of the chiral variables ${S}, G^a, T_\alpha$.
In addition, given that there are two kinds of triple intersection numbers surviving under the orientifold action, namely $k_{\alpha\beta\gamma}$ and $\hat{k}_{\alpha a b}$ defined in \eqref{eq:intersection}, we introduce the following shorthand notation to simplify intermediate computations,
\begin{align}
\label{eq:shorthands}
\hspace{1.5cm} k_0 &= k_{\alpha\beta\gamma} \, t^\alpha \, t^\beta \, t^\gamma= 6 \, {\cal V} \kom & k_\alpha &=k_{\alpha \beta} \, t^\beta  \kom  &k_{\alpha\beta} &= k_{\alpha\beta\gamma} \, t^\gamma\, ,\nn\\
\hat{k}_0 &= \hat{k}_{\alpha a b} \, t^\alpha\, b^a\, b^b  \kom & \hat{k}_{\alpha}& = \hat{k}_{\alpha a b} \,b^a\, b^b  \kom& \hat{k}_{a b} &= \hat{k}_{\alpha a b} \, t^\alpha \, , \\
\hat{\xi} &= s^{3/2}\, \,{\xi} \kom &\hat{k}_{a}& = \hat{k}_{\alpha a b} \,t^\alpha\, b^b\kom  &\hat{k}_{\alpha a} &= \hat{k}_{\alpha a b} \, b^b \,. \nonumber
\end{align}
Initially, we compute the relations, see Eq.~\eqref{eq:identitiesVariables} in App.~\ref{app:UsefulIdentities} for details,
\begin{equation}\label{eq:derY}
\frac{\partial {\cal Y}}{\partial {S}} = \frac{i\, \, \hat{k}_0}{8} - \frac{3\,i \, \hat{\xi}}{8\, s} \kom  \frac{\partial {\cal Y}}{\partial G^a} = -\, \frac{i\, \, \hat{{k}}_a}{4} \kom \frac{\partial {\cal Y}}{\partial T_\alpha}  = \frac{i\, \, t^\alpha}{4}\, ,
\end{equation}
which lead to the following derivatives of $K$,
\begin{align}
\label{eq:derK}
K_{S} = \frac{i}{2 \,s } + i \, {\cal G}_{ab} \, b^a \, b^b + \frac{3i\, \hat{\xi}}{4s\,{\cal Y}} \kom  K_{G^a} &= -2 \, i \, {\cal G}_{ab} \, b^b  \kom K_{T_\alpha} = -\frac{i \, t^\alpha}{2\, {\cal Y}} \, .
\end{align}
Here the $\alpha^\prime$-corrected moduli space metric and its inverse, ${\cal G}$ and ${\cal G}^{-1}$, are
\begin{align}
\label{eq:genMetrices}
\frac{{\cal G}_{\alpha \beta}}{36} &= \frac{k_\alpha \,k_\beta}{4\, {\cal Y}\, (k_0 - 2\, {\cal Y})} -\frac{k_{\alpha \beta}}{4\, {\cal Y}} \kom &{\cal G}^{ab} &= -4\, {\cal Y}\, \hat{k}^{ab}\, \\
36\,{\cal G}^{\alpha \beta} &= 2 \, t^\alpha \, t^\beta -4\, {\cal Y} \, k^{\alpha \beta} \kom &{\cal G}_{ab} &= - \frac{\hat{k}_{ab}}{4\, {\cal Y}}\,. \nonumber 
\end{align}
From the above,
we find that the various K\"ahler metric components can be written as
\begin{align}
\label{eq:simpK}
K_{{S} \ov{S}} &= \frac{1}{4\,s^2}\, \left(1 - \frac{3\, \hat{\xi}}{4\, {\cal Y}}+ \frac{9\, \hat{\xi}^2}{8\, {\cal Y}^2} \right) + \frac{{\cal G}_{ab}\, b^a\, b^b}{s} \left(1 + \frac{3\, \hat{\xi}}{4\, {\cal Y}}\right) + \frac{9\, {\cal G}^{\alpha\beta} \hat{{k}}_\alpha\, \hat{{k}}_\beta}{16\, {\cal Y}^2}\,, \nonumber\\[0.2em]
K_{G^a \, \ov{S}} &=  -\frac{{\cal G}_{ab}\, b^b}{s} \left(1 + \frac{3\, \hat{\xi}}{4\, {\cal Y}}\right) - \frac{9\, {\cal G}^{\alpha\beta} \hat{{k}}_{\alpha a}\, \hat{{k}}_\beta}{8\, {\cal Y}^2} =K_{{S}\,\ov{G}^a}, \nonumber\\[0.2em]
K_{T_\alpha \, \ov{S}} &= \frac{9\, {\cal G}^{\alpha\beta} \, \hat{{k}}_\beta}{8\, {\cal Y}^2} -\frac{3\, \hat{\xi}\, \, \, t^\alpha}{16\, s\, {\cal Y}^2}= K_{{S}\,\ov{T}_\alpha}, \\[0.2em]
K_{G^a \, \ov{G}^b} &= \frac{{\cal G}_{ab}}{s} + \frac{9\, {\cal G}^{\alpha\beta} \hat{{k}}_{\alpha a}\, \hat{{k}}_{\beta b}}{4\, {\cal Y}^2} , \nonumber\\[0.2em]
K_{T_\alpha \, \ov{G}^a} &= -\, \frac{9\, {\cal G}^{\alpha\beta}\, \hat{{k}}_{\beta a}}{4\, {\cal Y}^2} = K_{G^a\, \ov{T}_\alpha}, \nonumber\\[0.2em]
K_{T_\alpha \, \ov{T}_\beta} &= \frac{9}{4\, {\cal Y}^2}\, {\cal G}^{\alpha \beta}\,. \nonumber
\end{align}
This metric can be inverted to arrive at the inverse K\"ahler metric
\begin{align}
\label{eq:InvK}
K^{{S} \ov{S}} &=   \gamma_1\, ,\nn\\[0.2em]
 K^{G^a \, \ov{S}} &=  \gamma_1 \, b^a =K^{{S}\,\ov{G}^a}\, ,\nn\\[0.2em]
  K^{T_\alpha \, \ov{S}} &= \frac{\gamma_1 \, \hat{{k}}_{\alpha} + \gamma_2\, {k}_\alpha}{2} = K^{{S}\,\ov{T}_\alpha}\, , \\[0.2em]
K^{G^a \, \ov{G}^b} &= s\, {\cal G}^{ab} + \gamma_1\, b^a b^b \, ,\nn\\[0.2em]
 K^{T_\alpha \, \ov{G}^a} &= \,s \, {\cal G}^{ab} \, \hat{{k}}_{\alpha b} + \frac{(\gamma_1 \, \hat{{k}}_{\alpha}\,+ \gamma_2\, {k}_\alpha)\, b^a}{2} = K^{G^a\, \ov{T}_\alpha}\, ,\nn\\[0.2em]
K^{T_\alpha \, \ov{T}_\beta} &= \frac{4}{9}\, {\cal Y}^2\, {\cal G}_{\alpha \beta}  + s \, {\cal G}^{ab}\, {\hat{{k}}_{\alpha a}} \, {\hat{{k}}_{\beta b}} + \frac{(\gamma_1 \, \hat{{k}}_{\alpha} + \gamma_2\, {k}_\alpha)\,(\gamma_1 \, \hat{{k}}_{\beta} + \gamma_2\, {k}_\beta)}{4\, \gamma_1}\, , \nonumber
\end{align}
where the $\gamma_i$ are given by
\begin{equation}
\label{eq:gamma123}
 \gamma_1 = \frac{s^2\,\, \,(4 \,{\cal V}-\hat{\xi})}{({\cal V}-\hat{\xi})}\kom\quad \gamma_2 = \frac{3\, s\,\hat{\xi}}{\,({\cal V}-\hat{\xi})}\kom \quad\gamma_1 - s\, \gamma_2 = 4 \,s^2\, .
\end{equation}
Notice that, in the absence of the BBHL correction $\hat{\xi}= 0$, we find that the inverse metric components collected in Eq.~(\ref{eq:InvK}) reduce to the standard results of \cite{Grimm:2004uq} where $\gamma_1 = 4\, s^2$ and $\gamma_2 = 0$. Furthermore, in the absence of odd moduli, we recover the $(\alpha^{\prime})^{3}$-corrected K\"ahler metric components of \cite{Bobkov:2004cy}. Moreover, the inverse K\"ahler metric components have been computed in \cite{Hristov:2008if} in terms of rather lengthy expressions, whereas our results exhibit a surprisingly simple and compact structure.
This will come in handy in the subsequent section when working out powerful identities that are quintessential for deriving the master formulae in Sect.~\ref{sec:MasterFormulas}.

\subsection{No-scale structure and useful identities}\label{sec:NoScaleStrucIdentitiesTAPT} 

Ultimately, our aim is to streamline the computation of the $\cN=1$ $F$-term scalar potential \eqref{eq:Vtot} for general CY orientifold compactifications in the presence of perturbative quantum corrections.
Expanding the covariant derivatives in \eqref{eq:Vtot}, we find terms of the form ${K}_A\, {K}^{{A} \ov{B}}\p_{\ov B}\ov W$ or ${K}_A\, {K}^{{A} \ov{B}}K_{\ov B}$ for which certain tree level identities exist, see e.g. \cite{Grimm:2005fa}.
At higher order in $\alpha^{\prime}$,
we can utilise our expressions \eqref{eq:derK}, \eqref{eq:simpK} and \eqref{eq:InvK}
to compute equivalent relations.
Due to the length of the corresponding individual terms,
we listed the intermediate steps in Eqs.~(\ref{eq:identities0}), (\ref{eq:identities00}) in App.~\ref{app:UsefulIdentities}.
By applying appropriate summations,
we confirm that the following identities, which have already been known for the tree level K\"ahler potential (e.g. see \cite{Shukla:2015hpa}), still hold true even after including the BBHL $(\alpha^\prime)^3$-correction,
\begin{align}
\label{eq:identities1}
{K}_A\, {K}^{{A} \ov{S}} &= ({S} -\ov{S}) = - {K}^{{S} \ov {B}} \, {K}_{\ov B} \,,\nonumber\\
{K}_A\, {K}^{{A} \ov{G}^a} &= (G^a -\ov G^a) = - {K}^{{G^a} \ov {B}} \, {K}_{\ov B} \,,\\
{K}_A\, {K}^{{A} \ov{T}_\alpha} &= (T_\alpha -\ov T_\alpha) = - {K}^{{T_\alpha} \ov {B}} \,  {K}_{\ov B} \,. \nonumber
\end{align}
These relations significantly simplify the $F$-term scalar potential as we will demonstrate below in Sect.~\ref{sec:MasterFormulas}.
Moreover,
the above identities allow us to derive
\begin{align}
\label{eq:NoScaleIDProofExp} 
{K}_A\, {K}^{{A} \ov {B}}\, K_{\ov{B}} &= ({K}_A\, {K}^{{A} \ov{S}})\, K_{\ov{S}} + ({K}_A\, {K}^{{A} \ov{G}^a})\, K_{\ov{G}^a} + ({K}_A\, {K}^{{A} \ov{T}_\alpha})\, K_{\ov{T}_\alpha} \\
& = \biggl[1 + 2 \, s \, {\cal G}_{ab} \, b^a \, b^b + \frac{3\, \hat{\xi}}{2\,{\cal Y}}\biggr] + \biggl[-4 \, s \, {\cal G}_{ab} \, b^a \, b^b\, \biggr] + \biggl[ \frac{k_0}{2\, {\cal Y}} + 2 \, s \, {\cal G}_{ab} \, b^a \, b^b\biggr]\nonumber\\
&= 4 \,,\nonumber
\end{align}
%&= ({S} -\ov{S}) \biggl[\frac{(-i)}{2 \,s }\left(1 + 2 \, s \, {\cal G}_{ab} \, b^a \, b^b + \frac{3\, \hat{\xi}}{2\,{\cal Y}}\right)\biggr] \nonumber\\
%&\quad+ (G^a -\ov G^a)  \biggl[2 \, i \, {\cal G}_{ab} \, b^b\biggr] +(T_\alpha -\ov T_\alpha) \biggl[\frac{i \, t^\alpha}{2\, {\cal Y}}\biggr] \nonumber\\
where we used the relations $k_0 = {{k}}_\alpha \, t^\alpha =  6 {\cal V}$ and $\hat{k}_0 = \hat{{k}}_\alpha \, t^\alpha = - 4\, {\cal Y}\, {\cal G}_{ab}\, b^a\, b^b$ that follow from Eq.~(\ref{eq:genMetrices}).

Let us note that this no-scale structure relation, ${K}_A\, {K}^{{A} \ov {B}}\, K_{\ov{B}} = 4$, can also be anticipated, even after including tree level $(\alpha^{\prime})^{3}$-corrections, via homogeneity arguments \cite{Grimm:2005fa}.
That is, the K\"ahler potential \eqref{eq:KgenKGS} can be rewritten as
\bea
\label{eq:KqHomArg}
& & \hskip-2.0cm {K} =  -4 \ln g(s, {\cal V}^{2/3}) \kom g(s, {\cal V}^{2/3}) = \sqrt{\sqrt{2}\left({\cal V} \sqrt{s}+ \frac{\xi}{2}\, s^2\right)}\, ,
\eea
where $g(s, {\cal V}^{2/3})$ is a homogeneous function of degree one in the new coordinates $x^{i}\in \{s, {\cal V}^{2/3}\}$.
Indeed, we find that $g(\lambda \, x^i) = \lambda \, g(x^i)$ implies $x^i\, g_i = g$ and $x^i\, g_{ij} = 0$. Subsequently, one derives
\begin{align}
K_i &= -\frac{4 \, g_i}{g}\kom K_{ij} = -\frac{4 \, g_{ij}}{g} + \frac{4\, g_i\, g_j}{g^2} 
\end{align}
so that the no-scale identity is recovered from
\begin{align}
x^i\, \, K_{ij} &= -K_j\kom K^{ij}\, K_j = - x^i \kom K_i\, K^{ij}\, K_j = -x^j\, K_j = \frac{4\,x^j \, g_j}{g} = 4\,. 
\end{align}

To summarise, the explicit expression for the inverse K\"ahler metric together with the identities \eqref{eq:identities1} and \eqref{eq:NoScaleIDProofExp}
allows for a model independent reformulation of the $F$-term scalar potential which we provide in Sect.~\ref{sec:MasterFormulas}.
This is particularly useful for moduli stabilisation in generic CY orientifold compactifications with arbitrary numbers of K\"ahler moduli $T_\alpha$ and odd moduli $G^a$.

\subsection{No-scale breaking effects from $\mathrm{SL}(2,\mathbb{Z})$ invariance}\label{sec:LeadingOrderBrekingSL2Z} 
%\section{ $\mathrm{SL}(2,\mathbb{Z})$ invariance}\label{sec:LeadingOrderBrekingSL2Z} 

A natural question is how the no-scale identity \eqref{eq:NoScaleIDProofExp} is modified in the presence of further quantum effects.
Our previous homogeneity argument suggests that the no-scale structure is broken at higher order in the string loop expansion.
We can be even more precise by repeating the above analysis for a partially\footnote{Full modular invariance is only guaranteed by including also an additional $G$-dependent piece in \eqref{eq:ModVolume} as argued in \cite{Grimm:2007xm}, see \eqref{eq:YwithWS} below.} $\mathrm{SL}(2,\mathbb{Z})$ completed Kähler potential where $\cY$ in \eqref{eq:DefYTree} is replaced by\footnote{An F-theory analysis in \cite{Minasian:2015bxa} revealed that there exist further genuine $\cN=1$ corrections. At tree level, a single O7-plane wrapped on a divisor $D$ results in a shift of the Euler characteristic in \eqref{eq:DefYTree} by $\chi\rightarrow \chi+2\int_{X_{3}}\, D^{3}$. More generally, $f_{0}$ in \eqref{eq:ModVolume} would need to be replaced by a non-topological integral. We ignore such additional corrections subsequently.}
\begin{equation}\label{eq:ModVolume} 
{\cal Y} =  \cV + \frac{\zeta}{4}f_{0}(S,\bar{S})\kom \zeta=-\dfrac{\chi}{2(2\pi)^{3}}\, .
\end{equation}
Here,
$f_{0}(S,\bar{S})$ is the non-holomorphic Eisenstein series of weight $3/2$.
The modification \eqref{eq:ModVolume} is obtained from the $10$D $(\alpha^{\prime})^{3}$ correction $f_{0}R^{4}$ \cite{Green:1997as} as derived in $\cN=1$ CY orientifold compactifications \cite{Grimm:2007xm} under the assumption that a discrete subgroup $\Gamma_{S}\subset \mathrm{SL}(2,\mathbb{Z})$ survives in the $4$D theory.\footnote{For $\cN=2$ theories in 4D, it was conjectured in \cite{Robles-Llana:2006hby} that \eqref{eq:ModVolume} together with \eqref{eq:ModFunctionDef} is the correct modular completion.}
We focus here on the simpler case of the purely $S$-dependent part of the modular completion,
albeit D1-brane instanton corrections or, more precisely, $(p,q)$-strings give rise to a further $G^{a}$-dependent modular form in \eqref{eq:ModVolume} \cite{Grimm:2007xm}, cf. the discussion in Sect.~\ref{sec_conclusions}.\footnote{Another caveat in $\cN=1$ setups concerns the breaking of the product structure of moduli space as anticipated in \cite{Berg:2005ja}.
Due to the extended no-scale structure \cite{Cicoli:2007xp}, the effects of \cite{Berg:2005ja} appear at $1$-loop at order $(\alpha')^4$ making them sub-leading for our purposes.}

For the subsequent analysis, we define the modular functions
\begin{equation}\label{eq:ModFunctionDef} 
f_{k}(S,\bar{S})=\sum_{(\hat{l}_{1},\hat{l}_{2})\neq (0,0)}\, \dfrac{s^{\frac{3}{2}}}{(\hat{l}_{1}+S\hat{l}_{2})^{\frac{3}{2}+k}(\hat{l}_{1}+\bar{S}\hat{l}_{2})^{\frac{3}{2}-k}}\kom \bar{f}_{k}=f_{-k}\, .
\end{equation}
They transform covariantly under $\mathrm{SL}(2,\mathbb{Z})$
\begin{equation}
f_{k}\left (\dfrac{aS+b}{cS+d},\dfrac{a\bar{S}+b}{c\bar{S}+d}\right )=\left (\dfrac{cS+d}{c\bar{S}+d}\right )^{k}f_{k}(S,\bar{S})\, .
\end{equation}
Further, these functions satisfy
\begin{equation}
(S-\bar{S})\dfrac{\p}{\p S}f_{k}=\left (k+\dfrac{3}{2}\right )f_{k+1}-kf_{k}\quad ,\quad (S-\bar{S})\dfrac{\p}{\p\bar{S}}f_{k}=\left (k-\dfrac{3}{2}\right )f_{k-1}-kf_{k}
\end{equation}
which allows us to derive the following identities
\begin{align}
\p_{S}f_{0}&=\dfrac{3f_{1}}{2(S-\bar{S})}\kom
\p_{\bar{S}}f_{0}=-\dfrac{3f_{-1}}{2(S-\bar{S})}\kom
\p_{\bar{S}}\p_{S}f_{0}=-\dfrac{3f_{0}}{4(S-\bar{S})^{2}}\, .
\end{align}
This implies that
\begin{align}
\frac{\partial {\cal Y}}{\partial {S}} &= \frac{i\, \, \hat{k}_0}{8} +\dfrac{3\zeta}{8} \dfrac{f_{1}}{(S-\bar{S})}\kom\frac{\partial^{2} {\cal Y}}{\partial \bar{S}\partial {S}} = -\dfrac{3\zeta}{16} \dfrac{f_{0}}{(S-\bar{S})^{2}}\, .
\end{align}
We stress that $\p_{S}\p_{\bar{S}} \mathcal{Y}$ depends only on $f_{0}$ instead of $f_{\pm 2}$ due to aforementioned identities for $f_{k}$. Last but not least, we expand $f_{k}$ in the large $\mathrm{Im}(S)\gg 1$ (small string coupling) regime where
\begin{equation}\label{eq:ExpansionModFuncLargeImTauK} 
f_{k}(S,\bar{S})=a_{T}+\dfrac{a_{L}}{(1-4k^{2})}+\cO\left (\ee^{-\mathrm{Im}(S)}\right )
\end{equation}
in terms of
\begin{equation}\label{eq:TLCMFEXP} 
a_{T}=2\zeta(3)\mathrm{Im}(S)^{\frac{3}{2}}\kom a_{L}=\dfrac{2\pi^{2}}{3}\mathrm{Im}(S)^{-\frac{1}{2}}\, .
\end{equation}
The first term is associated with closed string tree level \cite{Gross:1986iv}, whereas the second term with $1$-loop effects \cite{Green:1981ya}. The final piece encodes contributions from non-perturbative D-instanton states \cite{Green:1997tv}.
For the lowest order modular functions, we can write
\begin{align}\label{eq:ExpansionModFuncTL} 
f_{0}(S,\bar{S})&=a_{T}+a_{L}+\cO(\mathrm{e}^{-\mathrm{Im}(S)})\kom f_{\pm 1}(S,\bar{S})=a_{T}-\dfrac{1}{3}a_{L}+\cO(\mathrm{e}^{-\mathrm{Im}(S)})\, .
\end{align}
To recover the convention of the tree level computation in the previous section,
we note that $ \zeta a_{T}=2\hat{\xi}$.

We now would like to compute the K\"ahler metric for the K\"ahler potential \eqref{eq:ModVolume} in CY orientifold compactifications.
In this context,
the sum in \eqref{eq:ModFunctionDef} needs to be appropriately restricted to orbits of $\Gamma_{S}$ \cite{Grimm:2007xm}
Proceeding as in Sect.~\ref{sec:InverseMetric},
we find that only the following components of \eqref{eq:simpK} are modified\footnote{We stress that \eqref{eq:KMSLExtended} is not modular invariant due to combinations of the form $f_{1}+f_{-1}$ or $f_{-1}$.
This is expected because one would need to include also the $G$-dependent piece in \eqref{eq:ModVolume}, see \eqref{eq:YwithWS} below.}
\begin{align}\label{eq:KMSLExtended} 
K_{{S} \ov{S}} &= \frac{1}{4\,s^2}\, \left(1 - \frac{3\, \zeta\, f_{0}}{8\, {\cal Y}}+ \frac{9\, \zeta^{2}\, f_{1}\, f_{-1}}{32\, {\cal Y}^2} \right) + \frac{{\cal G}_{ab}\, b^a\, b^b}{s} \left(1 + \frac{3\, \zeta}{16\, {\cal Y}} (f_{1}+f_{-1})\right) + \frac{9\, {\cal G}^{\alpha\beta} \hat{{k}}_\alpha\, \hat{{k}}_\beta}{16\, {\cal Y}^2}\,, \nonumber\\
K_{G^a \, \ov{S}} &=  -\frac{{\cal G}_{ab}\, b^b}{s} \left(1 + \frac{3\, \zeta}{8\, {\cal Y}} \, f_{-1}\right) - \frac{9\, {\cal G}^{\alpha\beta} \hat{{k}}_{\alpha a}\, \hat{{k}}_\beta}{8\, {\cal Y}^2} =\overline{K_{{S}\,\ov{G}^a}}, \nonumber\\
K_{T_\alpha \, \ov{S}} &= \frac{9\, {\cal G}^{\alpha\beta} \, \hat{{k}}_\beta}{8\, {\cal Y}^2} -\frac{3\, \zeta\, \, \, t^\alpha}{32\, s\, {\cal Y}^2}\, f_{-1}= \overline{K_{{S}\,\ov{T}_\alpha}}\,. 
\end{align}
The presence of the modular functions $f_{\pm 1}$ in $K_{G^a \, \ov{S}}$ and $K_{T_\alpha \, \ov{S}}$ imply that the components are complex.
The inverse Kähler metric can still be written in the form \eqref{eq:InvK} with the only difference being that the $\gamma_i$ are now defined as
\begin{align}\label{eq:GammaSLTZ} 
\gamma_{1}&=\dfrac{8s^{2}(8\mathcal{V}-\zeta f_{0})}{\gamma_{3}}\kom \gamma_{2}=\dfrac{24s\zeta f_{-1}}{\gamma_{3}}\kom \bar{\gamma}_{2}=\dfrac{24s\zeta f_{1}}{\gamma_{3}}\, ,\nn\\
\gamma_{3}&=8(2\mathcal{V}-\zeta f_{0})+9\zeta^{2}\frac{f_{0}^{2}-f_{1}f_{-1}}{4\mathcal{V}+\zeta f_{0}}\, .
\end{align}
In particular, $\gamma_{2}$ is now complex due to the presence of $f_{-1}$ in the numerator which requires appropriate complex conjugation in \eqref{eq:InvK}.
Interestingly, $K^{G^{a}\bar{S}}=\gamma_{1}b^{a}$ remains real.

Next,
we derive the corrected expressions of Sect.~\ref{sec:NoScaleStrucIdentitiesTAPT}.
First, the identities \eqref{eq:identities1} are modified as
\begin{align}\label{eq:IDKMKDSL} 
{K}_A\, {K}^{{A} \ov{S}}&=\dfrac{i}{2s\,\mathcal{Y}}\biggl \{\mathcal{Y}(\gamma_{1}-s\gamma_{2})+s(\cY-3\mathcal{V})(\gamma_{2}-\bar{\gamma}_{2})\biggl \}\nn\\
{K}_A\, {K}^{{A} \ov{T}_{\alpha}}&=\dfrac{-i}{128s^{2}\cY}\biggl \{64\sigma_{\alpha}\left [\cY(\gamma_{1}-s\gamma_{2})+s(\gamma_{2}-\bar{\gamma}_{2})(\cY-3\cV)\right ]\nn\\
&\hphantom{=\dfrac{-i}{128s^{2}\cY}\biggl \{}+\hat{k}_{\alpha}(\gamma_{2}-\bar{\gamma}_{2}) \left [\gamma_{3}\bar{\gamma}_{2}+96s\cV\right ] \biggl \}\nn\\
{K}_A\, {K}^{{A} \ov{G}^{a}}&=\dfrac{ib^{a}}{2s\cY}\biggl \{\cY(\gamma_{1}-s\gamma_{2})+s(\cY-3\mathcal{V})\left (\gamma_{2}-\bar{\gamma}_{2}\right ) \biggl \}
\end{align}
At tree level, we have\footnote{In fact, looking at \eqref{eq:ExpansionModFuncLargeImTauK}, $\gamma_{2}-\bar{\gamma}_{2}=0$ remains true even at the loop level which implies that $\gamma_{2}-\bar{\gamma}_{2}$ is associated with purely non-perturbative D-instanton corrections.}
\begin{equation}\label{eq:TreeLevelIDSGammaSLTZ} 
(\gamma_{1}-s\gamma_{2})\bigl |_{\text{tree}}=4s^{2}\kom (\gamma_{2}-\bar{\gamma}_{2})\bigl |_{\text{tree}}=0
\end{equation}
and thus recover \eqref{eq:identities1}, while including the 1-loop coefficient $a_{L}$ in the expansion \eqref{eq:ExpansionModFuncTL} leads to
\begin{align}
\dfrac{{K}_A\, {K}^{{A} \ov{S}}}{S-\ov S}&=\dfrac{{K}_A\, {K}^{{A} \ov{T}_{\alpha}}}{T_{\alpha}-\ov T_{\alpha}}=\dfrac{{K}_A\, {K}^{{A} \ov{G}^{a}}}{G^{a}-\ov G^{a}}= \dfrac{4\mathcal{V}+\zeta(a_{T}+a_{L})}{4\mathcal{V}+\zeta(a_{T}-a_{L})}\, .
\end{align}

Finally,
we aim at computing the former no-scale identity \eqref{eq:NoScaleIDProofExp} for the modified Kähler metric.
Overall, we obtain
\begin{equation}\label{eq:GenNoScaleIDBreaking} 
{K}_A\, {K}^{{A} \ov {B}}\, K_{\ov{B}}-4=-\dfrac{1}{2s^{2}}\biggl \{ -\gamma_{1}+s\dfrac{\gamma_{2}+\bar{\gamma}_{2}}{2}+4s^{2}+\dfrac{3\cV}{64\cY^{2}}\, \gamma_{3}\left (\gamma_{2}-\bar{\gamma}_{2}\right )^{2} \biggl \}\, .
\end{equation}
The right hand side is clearly real and vanishes at tree level due to \eqref{eq:TreeLevelIDSGammaSLTZ}.
To gauge the leading order breaking effect,
we restrict to linear order in $\zeta$ (any higher power would be modified by higher order $\alpha^{\prime}$ effects in $10$D) where
\begin{align}\label{eq:GenNoScaleIDBreakingLO} 
{K}_A\, {K}^{{A} \ov {B}}\, K_{\ov{B}}&=4+\dfrac{3}{8}\dfrac{\zeta (2f_{0}-f_{1}-f_{-1})}{\cV}=4+\dfrac{\zeta a_{L}}{\cV}+\cO\left (\ee^{-s}/\cV\right )\, .
\end{align}

Let us make the following two comments on these findings.
First,
the breaking of the generalised no-scale identity can already be anticipated from the previous homogeneity arguments.
Using the fact that at large $\mathrm{Im}(S)$ the $f_{k}$ enjoy an expansion of the form \eqref{eq:ExpansionModFuncLargeImTauK},
we have to modify \eqref{eq:KqHomArg} in such a way that
\begin{equation}
K=-4\log(g)\quad ,\quad g=\sqrt{\sqrt{2}\left (\mathcal{V}\sqrt{s}+\dfrac{\xi}{2}s^{2}+\dfrac{\zeta\pi^{2}}{6}+\mathcal{O}(\sqrt{s}\,\mathrm{e}^{-s})\right )}\, .
\end{equation}
Clearly, the $1$-loop and D-instanton contributions break the previous homogeneity argument $g(\lambda x^{A})\neq \lambda g(x^{A})$ which is why the no-scale identity must be broken.

Secondly and also more interestingly, this coefficient is directly related to $8$-derivative corrections in the $10$D effective action involving the R-R 3-form flux of the form $F_{3}^{2}R^{3}$.
Computing the $\alpha^{\prime}$-corrected flux scalar potential to leading order in $\zeta$ gives rise to
\begin{align}\label{eq:CorScalarPotAPTAPP} 
V_{\text{flux}}&=V_{\text{tree}}-\dfrac{\zeta f_{0}}{2\cV}V_{\text{tree}}+\dfrac{\mathrm{e}^{\mathbb{K}}}{2s}V_{\zeta}
\end{align}
where the second term comes from the standard Weyl rescaling of the $4$D metric.
In contrast,
the third term encodes the non-trivial effects of the $\mathrm{SL}(2,\mathbb{Z})$-completed $(\alpha^{\prime})^{3}$ corrections and reads\footnote{The fact that the $b^{a}$ dependence drops out completely is shown explicitly in Sect.~\ref{sec:MismatchDisc}.}
\begin{align}
V_{\zeta}&=\dfrac{3}{8}\;\dfrac{\zeta}{\cV^{3}}\biggl \{\left (2f_{0}-f_{1}-f_{-1}\right )\int_{X_{3}}\,  {G}_{3}\wedge\Omega\int_{X_{3}}\,  {{G}}_{3}\wedge\overline{\Omega}\nn\\
&\qquad\qquad+4 \mathrm{e}^{-2\phi_{0}} \left (2f_{0}+f_{1}+f_{-1}\right )\int_{X_{3}}\,  {H}_{3}\wedge\Omega\int_{X_{3}}\,  {{H}}_{3}\wedge\overline{\Omega} \biggl\}\, .
\end{align}
Notice that the R-R flux $F_{3}$ appears only in the first line with the same coefficient $\sim (3/8)\zeta (2f_{0}-f_{1}-f_{-1})/\cV$ as found in Eq.~\eqref{eq:GenNoScaleIDBreakingLO}.
In the large $\mathrm{Im}(S)$ expansion of $f_{k}$ \eqref{eq:ExpansionModFuncTL}, this then becomes
\begin{align}\label{eq:BBHLCorGVWIntH} 
V_{\zeta}&= \dfrac{3}{8}\;\dfrac{\zeta}{\cV^{3}}\biggl \{\left (\dfrac{8 a_{L}}{3}+\text{non-perturbative}\right )\int_{X_{3}}\, {G}_{3}\wedge\Omega\int_{X_{3}}\,  {{G}}_{3}\wedge\overline{\Omega}\\
&\hphantom{= \dfrac{3}{8}\;\dfrac{\zeta}{\cV^{3}}\biggl \{}+4 \mathrm{e}^{-2\phi_{0}} \left (4a_{T}+\dfrac{4 a_{L}}{3}+\text{non-perturbative}\right )\int_{X_{3}}\,  {H}_{3}\wedge\Omega\int_{X_{3}}\,  {{H}}_{3}\wedge\overline{\Omega} \biggl\}\nn\, .
\end{align}
At tree level, the only contribution comes from the second line through NS-NS 3-form flux as already observed in \cite{Becker:2002nn} which is the leading order no-scale breaking effect well-known from LVS.
Interestingly, the leading order contribution from $F_{3}$-flux is determined precisely by the coefficient derived in \eqref{eq:GenNoScaleIDBreakingLO}.
This interesting because all contributions in $V_{\zeta}$ are solely determined by zero mode Kaluza-Klein reductions of higher derivative corrections in the $10$D effective action such as $f_{0}|G_{3}|^{2}R^{3}$ and $f_{1}G_{3}^{2}R^{3}+f_{-1}\ov G_{3}^{2}R^{3}$ which were recently determined in \cite{Liu:2019ses}.
In this way,
the coefficient derived in \eqref{eq:GenNoScaleIDBreakingLO} can in principle be directly traced back to properties of the $10$D theory.

\section{Expressions for $\alpha^\prime$- and $g_s$-corrected $F$-term scalar potentials}\label{sec:GeneralScalarPotExpressions} 

In this section,
we apply the identities derived in the previous section to write down general master formulae for the $F$-term scalar potential at higher order in string perturbation theory.
Afterwards,
we briefly discuss moduli stabilisation in supersymmetric and non-supersymmetric settings.

\subsection{Three master formulae for general CY orientifold compactifications}\label{sec:MasterFormulas} 

The block diagonal nature of the total K\"ahler metric (and its inverse) facilitates the following splitting of the ${\cN} = 1$ $F$-term scalar potential
\begin{eqnarray}
\label{eq:V_gen}
& & \hskip-2cm V = \ee^{{\cal K}} \,\biggl[{\cal K}^{{\cal A} \ov {\cal B}} \, (D_{\cal A} W) \, (\ov D_{\ov {\cal B}} \ov{W}) -3 |W|^2 \biggr] \equiv V_{\text{cs}} + V_{k}\, ,
\end{eqnarray}

\vspace*{-0.25cm}
\noindent where
\begin{eqnarray}
\label{eq:VcsVk}
& & \hskip-1.5cm V_{\text{cs}} = \ee^{{\cal K}} \, {\mathbb K}^{{U^i} \ov {U^j}} \, (D_{U^i} W) \, (\ov D_{\ov {U^j}} \ov{W}), \quad V_{k} = \ee^{{\cal K}} \,\biggl(K^{{A} \ov {B}} \, (D_{A} W) \, (\ov D_{\ov {B}} \ov{W}) -3 |W|^2 \biggr)\,.
\end{eqnarray}

\vspace*{-0.25cm}
\noindent Recall that the indices $(i,j)$ count complex structure moduli $U^i$, while the remaining indices $(A,B)$ account for the rest of the chiral variables $\{{S}, G^a, T_\alpha\}$.
Given that our main focus will be mostly on moduli $S, T_\alpha$ and $G^a$, let us begin by looking at the various pieces in $V_k$ obtained from the Kähler potential \eqref{eq:DefYTree}.
We find that
\begin{align}
\label{eq:Vk_gen}
V_k &= \ee^{{\cal K}} \,\biggl[\left(K_A\, K^{{A} \ov {B}} \, K_{\ov B} -3 \right) |W|^2 + W_A\, K^{{A} \ov {B}} \, \ov W_{\ov B}+ \left(W\, K_A\, K^{{A} \ov {B}} \, \ov W_{\ov B} + \text{c.c.}\, \right) \biggr] \nonumber\\
& = \ee^{{\cal K}} \,\biggl(|W|^2 + W_A\, K^{{A} \ov {B}} \, \ov W_{\ov B} + \sum_{A \in \{{S}, G^a, T_\alpha\}} (A - \ov A) \, \, (W \, \ov W_{\ov A} - \ov W\, W_A) \biggr)
\end{align}
where we utilised the following two identities derived above (recall Eqs.~\eqref{eq:identities1} and \eqref{eq:NoScaleIDProofExp}),
\bea
\label{eq:identities}
& & K_A\, K^{{A} \ov {B}} \, K_{\ov B} = 4, \qquad  K_B\, K^{{B} \ov {A}} = (A - \ov A) = - \, K^{{A} \ov {B}} \, K_{\ov B}.
\eea
We stress again that these relations, while being naively true for the tree-level K\"ahler potential \cite{Grimm:2004uq}, remain correct even after including the BBHL correction \cite{Becker:2002nn}.
This gives rise to our ``first master formula" for the scalar potential,
\begin{equ}[First master formula]
\vspace*{-0.5cm}
\begin{equation}\label{eq:master2}
V = V_{\text{cs}} + \ee^{{\cal K}} \,\biggl(|W|^2 + W_A\, K^{{A} \ov {B}} \, \ov W_{\ov B} + \sum_{A \in \{{S}, G^a, T_\alpha\}} (A - \ov A) \, \, (W \, \ov W_{\ov A} - \ov W\, W_A) \biggr)\, .
\end{equation}
\end{equ}
Using the identities for the inverse metric in \eqref{eq:InvK},
the exact scalar potential is easily computed for any given model.
Contrary to commonly applied scenarios,
none of the moduli fields have been integrated out at this stage.
In particular,
it was imperative to treat $S$ on equal footing with $\{G^a, T_\alpha\}$ in deriving \eqref{eq:master2}.
The only undetermined input remains the superpotential which is easily plugged into the above expression, thereby making \eqref{eq:master2} highly attractive for moduli stabilisation purposes.

Although complex structure moduli are commonly studied separately from the remaining fields in the established Kähler moduli stabilisation procedures \cite{Kachru:2003aw,Balasubramanian:2005zx},
our exact identities in Sect.~\ref{sec_scalarpotential} allow for a compact expression for the complete $F$-term scalar potential in CY orientifold compactifications.
Indeed, using the K\"ahler derivatives in Eq.~(\ref{eq:derKcsSimp}) and the inverse K\"ahler metric in Eq.~(\ref{eq:invKcsSimp}), the complex structure piece $V_{\text{cs}}$ can be expressed as
\begin{align}
V_{\text{cs}} &= \ee^{{\cal K}} \biggl[ \left({\mathbb K}_{U^i}\, {\mathbb K}^{{U^i} \ov {U^j}} \, {\mathbb K}_{\ov{U^j}}\right) |W|^2 + \left({\mathbb K}_{U^i}\, {\mathbb K}^{{U^i} \ov {U^j}} \, W_{\ov{U^j}} W + \ov{W}\, W_{U^i} \, {\mathbb K}^{{U^i} \ov {U^j}} \, {\mathbb K}_{\ov{U^j}}\right) \nonumber\\
&\hphantom{=e^{{\cal K}} \biggl[}+ W_{U^i}\,{\mathbb K}^{{U^i} \ov {U^j}}\, \ov{W}_{\ov {U^j}})\biggr] \nn\\
&= \ee^{{\cal K}} \biggl[\frac{3\, l}{(l - 3\, \tilde{\xi})}\, |W|^2 +i\, \frac{(2\, l + 3\,\tilde{\xi})}{(l - 3\,\tilde{\xi})} \, u^i\, (W \ov{W}_{\ov U^i} - \ov{W} {W}_{U^i}) \nonumber\\
&\hphantom{=e^{{\cal K}} \biggl[}+  \frac{(2 \, l + 3 \, \tilde\xi)}{(l -3\, \tilde{\xi})} \left(u^i\, u^j - \frac{(l -3\, \tilde{\xi})}{3} \, l^{ij} \right) \,  W_{U^i} \, \ov{W}_{\ov U^j} \biggr],
\end{align}
where we have used the identities given in Eq.~(\ref{eq:id-KcsSimp}). Thus the most generic scalar potential for all moduli and axions can be read off from our ``second master-formula" for the scalar potential,
\begin{equ}[Second master formula]
\vspace*{-0.5cm}
\begin{align}
\label{eq:master1a}
V &= \ee^{{\cal K}} \,\biggl[\frac{4\, l -3\, \tilde\xi}{l - 3\, \tilde{\xi}}\, |W|^2 + W_{\cal A}\, K^{{\cal A} \ov {\cal B}} \, \ov W_{\ov {\cal B}} + \sum_{{\cal A} \in \{U^i, {S}, T_\alpha, G^a\}} ({\cal A} - \ov {\cal A}) \, \, (W \, \ov W_{\ov {\cal A}} - \ov W\, W_{\cal A})  \nonumber\\
&\quad + i\, \frac{9\, \tilde{\xi} \, u^i}{l - 3\,\tilde{\xi}} \, (W \ov{W}_{\ov U^i} - \ov{W} {W}_{U^i}) \biggr]\, ,
\end{align}
\end{equ}
where the summation $\cal A$ runs over all moduli $\lbrace U^i, {S}, T_\alpha, G^a\rbrace$. Notice that the last line arises because perturbative effects on the mirror type IIA side break the no-scale identity given in Eq.~(\ref{eq:id-KcsSimp}) as compared to Eq.~(\ref{eq:identities1}).
This is because $\hat{\xi}$ comes with a dilaton dependence in type IIB which ensures homogeneity of the $\alpha^{\prime}$-corrected Kähler potential,
while this is simply not the case for $\tilde{\xi}$. 
If this correction can be ignored via setting $\tilde\xi = 0$ in the limit of ``extremely'' large complex structure, the master formula \eqref{eq:master1a} reduces to an amazingly simple form
\bea
\label{eq:master1b}
& & \hskip-1cm V \simeq \ee^{{\cal K}} \,\biggl[4\, |W|^2 + W_{\cal A}\, K^{{\cal A} \ov {\cal B}} \, \ov W_{\ov {\cal B}} + \sum_{{\cal A} \in \{U^i, {S}, T_\alpha, G^a\}} ({\cal A} - \ov {\cal A}) \, \, (W \, \ov W_{\ov {\cal A}} - \ov W\, W_{\cal A})  \biggr]\, .
\eea

Before we continue,
we wish to point out that, although both the master formulae (\ref{eq:master2}) and (\ref{eq:master1a}) of the scalar potential are general, they are not fully equivalent. The only slight difference is the fact that the master formulae (\ref{eq:master2}) is also applicable to ``rigid" compactifications in which complex structure moduli are frozen or simply absent.
However, in arriving at the master formula (\ref{eq:master1a}), one implicitly assumes that there is at least one complex structure modulus present in the dynamics. For the so-called rigid compactification case, one simply has the following simplifications in the K\"ahler potential $({\mathbb K})$ of the complex structure sector \cite{Shukla:2015rua},
\bea
& & \left\{{\cal X}^0 = 1, \, \, {\cal F}_0 = -\,i \right\} \implies {\mathbb K} = -\ln\biggl[i\, \left(\ov {\cal X}^0 \, {\cal F}_0 - {\cal X}^0 \, \ov {\cal F}_0 \right)\biggr] = -\ln2.
\eea
In fact, one can choose a normalisation in the holomorphic three-form $\Omega_3$ by a factor of $\sqrt{2}$ to make ${\mathbb K} = 0$
which, in turn,
also needs to be incorporated in the superpotential,
thereby giving rise to $e^{\cal K} = \frac{1}{4\,s \, {\cal Y}^2}$. Thus, the rigid orientifold case leads to $V_{\text{cs}} = 0$ in the master formula (\ref{eq:master2}) making it $V = V_k$.
In contrast, (\ref{eq:master1b}) will have a factor of $4|W|^2$ in the first term instead of $|W|^2$, and the $U^i$ moduli will not appear in the summation over ${\cal A}$. Given that we are interested in studying models with a complex structure moduli dependence, for our purposes the two expressions are equivalent. Nevertheless it is worth to point out this subtlety in case the reader wants to apply the above to rigid compactifications, see for example \cite{Shukla:2019akv}.

\subsubsection*{A third Master formula for the $\mathrm{SL}(2,\mathbb{Z})$-completed Kähler potential}

As a final step,
we extend our toolkit by a third master formula that remains true even beyond string tree level.
Indeed,
we have seen that the no-scale structure is broken once higher string loop corrections are taken into account.
This modification leads to another master formula that is exact for general superpotentials and the Kähler metric derived from the Kähler potential depending on the $\mathrm{SL}(2,\mathbb{Z})$-completed volume \eqref{eq:ModVolume}.

Initially, we rewrite the identities \eqref{eq:IDKMKDSL} in the form
\begin{align}\label{eq:identitiesSL} 
{K}_A\, {K}^{{A} \ov{S}}=\dfrac{{K}_A\, {K}^{{A} \ov{G}^{a}}}{b^{a}}&=\dfrac{i}{2s} F_{1}\kom 
{K}_A\, {K}^{{A} \ov{T}_{\alpha}}=\dfrac{-i}{2s^{2}}\biggl \{\sigma_{\alpha} F_{1}+\hat{k}_{\alpha}F_{2}\biggl \}\, ,
\end{align}
as well as \eqref{eq:GenNoScaleIDBreaking} as
\begin{equation}
{K}_A\, {K}^{{A} \ov {B}}\, K_{\ov{B}}-3=1-\dfrac{F_{3}}{2s^{2}}
\end{equation}
where we introduced
\begin{align}
F_{1}&=(\gamma_{1}-s\gamma_{2})+\dfrac{s(\cY-3\mathcal{V})}{\cY}(\gamma_{2}-\bar{\gamma}_{2})\kom F_{2}=\dfrac{\gamma_{2}-\bar{\gamma}_{2}}{64\cY} \left [\gamma_{3}\bar{\gamma}_{2}+96s\cV\right ]\, ,\nn\\
F_{3}&= -\gamma_{1}+s\dfrac{\gamma_{2}+\bar{\gamma}_{2}}{2}+4s^{2}+\dfrac{3\cV}{64\cY^{2}}\, \gamma_{3}\left (\gamma_{2}-\bar{\gamma}_{2}\right )^{2}\, .
\end{align}
Here,
the $\gamma_{i}$ are the ones defined in Eq.~\eqref{eq:GammaSLTZ}.
Following the same steps as in \eqref{eq:Vk_gen}, we arrive at a third master formula given by
\begin{equ}[Third master formula]
\vspace*{-0.5cm}
\begin{align}
\label{eq:master3}
V &= V_{\text{cs}} + \ee^{{\cal K}} \,\biggl( \left (1-\dfrac{F_{3}}{2s^{2}}\right ) |W|^2 +\dfrac{1}{4s^{2}} \sum_{A \in \{{S}, G^a, T_\alpha\}} (A - \ov A) \, \, (F_{1}W \, \ov W_{\ov A} -\ov F_{1} \ov W\, W_A) \nn\\
&\hphantom{= V_{\text{cs}} + \ee^{{\cal K}} \,\biggl( } + W_A\, K^{{A} \ov {B}} \, \ov W_{\ov B}+\dfrac{-i\hat{k}_{\alpha}}{2s} \left(F_{2}W \, \ov W_{\ov T_{\alpha}} -\ov F_{2} W_{T_{\alpha}} \, \ov W\, \right)\biggr)\, .
\end{align}
\end{equ}
This expression provides the exact $F$-term scalar potential for $\cN=1$ orientifold compactifications including corrections from both string loop and D-instanton effects.
Recall that the combination $(\gamma_{2}-\bar{\gamma}_{2})$ vanishes at both string tree and $1$-loop level which implies that $F_{2}$ contributes only non-perturbatively via D-instanton contributions $\sim\ee^{-s}$.
By using the tree level identities \eqref{eq:TreeLevelIDSGammaSLTZ},
one easily verifies that $F_{1}=4s^{2}$ and $F_{2}=F_{3}=0$ leading us back to our first master formula \eqref{eq:master2}.
At string $1$-loop order,
we find
\begin{equation}
F_{1}=4s^{2} \dfrac{4\mathcal{V}+\zeta(a_{T}+a_{L})}{4\mathcal{V}+\zeta(a_{T}-a_{L})} \kom F_{2}=0\kom F_{3}=\dfrac{-8s^{2}\zeta a_{L}}{4\mathcal{V}+\zeta(a_{T}-a_{L})}
\end{equation}
%F_{3}=4s^{2}\left (1- \dfrac{4\mathcal{V}+\zeta(a_{T}+a_{L})}{4\mathcal{V}+\zeta(a_{T}-a_{L})} \right )
in terms of tree level and loop coefficient $a_{T}$ and $a_{L}$ defined in \eqref{eq:TLCMFEXP}.
Thus,
ignoring D-instanton effects,
we arrive at the $1$-loop expression
\begin{align}
\label{eq:master3a}
V\bigl |_{\text{1-loop}} &= V_{\text{cs}} + \ee^{{\cal K}} \,\biggl( \dfrac{4\mathcal{V}+\zeta(a_{T}+3a_{L})}{4\mathcal{V}+\zeta(a_{T}-a_{L})}|W|^2+ W_A\, K^{{A} \ov {B}} \, \ov W_{\ov B} \\[0.6em]
&\hphantom{= V_{\text{cs}} + \ee^{{\cal K}} \,\biggl( } +\dfrac{4\mathcal{V}+\zeta(a_{T}+a_{L})}{4\mathcal{V}+\zeta(a_{T}-a_{L})} \sum_{A \in \{{S}, G^a, T_\alpha\}} (A - \ov A) \, \, (W \, \ov W_{\ov A} - \ov W\, W_A) \nn\, .
\end{align}

In the remainder of this paper,
we focus on the two master formulae \eqref{eq:master2} and \eqref{eq:master1a} for simplicity.
Without too much effort,
everything that has been computed is easily generalisable by using \eqref{eq:master3} or \eqref{eq:master3a}.

\subsection{Supersymmetric moduli stabilisation}

The supersymmetric solutions can be obtained by imposing the vanishing of the covariant derivatives with respect to the chiral varibales,
\bea
\label{eq:Susy-conditions}
& & D_{U^i} W \, = 0 = \, \ov D_{\ov {U^j}} \ov{W}, \qquad \qquad D_{S} W \, = 0 = \, \ov D_{\ov{S}} \ov{W}, \\
& & D_{G^a} W \, = 0 = \, \ov D_{\ov{G}^a} \ov{W}, \qquad \qquad D_{T_\alpha} W \, = 0 = \, \ov D_{\ov{T}_\alpha} \ov{W}. \nonumber
\eea
Using the K\"ahler derivatives in Eqs. (\ref{eq:derKcsSimp}), (\ref{eq:derK}) and \eqref{eq:genMetrices}, these conditions read explicitly
\begin{align}
\label{eq:susy-USTG}
W_{U^i}  &= - \, \frac{6\, i\, l_i}{4\, l+ \tilde{\xi}} \, W\kom W_{G^a} = - \, i \,  \frac{\hat{k}_{a}}{2\, {\cal Y}} \,  W\kom W_{T_\alpha} = \frac{i \, t^\alpha}{2\, {\cal Y}} \, W\,,\nonumber\\
W_S &= - \, \frac{i}{2 \,s }\left(1 + 2 \, s \, {\cal G}_{ab} \, b^a \, b^b + \frac{3\, \hat{\xi}}{2\,{\cal Y}}\right) W\, .
\end{align}
For a supersymmetric minimum, these conditions (\ref{eq:susy-USTG}) must be imposed for all the chiral variables for a generic superpotential $W=W(U^i, S, T_\alpha, G^a)$.

For a superpotential $W=W(U^i, S, G^a)$ which does not depend on the $T_\alpha$ coordinates, the $F$-term conditions (\ref{eq:susy-USTG}) yield
\bea
& & W_{T_\alpha} = 0  = \ov{W}_{\ov T_\alpha}  \quad \implies \quad \langle W \rangle = 0 = \langle \ov{W} \rangle,
\eea
since the $t^\alpha$ cannot vanish altogether and the decompatfication limit ${\cal Y} \to \infty$ needs to be avoided.
This needs to be contrasted with the situation where the superpotential does not depend on the $G^{a}$, i.e.,  $W=W(U^i, S, T_\alpha)$.
In this case, \eqref{eq:susy-USTG} implies
\bea
& & \hskip-1cm W_{G^a} = 0  = \ov{W}_{\ov G^a}  \quad \implies \quad \Big\langle\frac{\hat{k}_{a}}{{\cal Y}}  \Big\rangle = 0 = \Big\langle \frac{\hat{k}_{\alpha a b}\, t^\alpha \, b^b}{{\cal Y}} \Bigr\rangle \quad {\rm for} \, \, \langle W \rangle \neq 0 \neq \langle \ov{W} \rangle,
\eea
which can be trivially satisfied for $\langle\hat{k}_{a}\rangle=\langle\hat{k}_{\alpha a b}\, t^\alpha \, b^b\rangle = 0$, or in particular for $\langle b^a \rangle = 0, \,\, \forall a$. This is because the saxionic partner of the R-R $C_{2}$-axions $c^{a}$ in the odd moduli $G^{a}$ are the NS-NS axions $b^{a}$ themselves.
For this reason, there are no issues with imposing $\langle b^a \rangle = 0$ at the extremum contrary to demanding the geometrical volume moduli to vanish $\langle t^\alpha \rangle = 0$ in the previous case which would oppose the SUGRA approximation.

To illustrate this point,
we look at the GVW superpotential $W= W_0(U^i,S)$ in \eqref{eq:WGVW} which is generated from NS-NS 3-form flux $(H_3)$ and the R-R 3-form flux $(F_3)$.
Subsequently,
we write
\begin{equation}
W_0 (U^i, S) =w_{F}(U^i) + S w_{H}(U^i)
\end{equation}
where $ w_{F} (U^i)$ and $ w_{H} (U^i)$ are some cubic polynomials in complex structure moduli $U^i$. In this case, the tree level SUSY conditions in (\ref{eq:susy-USTG}) with $\hat{\xi}=0$ reduce to (ignoring the condition for $T_{\alpha}$)
\bea
\label{eq:susy-USTG-tree}
W_{U^i}  = - \, \frac{3\, i\, l_i}{2\, l+3\tilde{\xi}} \, W \kom W_S = - \, \frac{i}{2 \,s }\left(1 + 2 \, s \, {\cal G}_{ab} \, b^a \, b^b \right) W\kom W_{G^a} = - \, i \,  \frac{\hat{k}_{a}}{2\, {\cal V}} \,  W = 0 \, . \nonumber
\eea
These conditions are collectively solved for $\langle W_0 \rangle \neq 0$ leading to the following relations determining the VEVs of the respective moduli/axions,
\bea
\label{eq:susy-USTG-tree1}
& & \hskip-0.5cm \langle c_0 \rangle = - \frac{\langle m \rangle \,\langle n \rangle + \langle {\frak m} \rangle\, \langle {\frak n} \rangle}{| w_{H} |^{2}}, \qquad \langle s \rangle = \frac{\langle {\frak m} \rangle\, \langle n \rangle - \langle m \rangle\, \langle {\frak n} \rangle}{| w_{H} |^{2}}\, , \\
& & \hskip-0.5cm m_i  = \frac{1}{| w_{H} |^{2}} \biggl[m \, \left( n \, n_i - {\frak n} \, {\frak n}_i \right) + {\frak m} \, \left({n}  \, {\frak n} _i + {\frak n}\, {n}_i \right) - \frac{12\, l_i}{4l+\tilde{\xi}}\, n \, \left(m \, {\frak n} - n \, {\frak m} \right)\biggr] \quad \forall \, \, i,  \nonumber\\
& & \hskip-0.5cm {\frak m}_i  = \frac{1}{| w_{H} |^{2}} \biggl[{\frak m} \, \left({\frak n} \, {\frak n}_i  - n \, n_i \right) + {m} \, \left({n}  \, {\frak n} _i + {\frak n}\, {n}_i \right) - \frac{12\, l_i}{4l+\tilde{\xi}}\, {\frak n} \, \left(m \, {\frak n} - n \, {\frak m} \right)\biggr] \quad \forall \, \, i,  \nonumber
\eea
where the complex structure moduli $\{u^i, v^i\}$ get their respective susy VEVs through the last two set of conditions. Here we have defined,
\begin{align}
\label{eq:ReMetc}
m &= \text{Re}( w_{F} )\kom {\frak m} = \text{Im}( w_{F} )\kom {m_i} = \text{Re}(\partial_{U^i} w_{F} )\kom {{\frak m}_i} = \text{Im}(\partial_{U^i} w_{F} )\, , \nonumber\\
n &= \text{Re}( w_{H} )\kom\,\,\, {\frak n} = \text{Im}( w_{H} )\kom\,\,\, {n_i} = \text{Re}(\partial_{U^i} w_{H} )\kom\,\,\, {{\frak n}_i} = \text{Im}(\partial_{U^i} w_{H} )\, .
\end{align}
In particular,
the SUSY minimum determines the $h^{1,1}_{-}$ conditions
\begin{equation}\label{eq:Susy-ba0} 
\langle \hat{k}_{a} \rangle=\langle\hat{k}_{\alpha a b}\, t^\alpha \, b^b\rangle = 0\, .
\end{equation}
That is,
the point along the flat direction of the NS-NS axions $b^a$ at which SUSY is restored is given by $\langle b^a \rangle = 0, \, \forall \, a$.
%even if the superpotential has no explicit dependence on any of the odd moduli $G^a$.
%Of course, the condition (\ref{eq:Susy-ba0}) always has a solution, namely 
It turns out that the above minimum determined by \eqref{eq:susy-USTG-tree1} and \eqref{eq:Susy-ba0} is Minkowskian,
\bea
\label{eq:VGVWsusytree}
& & \langle V_{GVW}^{tree} \rangle = 0
\eea
which breaks SUSY because $D_{T_{\alpha}}W\neq 0$.

We close this section with two comments.
First, as we will show momentarily,
stabilising the NS-NS axions $b^{a}$ through fluxes is apparently misleading.
The point is that, in the absence of a $T_{\alpha}$ dependence in $W$, the scalar potential in the basis $\lbrace t^{\alpha},b^{a}\rbrace$ is independent of $b^{a}$ (and trivially $c^{a}$).
This becomes obvious from our second master formula \eqref{eq:master1a} where, setting $W=W(U^{i},S)$, the only potential source of a $b^{a}$ dependence could arise from $W_{\cal A}\, K^{{\cal A} \ov {\cal B}} \, \ov W_{\ov {\cal B}}$.
However,
both $K^{S\bar{S}}$ and $\mathbb{K}^{i\bar{\jmath}}$ are independent of $b^{a}$.
This means that the $b^{a}$ dependence arises implicitly from working in the basis of chiral variables.

Secondly,
we notice that an explicit $b^{a}$ dependence in \eqref{eq:master1a} arises as soon as the superpotential also depends on $T_{\alpha}$, $W=W(U^{i},S,T_{\alpha})$.
In this case,
$K^{T_{\alpha}\ov T_{\beta}}$ and $K^{T_{\alpha}\ov S}$ induce terms like $\hat{k}_{a}$ etc.
as can be seen from \eqref{eq:InvK}.
In addition,
the chiral fields $T_{\alpha}$ in \eqref{eq:N1coords} include the $b^{a}$ explicitly.
As it turns out,
the minimum is still determined by $\langle b^a \rangle = 0, \, \forall \, a$.
In this sense, the axionic solution in Eq. (\ref{eq:Susy-ba0}) is a quite `special solution' which reappears in the subsequent scalar potential analysis in Sect.~\ref{sec:SPAnalysisWT}.
In fact,
this can lead to significant simplifications in the inverse K\"ahler metric,
thereby generically restoring many well known moduli stabilisation schemes, in particular (A)dS$_{4}$ vacua realised in KKLT and LVS models, from more general set-ups including the odd moduli.

\subsection{Non-supersymmetric moduli stabilisation}

For non-supersymmetric moduli stabilisation, one needs to consider the full scalar potential $V$ in Eq.~\ref{eq:V_gen}. Using the expressions in Eq.~(\ref{eq:master2}) and Eq.~(\ref{eq:master1a}), we now analyse a variety of scenarios by considering different forms of the superpotential.
Occasionally, we will also use the scalar potential formulation in Eq.~(\ref{eq:master1b}) which is valid in the large complex structure regime. Having these so-called ``master formulae" for the generic scalar potential applicable for arbitrary numbers of $S,  U^i, T_\alpha$ and $G^a$ moduli at hand allows us to easily read off the various pieces using the inverse K\"ahler metric and ansatz for the superpotential.

Indeed, for a general superpotential, our first master formula \eqref{eq:master2} gives rise to
\begin{align}\label{eq:master2GenExp} 
V &= V_{\text{cs}} + \ee^{{\cal K}} \,\biggl(|W|^2 + W_A\, K^{{A} \ov {B}} \, \ov W_{\ov B} - 4\,s \, \, \text{Im}(W \, \ov W_{\ov S}) - 4\,s \,b^a \, \text{Im}(W \, \ov W_{\ov G^a}) \nonumber\\
&\hphantom{= V_{\text{cs}} + \ee^{{\cal K}} \,\biggl(} + 2\, (k_\alpha - s\, \hat{k}_\alpha) \, \text{Im}(W \, \ov W_{\ov T_\alpha}) \biggr )\, .
\end{align}
Equivalently,
the second master formula (\ref{eq:master1a}) gives rise to
\begin{align}
V &=  \ee^{{\cal K}} \,\biggl[\dfrac{4\,l-3\,\tilde{\xi}}{l-3\,\tilde{\xi}}\, |W|^2 + W_{\cal A}\, K^{{\cal A} \ov {\cal B}} \, \ov W_{\ov {\cal B}} -{ 2\, u^i \left (2+\dfrac{9\tilde{\xi}}{l-3\,\tilde{\xi}}\right )} \,\text{Im}(W \, \ov W_{\ov U^i})  \nonumber\\
& \hphantom{= \ee^{{\cal K}} \,\biggl[} - 4\,s \, \, \text{Im}(W \, \ov W_{\ov S})- 4\,s \,b^a \, \text{Im}(W \, \ov W_{\ov G^a}) + 2\, (k_\alpha - s\, \hat{k}_\alpha) \, \text{Im}(W \, \ov W_{\ov T_\alpha}) \biggr]\, .
\end{align}
Given that we will usually ignore any explicit complex structure moduli dependence,
we mostly work with variants of \eqref{eq:master2GenExp}.
Furthermore, it is straight forward to generalise \eqref{eq:master2GenExp} to higher orders in the loop expansion by using our third master formula \eqref{eq:master3}.
Plugging in the inverse K\"ahler metric components of Eq. (\ref{eq:InvK}),
we find
\begin{equation}
V=V_{0}+V_{1}+V_{2}
\end{equation}
in terms of
\begin{align}\label{eq:ScalarPotGModGen} 
V_0 &= V_{\text{cs}} + \ee^{\cal K} \biggl[|W|^2 - 4 s \, {\rm Im}\bigl[W \ov {W}_{\ov{S}}\bigr] + \, {W}_{S} \,\gamma_1 \, \ov{W}_{\ov{S}}\nn \\
&\quad  + \, {W}_{T_\alpha}\, \left(\frac{4{\cal Y}^2 \, {\cal G}_{\alpha\beta}}{9} + \frac{\gamma_2^2\, k_\alpha\, k_\beta}{4\, \gamma_1}\right)\, \ov {W}_{\ov T_{\beta}} \nonumber\\
&\quad + 2 \, k_\alpha \, {\rm Im}\bigl[W \ov {W}_{\ov T_{\alpha}}\bigr] +\, \gamma_2 \, k_\alpha\, \, {\rm Re}[{W}_{S} \, \ov {W}_{\ov T_\alpha}] \biggr]\, , \nonumber\\
V_1 &= \ee^{\cal K} \, \biggl[\gamma_1\, \hat{k}_\alpha \, {\rm Re}[{W}_{S} \, \ov {W}_{\ov T_\alpha}] - \, 2 \, s \, \hat{k}_\alpha \, {\rm Im}\bigl[W \ov {W}_{\ov T_{\alpha}}\bigr]\nn\\
&\quad+\biggl(s\, {\cal G}^{ab}\, \hat{k}_{\alpha a}\, \hat{k}_{\beta b} + \frac{\gamma_1}{4} \, \hat{k}_\alpha \, \hat{k}_\beta\, +\frac{\gamma_2}{4}\, (k_\alpha \, \hat{k}_\beta + k_\beta \, \hat{k}_\alpha) \biggr)\, {W}_{T_\alpha} \, \ov {W}_{\ov T_{\beta}}\biggr]\, ,\\
V_{2}
&= \ee^{{\cal K}} \,\biggl[2 \gamma_1 \, b^a\text{Re}( W_S\, \ov W_{\ov G^{a}} ) +\left (s\, {\cal G}^{ab} + \gamma_1\, b^a b^b\right ) W_{G^{a}}\, \ov W_{\ov G^{b}}-4sb^{a} \;\text{Im}(W \, \ov W_{\ov G^{a}})\nn\\
&\hphantom{= \ee^{{\cal K}} \,\biggl(}+2\left ( \,s \, {\cal G}^{ab} \, \hat{{k}}_{\alpha b} + \frac{(\gamma_1 \, \hat{{k}}_{\alpha}\,+ \gamma_2\, {k}_\alpha)\, b^a}{2} \right ) \text{Re}(W_{T_{\alpha}}\, \ov W_{\ov G^{a}} ) \biggr]\nn\, .
\end{align}
We will utilise this formula to study a variety of scenarios based on the form of the superpotential.
Subsequently, we separate our analysis into three steps:
\begin{itemize}
\item{{\bf step-1:} We study the GVW superpotential $W= W(U^i, S)= W_0$ defined in \eqref{eq:WGVW}.}
\item{{\bf step-2:} We consider superpotentials of the form $W=W(U^i, S, T_\alpha)$ which also depend on the $T_\alpha$ moduli. Such corrections can be induced via the non-perturbative effects and subsequently used for Kaehler moduli stabilisation along the lines of KKLT \cite{Kachru:2003aw} or LVS \cite{Balasubramanian:2005zx}.}
\item{{\bf step-3:} In the final step, we examine a class of superpotentials which can generically depend on all the moduli, namely $W=W(U^i, S, T_\alpha, G^a)$. The various sources for inducing such superpotential terms are for instance (non-)geometric flux \cite{Aldazabal:2006up, Benmachiche:2006df, Robbins:2007yv, Guarino:2008ik, Blumenhagen:2013hva, Shukla:2015rua, Shukla:2015bca, Blumenhagen:2015kja, Blumenhagen:2015lta, Shukla:2015hpa, Shukla:2016hyy, Shukla:2016xdy, Plauschinn:2018wbo, Shukla:2019wfo}, (fluxed)$E3$-instantons \cite{Grimm:2007xm,Grimm:2011dj} or D5-gaugino condensation \cite{Grimm:2007xm,Grimm:2007hs,Ben-Dayan:2014zsa, Ben-Dayan:2014lca}.}
\end{itemize}
In the following sections, we will show how our master formula in the form of Eq.~(\ref{eq:ScalarPotGModGen}) can help us ``reading-off" the scalar potential pieces for the above three generic classes of models, which can be subsequently utilised for moduli stabilisation purposes.
Given that our expressions are exact and compact,
it facilitates the numerical implementation of both the scalar potential and stationary point conditions.
The problem of stabilising moduli is thus reduced to finding minima for given choices of parameters which can be achieved via sophisticated search optimisation algorithms \cite{Cole:2019enn,AbdusSalam:2020ywo,Krippendorf:2021uxu}.

\vskip0.2cm
\noindent
{\bf Note:} We stress that, after knowing the explicit form of the scalar potential $V$ in terms of the set of real moduli and axions, namely $\{s, c_0, t^\alpha, \rho_\alpha, b^a, c^a\}$, we plan to study moduli stabilisation using the set of real variables and not the complexified chiral variables.
This is because converting the $t^{\alpha}$ to $\tau_{\alpha}$ might not necessarily be possible analytically,
though it is always defined implicitly.
In particular, the map between the two choices is bijective in the Käher cone where the Kähler cone conditions pick out the unique solution in the quadratic equation relating the $\tau_{i}$ to the $t_{k}$.
The two choices are therefore equivalent and we will henceforth assume that e.g. the 2-cycle volume moduli $t^\alpha$ are independent variables in the minimisation process, which are otherwise implicit functions of the chiral variables $t^\alpha \equiv t^\alpha(S, T_\alpha, G^a; \ov{S}, \ov{T}_\alpha, \ov{G}^a)$.
When re-deriving the supersymmetry solutions from extrema of the scalar potential, one has to appropriately take this into account.

\section{Analysing the scalar potential for $W=W(U^i, {S})$}\label{sec_Loopholes}

In this section we illustrate the power of the master formulae we derived for the general scalar potential. For example, an immediate result of the formula (\ref{eq:master1a}) is the case of a constant superpotential $W= w_0$, which does not depend on any of the moduli and subsequently leads to a scalar potential of the following form, 
\bea
& & V = \ee^{{\cal K}} \, \frac{4\, l - 3\,\tilde\xi}{l - 3\,\tilde{\xi}}\,|w_0|^2 \quad \xrightarrow[\text{}]{\tilde\xi \to 0} \quad 4\, \ee^{{\cal K}} \,|w_0|^2.
\eea
Note that, in the large complex structure approximation, the $\tilde\xi \to 0$ limit on the mirror side can generically be a reasonable assumption.

To begin with,
we like to point out that in frequently studied set-ups of K\"ahler moduli stabilisation one assumes the complex-structure moduli and axio-dilaton to be stabilised supersymmetrically via $3$-form fluxes. For that purpose, we consider the usual GVW flux superpotential \eqref{eq:WGVW}
\bea
& & W(U^i, {S}) =  w_{F} (U^i) + {S} \, \,  w_{H} (U^i) \equiv W_0\, ,
\eea
where $ w_{F} (U^i)$ and $ w_{H} (U^i)$ are cubic polynomials in the complex structure moduli $U^i$ induced by $F_{3}$- and $H_{3}$-flux respectively.
Hence, the superpotential $W$ is still independent of any of the odd moduli $G^a$ and the $T_\alpha$ moduli. It is convenient to impose the $F$-term conditions
\bea
\label{eq:susyDW}
& & D_{U^i} W = 0 = \ov{D}_{\ov{U^i}}\, {\ov W}, \qquad D_{S} W = 0 = \ov{D}_{\ov{S}}\, {\ov W}\,.
\eea
However, we argue that this approach is actually misleading in the presence of odd moduli. This is because of the fact that imposing the SUSY stabilisation for the axio-dilaton induces some ``fictitious" terms in the scalar potential which are absent when one directly computes the full scalar potential.

\subsection{An apparent mismatch in the standard approach}\label{sec:MismatchDisc}

Let us first highlight the subtle issue that appears when considering moduli stabilisation by imposing the SUSY stabilisation of the axio-dilaton via $D_S W = 0 = D_{\ov S}\ov{W}$. For this purpose, we consider the $F$-term scalar potential in Eq.~(\ref{eq:V_gen}) which for $W \equiv W_0(U^i, {S})$ can be expressed in the following way,
\bea
\label{eq:Vgen0}
& & \hskip-0.5cm V = \ee^{{\cal K}} \, \biggl[{\mathbb K}^{{U^i} \ov {U^j}} \, (D_{U^i} W_0) \, (\ov D_{\ov {U^j}} \ov{W_0}) + K^{{S} \ov{S}} \, (D_{S} W_0) \, (\ov D_{\ov{S}} \ov{W_0}) \nonumber\\
& & \hskip0.3cm + \left(K^{{S} \ov{G}^a} \, K_{\ov{G}^a} + K^{{S} \ov{T}_\alpha} \, K_{\ov{T}_\alpha} \right) (D_{S} W_0)  \, \ov{W_0} + \left(K_{G^a} \,K^{{G^a} \ov{S}} + K_{T_\alpha} \,K^{{T_\alpha} \ov{S}} \right) (\ov D_{\ov{S}} \ov{W_0}) \,  W_0 \nonumber\\
& & \hskip0.3cm + \left(K_{A^\prime} \, K^{{A^\prime} \ov{B^\prime}} \, \ov K_{\ov{B^\prime}} -3 \right)|W_0|^2 \biggr].
\eea
Here, the indices $\{A^\prime, B^\prime\}$ run only over $G^a$ and $T_\alpha$ moduli, and do not include the axio-dilaton. Now using the identities given in the appendix~\ref{app:UsefulIdentities} leads to the simplified relations
\bea
\label{eq:useful-id}
& & K_{G^a} \,K^{{G^a} \ov{S}} + K_{T_\alpha} \,K^{{T_\alpha} \ov{S}} = - \frac{9\, i\, s\, \hat\xi\, {\cal V}}{(2{\cal V} + \hat\xi)\,({\cal V}- \hat\xi)} + \frac{\, i\, s^2\, \hat{k}_0\, (4{\cal V}-\hat\xi)}{2\,(2{\cal V} + \hat\xi)\,({\cal V}- \hat\xi)}\\\
& & K_{A^\prime} \, K^{{A^\prime} \ov{B^\prime}} \, \ov K_{\ov{B^\prime}} -3 = \frac{3\, \hat{\xi }\left({\cal V}^2+7 \hat{\xi } \, {\cal V} + \, \hat{\xi }^2\right)}{\left(\, {\cal V}-\hat{\xi }\right) \left(\hat{\xi }+2 \, {\cal V}\right)^2} + \frac{\hat{k}_0^2 \,s^2 \,\left(4 \, {\cal V}-\hat{\xi }\right)-36 \,\hat{k}_0 \,\hat{\xi } \,s \, {\cal V}}{4 \left(\, {\cal V}-\hat{\xi }\right) \left(\hat{\xi }+2 \, {\cal V}\right)^2},\nonumber
\eea
where the first pieces on the RHS of the above two relations correspond to the case of the absence of odd moduli.
They can be precisely matched with the well known BBHL scalar potential result, see for example Eq.~(3.31) of \cite{Becker:2002nn}.
On the contrary, the second terms involving $\hat{k}_0 = \hat{k}_{\alpha ab}t^\alpha b^ab^b$ appear due to the presence of odd moduli.

At this point, if we enforce the SUSY stabilisation conditions (\ref{eq:susyDW}) in the scalar potential (\ref{eq:Vgen0}) and use the relations in Eq.~(\ref{eq:useful-id}), the scalar potential reduces to
\bea
\label{eq:Vgen00}
& & \hskip-0.5cm V = \ee^{{\cal K}} \, |W_0|^2 \, \left(K_{A^\prime} \, K^{{A^\prime} \ov{B^\prime}} \, \ov K_{\ov{B^\prime}} -3 \right) \\
& & = \ee^{{\cal K}} \, |W_0|^2 \, \left(\frac{3\, \hat{\xi }\left({\cal V}^2+7 \hat{\xi } \, {\cal V} + \, \hat{\xi }^2\right)}{\left(\, {\cal V}-\hat{\xi }\right) \left(\hat{\xi }+2 \, {\cal V}\right)^2} + \frac{\hat{k}_0^2 \,s^2 \,\left(4 \, {\cal V}-\hat{\xi }\right)-36 \,\hat{k}_0 \,\hat{\xi } \,s \, {\cal V}}{4 \left(\, {\cal V}-\hat{\xi }\right) \left(\hat{\xi }+2 \, {\cal V}\right)^2} \right)\,. \nonumber
\eea
This form of scalar potential represents an odd moduli generalisation of the results of \cite{Becker:2002nn}.
For the moment, let us consider the tree level case $\hat{\xi} = 0$ in the presence of odd moduli which leads to
\bea
\label{eq:Vgen000}
& & \hskip-0.75cm V = \ee^{{\cal K}} \, |W_0|^2 \, \biggl[\frac{s^2\, (\hat{k}_{\alpha a b}\, t^\alpha \, b^a\, b^b)^2}{4\, {\cal V}^2}\biggr] = \frac{s\, \, \, \ee^{\mathbb{K}}\, (\hat{k}_{\alpha a b}\, t^\alpha \, b^a\, b^b)^2\, |W_0|^2}{8\, {\cal V}^4}\,. 
\eea 
This result suggests that,
\begin{itemize}
\item{The axionic shift symmetry for the NS-NS axion $b^a$ is broken by a quartic potential which is purely generated by the flux background
%depending only on the ${S}$ and $U^i$ moduli via $W_0(U^i,S)$.
which for instance has been observed in \cite{Hristov:2008if, Ben-Dayan:2014lca} in concrete set-ups.
In fact, most of the odd moduli literature\footnote{However, there are a couple of works in which the $b^4$-term was not found in the generic scalar potential. For example, see \cite{Flauger:2009ab} regarding axion-monodromy, and \cite{Shukla:2015rua,Shukla:2015hpa,Shukla:2016hyy,Shukla:2019wfo} regarding non-geometric flux compactifications where one needs to switch-off certain fluxes to make it manifest.} aiming to perform moduli stabilisation automatically assumed the SUSY stabilisation for the axio-dilaton which is why all those models suffer from this outcome.
Moreover, this also contradicts the general notion that continuous axionic shift symmetries are broken, usually to a discrete subgroup, only through non-perturbative effects.}
\item{However,
we note that the non-trivial axionic potential $V(b^a)$ in Eq.~(\ref{eq:Vgen000}) appears at the sub-leading order as compared to the usual KKLT and LVS potentials for the K\"ahler moduli.
This issue could remain hidden in the analysis where a large volume approximation is assumed before performing moduli stabilisation as, e.g. in the analysis of \cite{Gao:2013rra} which keeps only terms of ${\cal O}({\cal V}^{-3})$. However, the term can explicitly appear if one goes beyond this order, e.g. as in \cite{Ben-Dayan:2014lca}.
% can be seen in Eqn.~(4.11) (KKLT) and in Eqn.~(4.34) (LVS) of 
Even if the $t^\alpha$'s correspond to, say, the `large' 2-cycle volume moduli such that $t^{\alpha} \sim {\cal V}^{1/3}$ as e.g. in the single $T$-modulus case of KKLT models \cite{Kachru:2003aw} or $t^\alpha \sim t_b$ in the LVS models based on strong swiss-cheese CY compactifications \cite{Balasubramanian:2005zx}, the scalar potential for the $b^a$ axions scales as,
\bea
\label{eq:fakeVb}
& & V(b^a) = \frac{s\, \, \, \ee^{\mathbb{K}}\, (\hat{k}_{\alpha a b}\, t^\alpha \, b^a\, b^b)^2\, |W_0|^2}{8\, {\cal V}^4}\, \propto \frac{{(b^a)}^4}{{\cal V}^{10/3}}\,,
\eea 
which is suppressed by additional volume factors as compared to the usual KKLT and LVS potentials.}
\item{Nevertheless, such a factor of 2-cycle volume moduli appearing in the numerator along with $b^a$ dependence can be relevant for highly anisotropic compactifications.
This is for instance the case in the ``large fibre" or ``large base" limits of the $K3$-fibred CY orientifold models \cite{Cicoli:2011yy}, where the extra volume suppression might go away, making the term compete with the LVS or KKLT potential.
%For example, if we take the volume of the $K3$-fibred CY, which takes and form ${\cal V} \simeq t_b^2 \,t_f \simeq \tau_b \, \sqrt{\tau_f}$ where $\tau_f$ is the volume of the divisor $K3$, and consider the anisotropic compactificatins with say `large' fibre volume limit such that $\tau_f \gg \tau_b$ \cite{Cicoli:2011yy}, then ${\cal V} \simeq \sqrt{\tau_f}$ which makes $t_b \simeq \sqrt{\tau_s} \simeq {\cal V}$.
In such extreme scenarios, while studying the dynamics away from the $\langle b^a \rangle = 0$ minimum, such fictitious $b^4$-terms may have some significant impact.
}
\end{itemize}
Next, we will show that such a quartic-term is purely an artefact of imposing the SUSY extremisation conditions for the axio-dilaton, and it does not naturally arise in the scalar potential. For that we consider our master formula in the expression (\ref{eq:master2}), which gives us the following scalar potential,
\bea
\label{eq:loophole1}
& & \hskip-0.5cm V = V_{\text{cs}} + \ee^{\cal K}\, \biggl[| w_{F}  + {S} \,  w_{H} |^2 + K^{{S} \ov{S}} \, | w_{H} |^2 - 4\,s^2\, \, | w_{H} |^2 - 4\, s \, {\rm Im}\bigl( w_{F}  \, \ov{ w_{H} } \bigr) \biggr]\,.
\eea
Given that $ w_{F} (U^i)$ and $ w_{H} (U^i)$ are cubic polynomials only in the $U^i$, the only possible source\footnote{One can easily verify that similar arguments are true for the metric derived in Sect.~\ref{sec:LeadingOrderBrekingSL2Z} using the modified Kähler potential \eqref{eq:ModVolume} from $\mathrm{SL}(2,\mathbb{Z})$ invariance.
This leads to the scalar potential given in Eq.~\eqref{eq:CorScalarPotAPTAPP}.} of a dependence on the $G^a$ moduli (or the $b^a/c^a$ odd-axions) can be the factor $K^{{S} \bar{S}}$.
However, we found this component to be given by (cf. Eq.~(\ref{eq:InvK}))
\bea
K^{{S} \ov{S}} = \gamma_1 = \frac{s^2\,\, \,(4 \,{\cal V}-\hat{\xi})}{({\cal V}-\hat{\xi})} \,,
\eea
which is a surprisingly simple expression independent of the odd moduli despite having complicated couplings in the original K\"ahler metric component $K_{{S}\ov{S}}$, recall \eqref{eq:simpK}.
In particular, in spite of the $(\alpha^{\prime})^{3}$ correction $\hat{\xi}$ inducing a scalar potential dependence on the overall volume ${\cal V}$,
both of the odd moduli remain flat directions at this level.
Therefore, we conclude that:
\begin{itemize}
\item{Using the leading order tree level GVW flux superpotential \eqref{eq:WGVW} in the presence of general 3-form fluxes, the shift symmetry in both of the odd moduli axionic components $b^a$ and $c^a$ is unbroken, irrespective of the presence of the BBHL correction. This is because of the no-scale structure \eqref{eq:WeakNoScale} which holds in the presence of odd moduli.}
\item{Our general analysis shows that the observation of a quartic potential $V(b^a)$ in Eq.~(\ref{eq:fakeVb}) is actually misleading and arises simply because the SUSY stabilisation condition for the axio-dilaton (\ref{eq:susyDW}) was enforced.}
\item{This opposes the naive expectation because of the non-trivial dependence of the K\"ahler potential on the odd moduli $G^a$ when it is expressed in terms of the chiral variables ${S}, G^a$ and $T_\alpha$. In such cases, one would naively expect that the shift symmetry for the NS-NS axion $b^a$ would be broken, whereas the shift symmetry for the $c^a$ axion remains intact.}
\end{itemize}

\subsubsection*{Tracking the mismatch:}

Before we continue,
we would like to explicitly track the main cause for the mismatch between the two approaches;
one is obtained directly from Eq.~(\ref{eq:loophole1}), whereas the other one is obtained after imposing the SUSY conditions (\ref{eq:susyDW}) in the scalar potential (\ref{eq:Vgen0}).
This becomes evident by noting that Eq.~(\ref{eq:Vgen0}) can also be rewritten as
\bea
\label{eq:Vgen00000}
& & \hskip-0.1cm V = V_{\text{cs}} + \ee^{{\cal K}} \, \biggl[K^{{S} \ov{S}} \, (\partial_{S} W_0) \, (\ov \partial_{\ov{S}} \ov{W_0}) + K^{{S} \ov{A}} \, K_{\ov{A}}\, (\partial_{S} W_0)  \, \ov{W_0} +\, K_{A} \,K^{{A} \ov{S}}  \, (\ov \partial_{\ov{S}} \ov{W_0}) \,  W_0 \nonumber\\
& & \hskip0.3cm + \biggl\{ \left(K_{S} K^{{S} \ov{S}} K_{\ov{S}} + K_{S} K^{{S} \ov{G}^a} K_{\ov{G}^a} + K_{S} K^{{S} \ov{T}_\alpha} K_{\ov{T}_\alpha} + K_{G^a} K^{G^a \ov{S}} K_{\ov{S}} + K_{T_\alpha} K^{T_\alpha \ov{S}} K_{\ov{S}} \right) \nonumber\\
& & \hskip0.3cm + \left(K_{A^\prime} \, K^{{A^\prime} \ov{B^\prime}} \, \ov K_{\ov{B^\prime}} -3 \right) \biggr\} \, |W_0|^2 \biggr]\,,
\eea
where summation indices are such that $A=\{S, T_\alpha, G^a\}$ while $A^\prime=\{T_\alpha, G^a\}$.
In this expression,
the terms in the second line arise from expanding the Kähler covariant derivative.
Now, after using the explicit expressions for the inverse K\"ahler metric and the K\"ahler derivatives and the identities \eqref{eq:identities00}, we find that the second line in Eq.~(\ref{eq:Vgen00000}) sums up to the following pieces,
\bea
& & \hskip-0.5cm K_{S} K^{{S} \ov{S}} K_{\ov{S}} + K_{S} K^{{S} \ov{G}^a} K_{\ov{G}^a} + K_{S} K^{{S} \ov{T}_\alpha} K_{\ov{T}_\alpha} + K_{G^a} K^{G^a \ov{S}} K_{\ov{S}} + K_{T_\alpha} K^{T_\alpha \ov{S}} K_{\ov{S}}\nonumber\\
& & \hskip0.5cm = 1 - \, \frac{3\, \hat{\xi }\left({\cal V}^2+7 \hat{\xi } \, {\cal V} + \, \hat{\xi }^2\right)}{\left(\, {\cal V}-\hat{\xi }\right) \left(\hat{\xi }+2 \, {\cal V}\right)^2} - \frac{\hat{k}_0^2 \,s^2 \,\left(4 \, {\cal V}-\hat{\xi }\right)-36 \,\hat{k}_0 \,\hat{\xi } \,s \, {\cal V}}{4 \left(\, {\cal V}-\hat{\xi }\right) \left(\hat{\xi }+2 \, {\cal V}\right)^2} \nn\\
& & \hskip0.5cm = 1 -\left(K_{A^\prime} \, K^{{A^\prime} \ov{B^\prime}} \, \ov K_{\ov{B^\prime}} -3 \right)\, ,
\eea
which leads to a precise cancellation of moduli dependent terms in the third line of Eq.~(\ref{eq:Vgen00000}). In other words, this is simply due to the no-scale identity $K_{A} \, K^{{A} \ov{B}} \, \ov K_{\ov{B}}=4$ derived above, when $A$ and $B$ run in the set of closed string moduli ${S}, G^a$ and $T_\alpha$.

The punchline is that imposing the SUSY stabilisation condition may lead to misleading conclusions in the presence of odd moduli due to exact cancellations occurring at the sub-leading order after breaking the no-scale structure through $\alpha^{\prime}$ corrections. In such a case, if some moduli are stabilised by the leading-order no-scale breaking corrections,
then those results could still be consistent. However, the no-scale structure is usually not completely broken leaving some moduli as flat directions which require next-to-subleading effects. For example, any inflationary model realised within a LVS framework, the inflationary potential is suppressed by volume factors ${\cal V}^{-a}$ where $a>3$, e.g., in fibre inflation $a = 10/3$ \cite{Cicoli:2008gp}.
Hence,
when adding odd moduli in such a scenario,
these sub-leading terms become especially relevant.
Nevertheless, as we said before, if axion VEVs are not shifted too much from $b^a \simeq 0$ after including all possible corrections to fix them, they may not significantly affect the earlier minimum.
This is not to say though that situations away from the minimum e.g. in the regime where inflation occurs might be sensitive to our aforementioned observation.

\subsection{BBHL-corrected flux scalar potential}

At this point, we present the explicit form of the $(\alpha^{\prime})^{3}$-corrected GVW scalar potential which can be utilised to explore non-SUSY vacua.
In this scenario, generically one can hope to stabilise the complex structure moduli, namely the axions $v^i$ and their respective saxions $u^i$, as well as the universal axion $c_0$ along with the dilaton $s$.
Using the scalar potential in Eq. (\ref{eq:master1b}) for $\tilde{\xi}=0$ we have
%\bea
%& & \hskip-0.5cm V = \ee^{{\cal K}} \,\biggl[4\, | w_{F} +\,S w_{H} |^2 + ( w_{F} _{U^i}+S w_{H} _{U^i}) K^{{U^i} \ov{U^j}} \, (\ov w_{F} _{\ov{U^j}}+\ov{S}\,\ov{ w_{H} }_{\ov{U^j}}) +  K^{S \ov{S}} \, | w_{H} |^2 \nonumber\\
%& & + ({U^i} - \ov {U^i}) \, \, \left(( w_{F} +S w_{H} ) \, \ov  (\ov w_{F} _{\ov{U^i}}+\ov{S}\,\ov{ w_{H} }_{\ov{U^i}}) - (\ov w_{F} +\ov{S}\ov w_{H} )\,( w_{F} _{U^i}+S w_{H} _{U^i})\right) \nonumber\\
%& & + (S - \ov{S}) \, \, \left(( w_{F} +S w_{H} ) \, \ov w_{H}  - (\ov w_{F} +\ov{S}\ov w_{H} )\,  w_{H} \right) \biggr].
%\eea
%This can be simplified to take the following form,
\bea
\label{eq:VGVW}
& & \hskip-0.5cm V_{GVW} \equiv V_{GVW}(v^i, u^i, c_0, s, {\cal V}) \simeq \ee^{{\cal K}} \,\left(V_{GVW}^{(1)} + V_{GVW}^{(2)} + V_{GVW}^{(3)} \right),
\eea
where
\bea
\label{eq:VGVW1}
& & \hskip-0.2cm V_{GVW}^{(1)}  = 4\,\biggl[\left({m}+c_0 \, {n}\right)^2 +\left({{\frak m}} +c_0 \, {\frak n}\right)^2 + \frac{s^2\,  | w_{H} |^{2} \, (4\, {\cal V} - \hat\xi)}{4\, ({\cal V} - \hat\xi)} + s\, \left({{\frak m}} \, {n}-{m} \, {\frak n} \right) \biggr], \nonumber\\
& & \hskip-0.2cm V_{GVW}^{(2)}  = 4\, u^i \,\biggl[({m} + c_0 \, {n}) ({{\frak m}_i} + c_0 \, {{{\frak n}_i}}) - ({{\frak m}} + c_0 \, {\frak n}) ({m_i} + c_0 \, {{n_i}})  \nonumber\\
& & \hskip2cm + \left({m} {{n_i}} -{m_i} \, {n}  + {{\frak m}} \, {{{\frak n}_i}} - {{\frak m}_i} \, {\frak n} \right) s + \left({n} \, {{{\frak n}_i}} -{{n_i}}\, {\frak n}\right) s^2\biggr], \\
& & \hskip-0.2cm V_{GVW}^{(3)}  = \left(2 u^i\, u^j- \frac{2\, l\, l^{ij}}{3} \right)\,\biggl[({m_i} + c_0 \, {{n_i}})^2+({{\frak m}_i} + c_0 \, {{{\frak n}_i}})^2  - 2\,s\, ({m_i} {{{\frak n}_i}}-{{\frak m}_i} {{n_i}}) \nonumber\\
& & \hskip2cm + s^2\,\left(({{n_i}})^2 + ({{{\frak n}_i}})^2)\right) \biggr]. \nonumber
\eea
Here we have used the symbols for the real and imaginary parts of $ w_{F} (U^i)$, $ w_{H} (U^i)$ and their respective derivatives as defined in Eq. (\ref{eq:ReMetc}). %\bea
This potential depends on the real fields $\{v^i, u^i, c_0, s, {\cal V}\}$,
albeit not on the fields $b^a, c^a$ or any other $t^\alpha$.
Moreover, in the absence of the BBHL correction, the dependence on the overall volume ${\cal V}$ comes from $e^{\cal K}$.

Subsequently, we can separate out the tree level and the next sub-leading terms in the scalar potential (\ref{eq:VGVW}),
\bea
\label{eq:VGVW-tree+BBHL}
& & V_{GVW} \simeq V_{GVW}^{tree} + V_{GVW}^{\alpha^\prime}, 
\eea
where
\bea
\label{eq:VGVW-tree+BBHL1}
& & V_{GVW}^{tree} = \frac{e^{{\mathbb K}}}{2\, s\, {\cal V}^2} \,\biggl[4\, |W|^{2}+ 4\, s\, \left({{\frak m}} \, {n}-{m} \, {\frak n} \right) + V_{GVW}^{(2)} + V_{GVW}^{(3)} \biggr], \nonumber\\[0.6em]
& & V_{GVW}^{\alpha^\prime} =  - \frac{\hat\xi}{{\cal V}} \left(V_{GVW}^{tree} - \frac{e^{{\mathbb K}}}{2\, s\, {\cal V}^2} \times 3 \, s^2\,  | w_{H} |^{2}\right) + {\cal O}\left(\frac{1}{{\cal V}^4}\right)\, .
\eea
As expected from \eqref{eq:CorScalarPotAPTAPP}, $V_{GVW}^{\alpha^\prime}$ splits into a piece arising from the $4$D Weyl rescaling of $V_{GVW}^{tree}$ with the corrected volume $\cV+\hat{\xi}/2$ and a non-trivial piece.
The coefficient of the latter matches the expectation from \eqref{eq:BBHLCorGVWIntH}, see also \eqref{eq:VGVW-susy-uplift1} below.
%& & V_{GVW}^{\alpha^\prime} = -\,\frac{e^{{\mathbb K}}}{2\, s\, {\cal V}^3} \, \hat\xi\,\biggl[4\,\left({m}+c_0 \, {n}\right)^2 + 4\, \left({{\frak m}} +c_0 \, {\frak n}\right)^2 + \, s^2\, ({n}^2+{\frak n}^2) \nonumber\\
%& & \hskip1cm + 4\, s\, \left({{\frak m}} \, {n}-{m} \, {\frak n} \right) + V_{GVW}^{(2)} + V_{GVW}^{(3)} \biggr] + {\cal O}\left(\frac{1}{{\cal V}^4}\right).

Clearly, at the tree level SUSY minimum, we have $\langle V_{GVW}^{tree}\rangle_{susy} = 0$ from Eq. (\ref{eq:VGVWsusytree}).
This can also be re-derived from the scalar potential given in Eqs. (\ref{eq:VGVW-tree+BBHL})--(\ref{eq:VGVW-tree+BBHL1}) after imposing the conditions (\ref{eq:susy-USTG-tree1}) in the three pieces of (\ref{eq:VGVW1}). To appreciate the non-trivial cancellations among the various terms, we mention that
\begin{align}
\langle V_{GVW}^{(1)} \rangle &=\langle|W_{0}|^{2}\rangle\left (3 + \frac{3\, \langle\hat\xi \rangle}{4({\cal V} - \langle \hat\xi \rangle)}\right ) \kom\langle V_{GVW}^{(2)} \rangle=-2\langle V_{GVW}^{(3)} \rangle  = - 6\langle|W_{0}|^{2}\rangle \, ,
\end{align}
where we used that at the leading order supersymmetric minimum
\bea\label{eq:CondWZeroSN} 
& & 4\, \langle s \rangle^2 \, \langle | w_{H} |^2 \rangle = 4\, \langle s \rangle^2 \left[\langle {n}\rangle^2+\langle{\frak n}\rangle^2\right] =\langle|W_{0}|^{2}\rangle\, .
\eea
Overall, this implies 
\bea
\label{eq:VGVW-susy-uplift1}
& & \langle V_{GVW}^{\alpha^\prime} \rangle = \frac{3\, \langle \hat\xi \rangle \, \ee^{\langle {\mathbb K}\rangle} \langle |W_0|^2 \rangle}{8\, \langle s\rangle\, {\cal V}^3} + {\cal O}\left(\frac{1}{{\cal V}^4}\right) \geq 0\, .
\eea 
This positive semidefinite term is well-known from LVS models \cite{Balasubramanian:2005zx}. Thus, our master formula for the generic treatment of the scalar potential can elegantly and efficiently re-derive those results.

Note that given the form of the superpotenial, our generic formulation of the BBHL-corrected GVW scalar potential in Eqns. (\ref{eq:VGVW})--(\ref{eq:VGVW1}) enables one to not only perform the analytic computations, but also to proceed with numerical implementations. Having such a general form can help in directly exploring the non-supersymmetric vacua of the complex structure moduli and the axio-dilaton which could result in subleading terms that can be useful for uplifting purposes \cite{Gallego:2017dvd}.

\subsection{Axio-dilaton mass splitting via the BBHL correction}\label{sec:MassSplittingBBHL} 

In this subsection, we will first re-derive a well known no-go results of \cite{Blumenhagen:2014nba,Gao:2015nra} about the obstacles in creating a mass splitting between the masses of the universal axion $(m_{c_0})$ and the dilaton $(m_s)$ using the GVW superpotential. For that purpose, assuming the large volume and large complex structure limit, we consider the GVW scalar potential given in eqns.~(\ref{eq:VGVW-tree+BBHL})--(\ref{eq:VGVW-tree+BBHL1}) by ignoring the BBHL correction, which leads to
\bea
\label{eq:V-onlyH3simp}
& & V_{GVW}^{tree} \equiv V(c_0, s) = \frac{a_1}{s} \left(1+ {a_2\, c_0} + a_3\, s + a_4\, c_0^2 + a_5\, s^2 \right),
\eea
where $a_i$'s are some complex structure moduli dependent functions while $a_1$, in addition, has a dependence on the overall volume ${\cal V}$ as well. To be more specific, we have the following explicit expressions for the $a_i$ coefficients,
\bea
\label{eq:ais-tree}
& & a_1 = \frac{e^{\mathbb K}}{2\,{\cal V}^2} \, a_0, \\
& & a_2 = \frac{1}{a_0} \biggl[8\, (m\, n + {\frak m} \, {\frak n}) + {\mathbb K}^{U^i \ov{U}^j} \left({m}_i \, {n}_j + {n}_i \,  {m}_j  + {\frak m}_i \, {\frak n}_j + {\frak n}_i \, {\frak m}_j \right) \nonumber\\
& & \hskip2cm + 4 u^i \left(m \, {\frak n}_i - m_i \, {\frak n} + n \, {\frak m}_i - n_i \, {\frak m} \right)\biggr], \nonumber\\
& & a_3 = \frac{1}{a_0} \biggl[4 ({\frak m}\, n - {\frak n} \, m) + 4 u^i \left({\frak m} \, {\frak n}_i - {\frak m}_i \, {\frak n} -  \, {m}_i \, n + m\,  n_i \right) + 2\, {\mathbb K}^{U^i \ov{U}^j} \left({\frak m}_i \, n_j - m_i \, {\frak n}_j \right)\biggr], \nonumber\\
& & a_4 = \frac{1}{a_0} \biggl[4\, | w_{H} |^{2} + 4\, u^i \left({n} \, {{\frak n}_i} - \,{{\frak n}} \, {n_i} \right) + {n}_i \, {\mathbb K}^{U^i \ov{U}^j} \,  {n}_j + {\frak n}_i \, {\mathbb K}^{U^i \ov{U}^j} \,  {\frak n}_j\biggr] = a_5, \nonumber\\
& & \hskip-0.75cm {\rm where} \nonumber\\
& & a_0 = 4\, | w_{F} |^{2} + 4\, u^i \left({m} \, {{\frak m}_i} - \,{{\frak m}} \, {m_i} \right) + {m}_i \, {\mathbb K}^{U^i \ov{U}^j} \,  {m}_j + {\frak m}_i \, {\mathbb K}^{U^i \ov{U}^j} \,  {\frak m}_j. \nonumber
\eea
This tree level formulation of the scalar potential produces the results of \cite{Shukla:2019wfo} by considering $F_3/H_3$ fluxes only, and switching-off the (non-)geometric fluxes. Now, using the scalar potential (\ref{eq:V-onlyH3simp}) for the two-field $\{c_0, s\}$ dynamics, the solutions of the extremisation conditions, and scalar potential as well as the Hessian evaluated at the possible extrema amounts to
\bea
\label{eq:sol-c0-s}
& & \langle c_0 \rangle = -\frac{\langle a_2 \rangle}{2\, \langle a_4 \rangle}, \quad \langle s \rangle = \pm \frac{\sqrt{4 \, \langle a_4 \rangle -\langle a_2\rangle^2}}{2\sqrt{\langle a_4 \rangle \, \langle a_5 \rangle}}, \quad \langle V \rangle = \langle a_1 \rangle \left(\langle a_3 \rangle \pm 2 \langle s \rangle \langle a_5 \rangle \right), \nonumber\\
& & \langle V_{ij} \rangle = {\rm Diag} \left\{\pm\frac{2\, \langle a_1 \rangle\, \langle a_4\rangle}{\langle s \rangle}, \, \, \pm\frac{2\, \langle a_1 \rangle \, \langle a_5 \rangle}{\langle s \rangle} \right\}; \quad \frac{\langle V_{ss}\rangle}{\langle V_{c_0c_0}\rangle}  = \frac{\langle a_5 \rangle }{\langle a_4 \rangle} = 1.
\eea
Here although we do not discuss the complex structure moduli explicitly, we anticipate that the two-field analysis will continue to hold after fixing those moduli.
%and the after effect of the same would simply be the fact that all the complex structure moduli dependent coefficients $a_i$'s would receive a VEV leading to what we call as $\langle a_i \rangle$.
Now given that dilaton $s > 0$, only the positive solution of (\ref{eq:sol-c0-s}) is physical. 

Note that we have simply collected the coefficients (\ref{eq:ais-tree}) for the scalar potential in Eq. (\ref{eq:V-onlyH3simp}) where it automatically turned out that $a_4 = a_5$ for the quadratic coefficients of $c_{0}$ and $s$ irrespective of whether the complex structure moduli are at their minimum or not. This two-field analysis implies that masses of the universal axion $c_0$ and the dilaton $s$ remain exactly the same. Moreover, given the fact that both fields are part of the same chiral variable, namely $S$, and their source of potential is the same, it is not surprising that a mass hierarchy cannot be realised in this flux scenarios. This challenge was first observed for $3$-form flux backgrounds in \cite{Blumenhagen:2014nba} and was subsequently generalised for non-geometric models having two pairs of the $S$-dual fluxes, namely the $(F_3, H_3)$ fluxes and the non-geometric $(Q, P)$ fluxes in \cite{Gao:2015nra}.

\subsubsection*{Mass splitting via the BBHL correction}

Thus far,
we ignored the $(\alpha^\prime)^3$-effects which lead to a mass splitting between $c_{0}$ and $s$.
Indeed,
we can consider our generic form for the GVW scalar potential as given in Eq. (\ref{eq:VGVW}) with induced dilaton dependence through ${\cal Y}(s)$ where ${\cal Y}(s) = {\cal V} + \frac{1}{2}\hat\xi$ and $\hat\xi = s^{3/2} \xi$.
While the extremisation condition for $c_0$ still results in a linear polynomial,
the one for the dilaton results in a degree 13 polynomial in $s$ which cannot be solved analytically.
However, we can make progress by using the possible expansions, for example in the large volume limit ${\cal V} \gg \hat\xi$. %or the weak coupling limit, i.e., via expanding the scalar potential (\ref{eq:V-onlyH31}) around $g_s=1/s \sim 0$.

Nevertheless it turns out that the generic BBHL corrected GVW potential (\ref{eq:VGVW}) can still be expressed in the form as given in Eq. (\ref{eq:V-onlyH3simp}), where the coefficients $a_i$'s will however develop a dilaton dependence along with the complex structure moduli. The new set of $a_i$ parameters are
\bea
\label{eq:ais-gen}
a_1(s) = \frac{e^{\mathbb K} \, a_0}{2\left({\cal V} + \frac{1}{2}\hat{\xi}\right)^2} \kom a_5(s) = a_4 + \frac{3 \, \hat{\xi}\, | w_{H} |^{2}}{a_0 \,({\cal V} - \hat{\xi})}  \, , 
\eea
with all other $a_i$ as in \eqref{eq:ais-tree}. Now the extremisation of the $c_0$ axion remains the same, the dilaton VEV is changed giving rise to
\bea
\label{eq:Vc0s-BBHL-min}
& & \langle c_0 \rangle : \quad \langle c_0 \rangle = -\frac{ \langle a_2 \rangle}{2\,\langle a_4 \rangle}\\
& & \langle s \rangle : \quad \langle a_1\rangle \left(1 - \frac{\langle a_2 \rangle^2}{4 \langle a_4 \rangle} - \langle s \rangle^2 \, \langle a_5\rangle \right) \nonumber\\
& & \hskip1cm = \langle a_1\rangle \langle s\rangle^3 \langle \partial_s a_5\rangle + \langle s \rangle \left(1 - \frac{\langle a_2 \rangle^2}{4 \langle a_4 \rangle} + \langle s \rangle \, \langle a_3 \rangle + \langle s \rangle^2 \langle a_5\rangle\right) \langle \partial_s  a_1\rangle,  \nonumber\\
& & \langle V \rangle : \quad \langle V \rangle = \frac{\langle a_1\rangle}{\langle s \rangle} \left(1 - \frac{\langle a_2 \rangle^2}{4 \langle a_4 \rangle} + \langle s \rangle \langle a_3 \rangle + \langle s \rangle^2 \, \langle a_5\rangle\right), \nonumber\\
& & \langle V_{ij} \rangle : \quad \langle V_{ij} \rangle = {\rm Diag}\left\{\frac{2\, \langle a_1 \rangle \langle a_4\rangle}{\langle s \rangle}, \frac{2\, \langle a_1 \rangle \langle a_5 \rangle}{\langle s \rangle} + V_{ss}^{\rm corr}\right\},\nonumber\\
& & \hskip-0.95cm {\rm where} \quad V_{ss}^{\rm corr} =  \langle V \rangle \left(\frac{2 \langle \partial_s a_1 \rangle}{\langle s \rangle \langle a_1\rangle} -\frac{2 \langle \partial_s a_1\rangle^2}{\langle a_1\rangle^2} + \frac{\langle \partial_{ss} a_1\rangle}{\langle a_1\rangle} \right) + \langle a_1\rangle \left(4 \langle \partial_s a_5 \rangle + \langle s \rangle \langle \partial_{ss} a_5 \rangle\right). \nonumber
\eea
For knowing the dilaton VEV one would need to solve the corresponding polynomial expressed in the second relation of (\ref{eq:Vc0s-BBHL-min}).
We notice immediately that now there is a possibility of generating a mass-splitting between the masses of $c_0$ and $s$ due to $a_4 \neq a_5$. The ratio of the Hessian eigenvalues is
\bea
& & \hskip-2.5cm \frac{\langle V_{ss}\rangle}{\langle V_{c_0c_0}\rangle}  = \frac{\langle a_5 \rangle }{\langle a_4 \rangle} + \frac{\langle s \rangle}{2\, \langle a_1 \rangle \langle a_4\rangle} \, \langle V_{ss}^{\rm corr} \rangle 
\eea
which can be expanded as
\bea
&& \frac{\langle V_{ss}\rangle}{\langle V_{c_0c_0}\rangle}   = 1 + \biggl\{\frac{\langle\hat\xi \rangle}{{\cal V}} \times \frac{15\,\left[28 \, \langle a_4 \rangle \, \langle s\rangle^2 + \langle a_0\rangle \left(\langle a_2\rangle^2 - 4\, \langle a_4\rangle (1 + \langle a_3\rangle \langle s\rangle + \langle a_4\rangle \langle s\rangle^2)\right)\right]}{32 \, \langle a_0\rangle \langle a_4\rangle^2 \langle s\rangle^2} \biggr\}\nonumber\\
& & \hskip2cm + {\cal O}\left(\frac{1}{{\cal V}^{2}}\right)\, .
\eea
This shows the possibility of a mass-splitting being developed between the masses of the universal axion $(c_0)$ and the dilaton $(s)$ due to $(\alpha^{\prime})^{3}$ corrections. However, one has to provide the VEVs of the complex structure moduli and the dilaton in order to get the $\langle a_i \rangle$'s in the above estimates. Moreover, in realistic models, one also needs to take the dynamics of all other moduli into account in order to gauge the effects of the off-diagonal mixing terms in the mass matrix, especially for the cases when the superpotential depends on the $T_\alpha$-moduli.
Nevertheless, given that we have evaded the no-go result of the two-field analysis using $(\alpha^{\prime})^{3}$ corrections, this is quite a significant step via an analytical approach.

\section{Analysing the scalar potential for $W=W(U^i, {S}, T_\alpha)$}\label{sec:SPAnalysisWT} 

In this section,
we study different scenarios involving superpotentials of the form $W=W(U^i, {S}, T_\alpha)$.
The associated scalar potentials explicitly depend on the NS-NS axions $b^{a}$,
while the shift symmetry of the R-R axions $c^{a}$ remains intact.
We derive an expression for the Hessian of the $b^{a}$ which reduces to the results of \cite{Conlon:2006tq} in the SUSY case, while improving upon previous approximate estimates of \cite{Hristov:2008if} for non-SUSY minima.

\subsection{General remarks on moduli stabilisation}

From our above observations,
one could be tempted to conclude that the axionic shift symmetry of $b^{a}$ remains unbroken simply because the $G^{a}$ are not explicitly featured in the superpotential.
In this section, we show that this is actually not the case, and after including a $T_{\alpha}$-moduli dependence in the superpotential, the $b^a$ axions can indeed receive a contribution in the scalar potential. This makes odd moduli rather special\footnote{Recall that usually the partners of the R-R axions in the complexified chiral variables are called `saxions', but the partner of R-R $C_{2}$-axions $c^a$ appearing in the $G^a$ multiplet are also axions, namely the $b^a$ axions.} as compared to other moduli.
%In fact, there are quite involved cross-terms in the scalar potential which can induce a potential for $b^a$ despite the superpotential being independent of $G^a$.

Let us consider a superpotential of the form
\bea\label{eq:SuperpotentialTUS} 
& & W = W_0(U^i,S) + W_1(U^i, S,T_\alpha).
\eea
%\todoin{I also included a $S$ dependence in $W_{1}$ because e.g. the Pfaffian prefactors in the non-perturbative superpotential depend on $S$ as discussed e.g. in \cite{Demirtas:2021nlu}.}
The second piece of the superpotential can be generated by non-perturbative effects \cite{Witten:1996bn,Bianchi:2012pn,Bianchi:2011qh,Blumenhagen:2012kz,Demirtas:2021nlu} with an exponential $T_{\alpha}$-moduli dependence or in the presence of $S$-dual pair of $(Q, P)$ non-geometric fluxes \cite{Aldazabal:2006up,Blumenhagen:2015kja,Shukla:2016xdy,Shukla:2016hyy,Shukla:2015bca} inducing a polynomial dependence on the $T_{\alpha}$-moduli.
Subsequently, from Eq.~(\ref{eq:master2}) we obtain,
\bea
\label{eq:VtauT0}
& & \hskip-0.75cm V = V_{\text{cs}} + \ee^{\cal K} \, \biggl[|W|^2 + {W}_{S}\, K^{{S} \ov{S}} \, \ov {W}_{\ov{S}} + {(W_1)}_{T_\alpha}\, K^{T_{\alpha} \ov T_{\beta}} \, \ov {(W_1)}_{\ov T_{\beta}} \\
& & \hskip0.5cm  - 4\, s \, {\rm Im}\bigl[W \, \ov {W}_{\ov{S}}\bigr] - 4\, ({\rm Im}T_\alpha) \,\,  {\rm Im}\bigl[W \, \ov {(W_1)}_{\ov T_{\alpha}}\bigr]  + \, 2 \, K^{{S} \ov{T}_\alpha} {\rm Re}[{W}_{S} \, \ov {(W_1)}_{\ov T_\alpha}] \biggr] \,, \nonumber
\eea
or equivalently using the master formula in Eq. (\ref{eq:master1a}) this can be written as,
\bea
\label{eq:VtauT0AS}
& & \hskip-0.5cm V = \ee^{\cal K} \, \biggl[{\dfrac{4l-3\tilde{\xi}}{l-3\tilde{\xi}} }|W|^2 + {W}_{U^i} K^{{U^i} \ov{U^j} } \,\ov{W}_{\ov{U^j}}+{W}_{S}\, K^{{S} \ov{S}} \, \ov {W}_{\ov{S}}+ {(W_1)}_{T_{\alpha}}\, K^{T_{\alpha} \ov T_{\beta}} \, \ov {(W_1)}_{\ov T_{\beta}} \nn\\
& &  \hphantom{ = \ee^{\cal K} \, \biggl[ }  + \, 2 \, K^{{S} \ov{T}_\alpha} {\rm Re}[{W}_S\, \ov {(W_1)}_{\ov T_\alpha}]-{ 2\, u^i \left (2+\dfrac{9\tilde{\xi}}{l-3\tilde{\xi}}\right )} \, {\rm Im}\bigl[W \, \ov{W}_{\ov{U^j}}\bigr] \\
& &  \hphantom{ = \ee^{\cal K} \, \biggl[ }   - 4\, s \, {\rm Im}\bigl[W \, \ov {W}_{\ov{S}}\bigr]- 4\, ({\rm Im}T_\alpha) \,\,  {\rm Im}\bigl[W \, \ov {(W_1)}_{\ov T_{\alpha}}\bigr]    \biggr]\,, \nonumber
\eea
From the inverse K\"ahler metric components in Eq.~(\ref{eq:InvK}), we find that, in contrast to the previous case where $K^{{S} \ov{S}}$ is independent of odd moduli, the components $K^{{S} \ov{T}_\alpha}$ and $K^{T_{\alpha} \ov T_{\beta}}$ do indeed depend on the $B_{2}$-axions $b^a$, but not on the $C_2$-axions $c^a$. Also notice that with our definition of the chiral variable we have ${\rm Im}[T_\alpha] = \frac{1}{2} (s\, \hat{k}_\alpha - k_\alpha)$ (recall Eq.~\eqref{eq:N1coords}), and therefore one can consider the following two approaches for odd moduli stabilisation:
\begin{itemize}
\item{If we define a set of real moduli as $\tau_\alpha = -\, {\rm Im}[T_\alpha]$ and try to eliminate all the $t^\alpha$ dependence in terms of $\tau_\alpha$ and $b^a$, then subsequently the scalar potential (\ref{eq:VtauT0}) can receive $b^a$ dependence through the inverse K\"ahler metric components; e.g. via replacing $k_\alpha = 2 \tau_\alpha + s\, \hat{k}_\alpha$. In this case, there would be no $b^a$ dependence arising from the superpotential $W\equiv W(U^i, {\rm Re}(T_\alpha), {\rm Im}(T_\alpha)) = W(U^i, \tau_\alpha, \tilde\rho_\alpha)$ where $\tilde{\rho}_\alpha$ is the collection of all R-R axions. Moreover due to the need for replacing $k_\alpha = 2 \tau_\alpha + s\, \hat{k}_\alpha$ in the inverse metric components, some additional $b^a$ dependences would also be introduced in the scalar potential. 

However after looking at the inverse K\"ahler metric components, one observes that, apart from eliminating $k_{\alpha}$ in favour of $\tau_\alpha$, there remain pieces of the type $k_{\alpha\beta}$ and $\hat{k}_{ab}$ with linear dependence on the 2-cycle volumes $t^\alpha$. This can be a key obstacle when writing down the scalar potential as an explicit function of $\{\tau_\alpha, b^a\}$ only. 
}
\item{If we define $\tau_\alpha = \frac{1}{2} k_\alpha = \frac{1}{2} k_{\alpha\beta\gamma} t^\beta t^\gamma$ as the usual geometric volume of the four-cycles of the CY threefold, then we need to replace ${\rm Im}[T_\alpha] = \frac{s}{2} \hat{k}_\alpha - \tau_\alpha$.
Then, some explicit $b^a$ dependence will be introduced through the K\"ahler potential as well as the superpotential despite it not depending on the odd moduli $G^a$. However, in this case one can stabilise moduli working in the basis $\{t^\alpha , b^a\}$ and therefore there would be no need to make the conversion $t^\alpha \to \tau_\alpha$.}
\end{itemize}
Let us note that for both approaches, there is no scalar potential piece induced for the $c^a$ axions through the superpotential\footnote{In this statement, we use that the $C_4$-axions only enter through the combination $\tilde\rho_\alpha$ as defined in \eqref{eq:ConventionModAx} which serves as a single variable to avoid fictitious dependence on $c_0, b^a$ and $c^a$ axions.}.
The second approach of working in the basis $\{t^\alpha, b^a\}$ works well for arbitrary number of moduli, and hence is usually preferred.
However,
limitations arise when computing the mass matrix for which a combination of Hessian components and the inverse K\"ahler metric has to be considered. Thus, either of the two approaches might not be always be applicable for generic CY orientifolds.

As the problem is inherited from the fact that it is hard to convert $t^\alpha$ into the $\tau_\alpha$, it is more convenient to consider the scalar potential as a function of the 2-cycle volumes $t^\alpha$ and the axions $b^a$ for many moduli stabilisation or minimisation purposes.
While this approach is not fully established yet in the literature,
it has already been proposed in \cite{AbdusSalam:2020ywo}.
For that purpose, we rewrite the scalar potential (\ref{eq:VtauT0}) as
\bea
\label{eq:eq:VtauT00}
& & \hskip-0.2cm V = V_0 + V_1 \,, 
\eea
in terms of $V_{0}$ and $V_{1}$ defined in \eqref{eq:ScalarPotGModGen}.
Here $V_0$ can have an explicit $b^a$ dependence only through the superpotential piece $W_1$ (recall that $\gamma_1,  \gamma_2, k_\alpha$ as well as the metric ${\cal G}_{\alpha\beta}$ are independent of the $b^a$ axions). Further, the pieces $V_1$ and $V_2$ may have an explicit $b^a$ dependence also through the K\"ahler metric.\footnote{Recall that ${\cal G}^{ab}$ only has an explicit dependence on the $t^\alpha$ moduli, but not the $b^a$ axions.}
However, given that the moduli dependence of the superpotential is
\begin{equation}
W_1\equiv W_1(U^i, S,T_\alpha) = W_1(u^i, v^i, c_{0},s,k_\alpha, \hat{k}_\alpha, \tilde\rho_\alpha)
\end{equation}
where $\tilde{\rho}_\alpha$ is again some collection of all R-R axions, the form of \eqref{eq:ScalarPotGModGen} suggests that the $b^a$ dependence of the scalar potential can always be collected as
\bea
\label{eq:Vba}
& & V(b^a) \equiv V(\hat{k}_\alpha, \, K_{\alpha a\beta b})
\eea
where
\begin{equation}
K_{\alpha a\beta b}=\hat{k}_{\alpha a} \hat{k}_{\beta b}\, .
\end{equation}

\subsection{Non-perturbative superpotentials}\label{sec:NonPertTSup} 

Another commonly studied scenario is that of a non-perturbative superpotential induced by E3-instantons or D7 gaugino condensation.
In this case,
we write
\begin{equation}\label{eq:SuperPotentialE3ETC} 
W_{0}=w_{F}(U^{i})+Sw_{H}(U^{i})\kom W_{1}(U^{i},S,T_{\alpha})=\sum_{\alpha=1}^{h_{+}^{1,1}}\, A_{\alpha}(U^{i},S)\ee^{-ia^{\alpha}T_{\alpha}}\, .
\end{equation}
%\todoin{Do we want to add another label here so that $a_{\alpha}^{\gamma}T_{\gamma}$? Otherwise, some repeated indices will appear below, especially in the Hessian.}
%\todoin{Nore sure, though can be relevant for generating potentials when say the big divisor of LVS is rigidified to result in racetrack model with non-diagonal basis see, e.g. Roberto's explicit de-Sitter paper of 2012. }
Here,
the Pfaffian prefactors $A_{\alpha}(U^{i},S)$ is associated with the partition function of the D3-brane wrapping a certain divisor $D$ in the orientifold.
As argued recently in \cite{Demirtas:2021nlu},
for many CY orientifolds the $A_{\alpha}(U^{i},S)$ can be treated as constant numbers, so-called \emph{Pfaffian numbers}, since the divisor $D$ becomes a pure rigid divisor in the F-theory uplift.
Thus, we set $A_{\alpha}(U^{i},S)=A_{\alpha}$ in the subsequent discussion which is commonly used in the literature.
We note, however,
that our general expressions in Eq.~\eqref{eq:VtauT0} and Eq.~\eqref{eq:VtauT0AS} make it straight forward to include such corrections systematically.

Plugging \eqref{eq:SuperPotentialE3ETC} into \eqref{eq:VtauT0},
we write the scalar potential as
\begin{equation}
V=V^{\text{pert}}+V^{\text{np1}}+V^{\text{np2}}
\end{equation}
where the individual terms are given by
\begin{align}\label{eq:ScalarPotED3} 
V^{\text{pert}} &= V_{\text{cs}} +\ee^{\cK}\left (|W_{0}|^{2}- 4 s \, {\rm Im}\bigl[W_{0}\, \ov {w}_H\bigr]+ \, w_H \,\gamma_1 \,  \ov {w}_H\right ) \nn\\
V^{\text{np1}}&=\ee^{\cK}\sum_{\alpha=1}^{h_{+}^{1,1}}\; \biggl[  2(1+a^{\alpha}(k_{\alpha}-s\hat{k}_{\alpha}))\text{Re}\left [(w_{F}+c_{0}w_{H})\bar{A}_{\alpha}\ee^{ia^{\alpha}\ov{T}_{\alpha}}\right ]\nn\\
&\hphantom{=\sum_{\alpha=1}^{h_{+}^{1,1}}\; \biggl[ }+ (2 s+a^{\alpha}(k_{\alpha}(\gamma_2-2s)+\hat{k}_{\alpha}(\gamma_{1}+2s^{2}))) \, {\rm Im}\bigl[w_{H}\bar{A}_{\alpha}\ee^{ia^{\alpha}\ov{T}_{\alpha}} \bigr]\biggr ] \\
V^{\text{np2}}&=\ee^{\cK}\sum_{\alpha,\beta=1}^{h_{+}^{1,1}}\; A_{\alpha}\bar{A}_{\beta}\ee^{-ia^{\alpha}T_{\alpha}+ia^{\beta}\ov{T}_{\beta}}\biggl [1+a^{\alpha}(k_{\alpha}-s\hat{k}_{\alpha})+a^{\beta}(k_\beta-s\hat{k}_{\beta})\nn\\
&\quad +a^{\alpha}a^{\beta}\, \biggl (\frac{4{\cal Y}^2 \, {\cal G}_{\alpha\beta}}{9} +\frac{\gamma_2^2\,  k_\alpha\, k_\beta}{4\, \gamma_1}+s\, {\cal G}^{ab}\, \hat{k}_{\alpha a}\, \hat{k}_{\beta b} + \frac{\gamma_1}{4} \, \hat{k}_\alpha \, \hat{k}_\beta\, +\frac{\gamma_2}{4}\, (k_\alpha \, \hat{k}_\beta + k_\beta \, \hat{k}_\alpha)\biggl)\biggl ]\nn\, .
\end{align}
This generalises the master formula of \cite{AbdusSalam:2020ywo} in the presence of odd moduli and without enforcing the SUSY stabilisation of the axio-dilaton $S$.
Clearly,
the shift symmetry of $b^{a}$ is broken by the presence of terms $\sim \hat{k}_{\alpha}$ etc.
%This needs to be contrasted with our observation in Sect.~\ref{sec:MismatchDisc} for $W=W_{0}(U^{i},S)$.
%The main difference is that 
due to the way the chiral variables $T_{\alpha}$ and $G^{a}$ are coupled in the Kähler potential Eq.~\eqref{eq:YExplicit}.
% and thence from the inverse Kähler metric \eqref{eq:InvK}.
%Consequently, a $T_{\alpha}$-dependent superpotential automatically induces a dependence on the $b^{a}$ in the scalar potential through the inverse K\"ahler metric in \eqref{eq:master2}.

Finally, we solve the stationary point conditions for the axions in the exact scalar potential.\footnote{While this is a self consistent class of solutions,
there can be other sets of solutions for the axions $\tilde{\rho}_{\alpha}$.}
To this end, we set for the complex parameters
\begin{equation}
w_{F}=|w_{F}|\ee^{i\lambda_{F}}\kom w_{H}=|w_{H}|\ee^{i\lambda_{H}}\kom A_{\alpha}=|A_{\alpha}|\ee^{i\lambda_{\alpha}}
\end{equation}
which allows us to determine that a solution to $\p_{c_{0}}V=0$ and $\p_{\tilde{\rho}_{\alpha}}V=0$ is
\begin{equation}\label{eq:LVSAxionMinima} 
\langle c_{0}\rangle=0\kom \lambda_{F}-\lambda_{H}=\dfrac{\pi}{2}\kom a^{\alpha}\langle\tilde{\rho}_{\alpha}\rangle-\lambda_{\alpha}+\lambda_{F}=\pi\, .
\end{equation}
%a^{\alpha}\langle\rho_{\alpha}\rangle-\lambda^{\alpha}+\lambda_{H}=\dfrac{\pi}{2}\kom
At this minimum, we have
\begin{align}
\text{Re}\left [(w_{F}+c_{0}w_{H})\bar{A}_{\gamma}\ee^{ia^{\gamma}\ov{T}_{\gamma}}\right ]&=-|w_{F}||A_{\gamma}|\ee^{-a^{\gamma}\sigma_{\gamma}}\, ,\nn\\
 \text{Im}\left [w_{H}\bar{A}_{\gamma}\ee^{ia^{\gamma}\ov{T}_{\gamma}}\right ]&=-|w_{H}||A_{\gamma}|\ee^{-a^{\gamma}\sigma_{\gamma}}\, ,\nn\\
\text{Re}\left [ A_{\gamma}\ov A_{\alpha}\ee^{-ia^{\gamma}T_{\gamma}+ia^{\alpha}\ov T_{\alpha}}\right ]&=|A_{\gamma}|\, |A_{\alpha}|\, \ee^{-a^{\gamma}\sigma_{\gamma}-a^{\alpha}\sigma_{\alpha}}\, . 
\end{align}
Similarly,
it is easy to see that $\p_{b^{a}}V=0$ is solved by (see also Sect.~\ref{sec:TachyonsTSUP})
\begin{equation}
\langle \hat{k}_{\alpha ab}b^{b}\rangle=0
\end{equation}
which is again solved for e.g. $\langle b^{a}\rangle=0\;\forall a$.
Below, we derive an exact expression for the Hessian at these non-SUSY minima for the NS-NS axions $b^{a}$.
By studying the simplest LVS set-up with an arbitrary number of odd moduli,
we show that tachyons can be avoided in most instances
which needs to be contrasted with the tachyonic no-go of \cite{Conlon:2006tq} for SUSY minima.

\subsection{A tree level superpotential from the non-geometric fluxes}\label{sec:NonGeoFluxSup} 

At tree level, there can be several possibilities to induce the K\"ahler moduli dependent pieces in the scalar potential, especially after including non-geometric fluxes \cite{Aldazabal:2006up,Benmachiche:2006df, Robbins:2007yv,Blumenhagen:2015kja,Shukla:2016xdy,Shukla:2016hyy}. To illustrate the form of the axionic potential $V(b^a)$ let us consider a superpotential of the following form,
\begin{equation}
W_0 =  w_{F} (U^i) + {S}\,  w_{H} (U^i)\kom W_1 = {\cal Q}^\alpha(U^i) \, T_\alpha \, ,
\end{equation}
where ${\cal Q}^\alpha$ can generically depend on the complex structure or axio-dilaton moduli, but we ignore this dependence subsequently. Such a superpotential with a linear dependence on the $T_{\alpha}$-moduli naturally arises in non-geometric flux compactifications \cite{Aldazabal:2006up,Robbins:2007yv,Blumenhagen:2013hva,Blumenhagen:2015kja,Shukla:2015hpa}, and their $S$-dual completions can also induce some $(S\, T_{\alpha})$-type superpotential couplings \cite{Aldazabal:2006up, Guarino:2008ik, Shukla:2015rua, Gao:2015nra, Shukla:2016hyy}.
Plugging the superpotential into \eqref{eq:VtauT0},
the $F$-term scalar potential reads
\begin{align}
V &= V_{\text{cs}} + \ee^{\cal K} \, \biggl[| w_{F}  + {S}\,  w_{H}  + T_\alpha{\cal Q}^\alpha|^2 + \gamma_{1} \, | w_{H} |^2 + (\gamma_1 \, \hat{{k}}_{\alpha} + \gamma_2\, {k}_\alpha) {\rm Re}[ w_{H} \, \ov {\cal Q}^\alpha]\nn\\
&\quad+ {\cal Q}^\alpha\, \biggl (\frac{4}{9}\, {\cal Y}^2\, {\cal G}_{\alpha \beta}  + s \, {\cal G}^{ab}\, {\hat{{k}}_{\alpha a}} \, {\hat{{k}}_{\beta b}} + \frac{(\gamma_1 \, \hat{{k}}_{\alpha} + \gamma_2\, {k}_\alpha)\,(\gamma_1 \, \hat{{k}}_{\beta} + \gamma_2\, {k}_\beta)}{4\, \gamma_1}\biggl ) \, \ov {\cal Q}^\beta  \\
&\quad - 4\, s \, {\rm Im}\bigl[( w_{F}  + {S}\,  w_{H}  + {T}_\alpha {\cal Q}^\alpha) \, \ov  w_{H} \bigr] - 4\, {\rm Im}(T_\alpha) \,\,  {\rm Im}\bigl[( w_{F}  + {S}\,  w_{H}  + T_\beta{\cal Q}^\beta) \, \ov {\cal Q}^\alpha\bigr]    \biggr]\,,\nn
\end{align}
The extremisation conditions for the $b^{a}$ are given as,
\begin{align}
\frac{\partial V}{\partial b^c}&= \ee^{\cal K} \, \biggl[ 2(\gamma_1-2s^{2}) \, \hat{{k}}_{\alpha c} {\rm Re}[ w_{H} \, \ov {\cal Q}^\alpha]+ {\cal Q}^\alpha\, \biggl (s \, {\cal G}^{ab}\,  ({\hat{{k}}_{\alpha a c}} \, {\hat{{k}}_{\beta b}}+{\hat{{k}}_{\alpha a}} \, {\hat{{k}}_{\beta b c}} )\nn\\
&\hphantom{= \ee^{\cal K} \, \biggl[} + \frac{\hat{{k}}_{\alpha c}(\gamma_1 \, \hat{{k}}_{\beta} + \gamma_2\, {k}_\beta)+(\gamma_1 \, \hat{{k}}_{\alpha} + \gamma_2\, {k}_\alpha)\,\hat{{k}}_{\beta c}}{2}\biggl ) \, \ov {\cal Q}^\beta \nn\\
&\hphantom{= \ee^{\cal K} \, \biggl[}  - 4s\,  \hat{k}_{\alpha c}\,  {\rm Im}\bigl[( w_{F}  + {S}\,  w_{H}  + T_\beta{\cal Q}^\beta) \, \ov {\cal Q}^\alpha\bigr]   - 4s\, {\rm Im}(T_\alpha)\hat{k}_{\beta c} \,\,  {\rm Re}\bigl[{\cal Q}^\beta \, \ov {\cal Q}^\alpha\bigr] \biggr]\, .
\end{align}
Hence, $\hat{k}_{\alpha a} = 0$ is always a solution and, in particular, $b^a = 0$. Although there can be other extrema, finding those analytically is a challenging task without specifying a particular model.
At the above stationary points,
one finds that the Hessian\footnote{The expression \eqref{eq:HessianNonGeoFlux} can easily be determined from our general expression \eqref{eq:HessianOddAxWTFull} for the Hessian to be derived below.}
\begin{align}\label{eq:HessianNonGeoFlux} 
\ee^{-\cK}\, \p_{b^{d}}\p_{b^{c}}V\bigl |_{b^{a}=0}&= \hat{k}_{\gamma c d}\biggl (2s\text{Im}\bigl[ {\cal Q}^{\gamma}(\ov w_{F}+c_{0}\ov w_{H})\bigl] +2(\gamma_{1}- 3 s^{2}) \, {\rm Re}\bigl[ {\cal Q}^{\gamma}  \ov {w}_{H}\bigr]\nn\\
&\hphantom{= \hat{k}_{\gamma c d}\biggl (}+ (2s+\gamma_{2}) \, k_\alpha \, {\rm Re}\bigl[{\cal Q}^{\gamma} \ov {\cal Q}^{\alpha}\bigr] \biggl )+2s\hat{k}_{\gamma ec}\,\cG^{ef}\, \hat{k}_{\alpha fd} {\cal Q}^{\gamma}\ov {\cal Q}^{\alpha}\, .
\end{align}
From this expression, we deduce that the presence of tachyons is determined by the intersection numbers $\hat{k}_{\gamma c d}$ which are highly model dependent.
Once an explicit model is specified, Eq.~\eqref{eq:HessianNonGeoFlux} is readily applicable to derive the mass contribution to the $b^{a}$ axions.
Again, our analysis shows that, in order to (partially) fix the odd moduli, in particular the $b^a$ axions, a superpotential dependence on $G^a$-moduli is no prerequisite and the presence of $T_{\alpha}$ moduli could do the job.

\subsection{Tachyons in SUSY and non-SUSY odd moduli stabilisation}\label{sec:TachyonsTSUP} 

An imperative question concerns the stabilisation of NS-NS axions $b^{a}$ in the absence of a $G^{a}$-dependent superpotential.
It is a well-known fact that each unfixed axion in a SUSY AdS minimum comes with a tachyonic superpartner \cite{Conlon:2006tq}.
The situation for non-SUSY minima is far from clear, see however \cite{Hristov:2008if} for earlier attempts.
Below,
we derive a general expression for the Hessian for $b^{a}$-axions for general $T_{\alpha}$-dependent superpotentials which is afterwards applied to two separate set-ups,
namely SUSY AdS minima reproducing \cite{Conlon:2006tq}
and non-SUSY minima from non-perturbative superpotentials of Sect.~\ref{sec:NonPertTSup}.

In what follows,
we work in the basis of $\lbrace t^{\alpha},b^{a},\tilde{\rho}_{\alpha},\hat{c}^{a}\rbrace$ so that e.g. $\p_{b^{a}}\cK=0$.
Further,
we assume that the superpotential is of the form \eqref{eq:SuperpotentialTUS} ignoring the $U^{i}$ and $S$ dependence in $W_{1}=W_{1}(T_{\alpha})$ for simplicity.
Since the $b^{a}$ dependence in $V$ is given as in \eqref{eq:Vba},
we can write
\begin{equation}
\p_{b^{c}}V=2\hat{k}_{\gamma c}\, \p_{\hat{k}_{\gamma}}V+\left (\hat{k}_{\gamma ec}\hat{k}_{\delta f}+\hat{k}_{\gamma e}\hat{k}_{\delta f c}\right )\p_{K_{\gamma e\delta f}}V\, .
\end{equation}
Clearly, $\hat{k}_{\gamma c}=0$ is a solution of $\p_{b^{c}}V=0$ which can be solved for $b^{a}=0$.
Then we obtain for the Hessian
\begin{align}
\p_{b^{d}}\p_{b^{c}}V&=2\hat{k}_{\gamma c d}\, \p_{\hat{k}_{\gamma}}V+4\hat{k}_{\gamma c}\hat{k}_{\delta d}\, \p_{\hat{k}_{\delta}}\p_{\hat{k}_{\gamma}}V+2\hat{k}_{\gamma c d} (\p_{b^{d}}K_{\lambda e\delta f}) \, \p_{\hat{k}_{\gamma}}\p_{K_{\lambda e\delta f}}V\nn\\
&\quad+\left (\hat{k}_{\gamma ec}\hat{k}_{\delta f d}+\hat{k}_{\gamma e d}\hat{k}_{\delta f c}\right )\p_{K_{\gamma e\delta f}}V+(\p_{b^{c}}K_{\gamma e\delta f}) (\p_{b^{d}}K_{\lambda g\rho h}) \p_{K_{\lambda g\rho h}}\p_{K_{\gamma e\delta f}}V\nn\\
&\quad+2\hat{k}_{\lambda d}(\p_{b^{c}}K_{\gamma e\delta f}) (\p_{\hat{k}_{\lambda}}\p_{K_{\gamma e\delta f}}V)
\end{align}
At the minimum $\hat{k}_{\alpha a}=0$,
we are left with
\begin{align}
\p_{b^{d}}\p_{b^{c}}V\bigl |_{\hat{k}_{\alpha a}=0}&=2\hat{k}_{\gamma c d}\, \p_{\hat{k}_{\gamma}}V\bigl |_{\hat{k}_{\alpha a}=0}+2s\ee^{\cK}\hat{k}_{\gamma e(c|}\hat{k}_{\delta f |d)} \cG^{ef}\, (W_{1})_{T_{\gamma}}(\ov W_{1})_{\ov T_{\delta}}\bigl |_{\hat{k}_{\alpha a}=0}\, .
\end{align}

To continue,
we use \eqref{eq:N1coords} to find
\begin{equation}
T_\alpha =\tilde{\rho}_\alpha + \frac{i}{2} \, \left(s\, \hat{{k}}_{\alpha}  - \, {k}_{\alpha}\right)\quad\Rightarrow\quad \p_{\hat{k}_{\gamma}}W=\dfrac{is}{2}\p_{T_{\gamma}}W\, .
\end{equation}
Using our result \eqref{eq:eq:VtauT00},
\begin{comment}
we obtain
\begin{align}\label{eq:HessianOddAxWT} 
2\p_{\hat{k}_{\gamma}}V_{0}\ee^{-\cK}\bigl |_{\hat{k}_{\alpha a}=0}&= -2s\text{Im}\bigl[(W_{1})_{T_{\gamma}}\overline{W}\,\bigl] - 4 s^{2} \, {\rm Re}\bigl[( W_1)_{T_{\gamma}} \ov {(W_0)}_{\ov{S}}\bigr]  \\
&\quad  - 2s\left(\frac{4{\cal Y}^2 \, {\cal G}_{\alpha\beta}}{9} + \frac{\gamma_2^2\, k_\alpha\, k_\beta}{4\, \gamma_1}\right)\,  \text{Im}\bigl [{(W_1)}_{T_\alpha\, T_{\gamma}}\, \ov {(W_1)}_{\ov T_{\beta}}\bigl ] \nonumber\\
&\quad+ 2s \, k_\alpha \, {\rm Re}\bigl[(W_1)_{T_{\gamma}} \ov {(W_1)}_{\ov T_{\alpha}}\bigr]- 2s \, k_\alpha \, {\rm Re}\bigl[W \ov {(W_1)}_{\ov T_{\alpha} \ov T_{\gamma}}\bigr] \nn\\
&\quad+ s\, \gamma_2 \, k_\alpha\, \, {\rm Im}\bigl[{(W_0)}_{S} \, \ov {(W_1)}_{\ov T_{\alpha} \ov T_{\gamma}}\bigl] \, ,\nonumber\\
\p_{\hat{k}_{\gamma}}V_{1}\ee^{-\cK}\bigl |_{\hat{k}_{\alpha a}=0} &= \gamma_1 \, {\rm Re}\bigl [{(W_0)}_{S} \, \ov {(W_1)}_{\ov T_\gamma}\bigl] - \, 2 \, s \, {\rm Im}\bigl[W \ov {(W_1)}_{\ov T_{\gamma}}\bigr]\nonumber\\
\p_{\hat{k}_{\gamma}}V_{2}\ee^{-\cK}\bigl |_{\hat{k}_{\alpha a}=0} &= \frac{\gamma_2}{2}\,  k_\alpha \text{Re}\bigl [{(W_1)}_{T_\gamma} \, \ov {(W_1)}_{\ov T_{\alpha}}\bigl ]\, .
\end{align}
Altogether,
\end{comment}
the Hessian can be written as
\begin{align}\label{eq:HessianOddAxWTFull} 
\ee^{-\cK}\, \p_{b^{d}}\p_{b^{c}}V\bigl |_{\hat{k}_{\alpha a}=0}&=\hat{k}_{\gamma c d}\biggl \{2s\text{Im}\bigl[(W_{1})_{T_{\gamma}}\overline{W}\,\bigl] + 2(\gamma_{1}-2 s^{2}) \, {\rm Re}\bigl[( W_1)_{T_{\gamma}} \ov {(W_0)}_{\ov{S}}\bigr] \nn \\
&\quad+ (2s+\gamma_{2}) \, k_\alpha \, {\rm Re}\bigl[(W_1)_{T_{\gamma}} \ov {(W_1)}_{\ov T_{\alpha}}\bigr]\nn\\
&\quad  - 2s\left(\frac{4{\cal Y}^2 \, {\cal G}_{\alpha\beta}}{9} + \frac{\gamma_2^2\, k_\alpha\, k_\beta}{4\, \gamma_1}\right)\,  \text{Im}\bigl [{(W_1)}_{T_\alpha\, T_{\gamma}}\, \ov {(W_1)}_{\ov T_{\beta}}\bigl ] \nonumber\\
&\quad- 2s \, k_\alpha \, {\rm Re}\bigl[W \ov {(W_1)}_{\ov T_{\alpha} \ov T_{\gamma}}\bigr] + s\, \gamma_2 \, k_\alpha\, \, {\rm Im}\bigl[{(W_0)}_{S} \, \ov {(W_1)}_{\ov T_{\alpha} \ov T_{\gamma}}\bigl]\biggl\} \nn\\
&\quad+2s\hat{k}_{\gamma e(c|}\hat{k}_{\delta f |d)} \cG^{ef}\, (W_{1})_{T_{\gamma}}(\ov W_{1})_{\ov T_{\delta}}\, .
\end{align}
Up to this point,
we have not made any assumptions about the VEVs for the other moduli which typically cannot be solved for analytically.
Nonetheless,
we can apply \eqref{eq:HessianOddAxWTFull} to special scenarios that allow us to fix certain subsets of fields exactly.
%In the remainder of this section,
%we distinguish three set-ups to gauge the presence of potential tachyonic directions.

\subsubsection*{SUSY minima}

As a first test of our result,
we look at the supersymmetric case where for $\langle W\rangle\neq 0$
\begin{equation}
W_{T_{\alpha}}=-K_{T_{\alpha}}W\kom W_{S}=-K_{S}W\kom K_{G^{a}}=0\, .
\end{equation}
Then, one verifies that
\begin{align}
- 2s\left(\frac{4{\cal Y}^2 \, {\cal G}_{\alpha\beta}}{9} + \frac{\gamma_2^2\, k_\alpha\, k_\beta}{4\, \gamma_1}\right)\,  \text{Im}\bigl [{(W_1)}_{T_\alpha\, T_{\gamma}}\, \ov {(W_1)}_{\ov T_{\beta}}\bigl ]&= 2s \, k_\alpha \, {\rm Re}\bigl[W \ov {(W_1)}_{\ov T_{\alpha} \ov T_{\gamma}}\bigr]\\
&\quad- s\, \gamma_2 \, k_\alpha\, \, {\rm Im}\bigl[{(W_0)}_{S} \, \ov {(W_1)}_{\ov T_{\alpha} \ov T_{\gamma}}\bigl]\nn
\end{align}
The remaining terms in \eqref{eq:HessianOddAxWTFull} simplify to
\begin{align}
\langle\p_{b^{d}}\p_{b^{c}}V\rangle_{\text{SUSY}} &=2|W|^{2}\ee^{\cK}\biggl \{\hat{k}_{ c d} \biggl[ \dfrac{3s}{2\cY}-  3\left (1-\dfrac{\hat{\xi}}{2\cY}\right ) \frac{\gamma_{2}}{4{\cal Y}} +\gamma_2\, \dfrac{3\cV }{4\cY^{2}} \biggl ]+\hat{k}_{ ec}\hat{k}_{ f d} \,  \frac{s\cG^{ef}}{4\, {\cal Y}^{2}} \biggl \}  \nn\\[0.3em]
&=-\dfrac{16s}{9}\cG_{ c d}   |m_{BF}|^{2}
\end{align}
in terms of the Breitenlohner-Freedman bound\footnote{More generally,
the Breitenlohner-Freedman bound asserts stability of the AdS vacuum provided fluctuations of scalar fields satisfy
\begin{equation*}
m^{2}>\dfrac{D-1}{2(D-2)}\kappa_{D}^{2}\langle V\rangle\xrightarrow{\;\; D=4\; \;}\dfrac{3}{4}\kappa_{4}^{2}\langle V\rangle=-\dfrac{9\kappa_{4}^{2}}{4}\ee^{\cK}|W|^{2}\, .
\end{equation*}}
\cite{Breitenlohner:1982jf}
\begin{equation}
|m_{BF}|^{2}=\dfrac{9}{4}|W|^{2}\ee^{\cK}\, .
\end{equation}
Since $\cG_{ c d}$ is positive definite, we find $h^{1,1}_{-}$ tachyonic directions in agreement with \cite{Conlon:2006tq}.
This is because each unstabilised axion, here $\hat{c}^{a}$, comes with a tachyonic superpartner in a SUSY AdS vacuum.
Clearly,
this no-go result is generically avoided once non-perturbative effects in $\cK$ are included that lift the flat directions.
Similarly, explicit $G^{a}$-dependent superpotentials lead to a potential for $\hat{c}^{a}$ that break the continuous shift symmetry as discussed in Sect.~\ref{sec:GDepSup}.

\subsubsection*{Non-SUSY minima for non-perturbative superpotentials}

For SUSY breaking minima,
making any statements for general superpotentials seems impossible.
To continue,
we make the convenient ansatz \eqref{eq:SuperPotentialE3ETC} for non-perturbative E3/D7 superpotentials
treating the $A_{\alpha}$ again as constant numbers.
In this case, we obtain
\begin{align}
2\hat{k}_{\gamma c d}\, \p_{\hat{k}_{\gamma}}V\bigl |_{\hat{k}_{\alpha a}=0}&=\ee^{\cK}a^{\gamma}\hat{k}_{\gamma c d}\biggl \{ 2 s \left [-1+a^{\gamma} \, k_\gamma \right ]\text{Re}\left [(w_{F}+c_{0}w_{H})\bar{A}_{\gamma}\ee^{ia^{\gamma}\ov{T}_{\gamma}}\right ]\\
&\quad- \left ( 2s^{2}\left (1+a^{\gamma}k_{\gamma}\right )+s\gamma_2 \left (2+a^{\gamma}\, k_\gamma\right )\right ) \text{Im}\left [w_{H}\bar{A}_{\gamma}\ee^{ia^{\gamma}\ov{T}_{\gamma}}\right ]\nn\\
&\quad+2sa^{\gamma}a^{\alpha}\left(\frac{4{\cal Y}^2 \, {\cal G}_{\gamma\alpha}}{9} + \frac{\gamma_2^2\, k_\gamma\, k_\alpha}{4\, \gamma_1}\right) \text{Re}\bigl [A_{\gamma}\ov A_{\alpha}\ee^{-ia^{\gamma}T_{\gamma}+ia^{\alpha}\ov T_{\alpha}}\bigl ]\nn\\
&\quad+\biggl ( -2s \left (1+a^{\gamma}k_{\gamma}-a^{\alpha} k_\alpha \right ) +\gamma_2\,  k_\alpha a^{\alpha}\biggl ){\rm Re}\bigl[A_{\alpha} \overline{A}_{\gamma}\ee^{-ia^{\alpha}T_{\alpha}+ia^{\gamma}\ov T_{\gamma}} \bigr]  \biggl \}\nn\, .
\end{align}
Using the VEVs \eqref{eq:LVSAxionMinima} for the $c_{0}$ and $\tilde{\rho}_{\alpha}$ axions,
the Hessian can be written as
\begin{align}\label{eq:NonSUSYHessT} 
\langle\ee^{-\cK}\p_{b^{d}}\p_{b^{c}}V\rangle_{\text{axions}}&=-sa^{\gamma}\hat{k}_{\gamma c d}|A_{\gamma}|\ee^{-a^{\gamma}\sigma_{\gamma}} \biggl \{ 2  \left [-1+a^{\gamma} \, k_\gamma \right ]|w_{F}|  \nn\\
&\quad-\left ( 2s\left (1+a^{\gamma}k_{\gamma}\right )+\gamma_2 \left (2+a^{\gamma}\, k_\gamma\right )\right ) |w_{H}| \biggl \}\nn\\
&\quad+
a^{\gamma}\biggl \{2\hat{k}_{\gamma c d}sa^{\gamma}a^{\alpha}\left(\frac{4{\cal Y}^2 \, {\cal G}_{\gamma\alpha}}{9} + \frac{\gamma_2^2\, k_\gamma\, k_\alpha}{4\, \gamma_1}\right)\nn\\
&\hphantom{=+\ee^{\cK}a^{\gamma}\biggl \{}+\hat{k}_{\gamma c d}\, \biggl (-2s\left (1+a^{\gamma}k_{\gamma}-a^{\alpha} k_\alpha \right ) +\gamma_2\,  k_\alpha a^{\alpha}\biggl ) \nn\\
&\hphantom{=+\ee^{\cK}a^{\gamma}\biggl \{}+4sa^{\alpha}\hat{k}_{\gamma e(c|}\cG^{ef}\hat{k}_{\alpha f |d)}\biggl \}|A_{\gamma}|\, |A_{\alpha}|\, \ee^{-a^{\gamma}\sigma_{\gamma}-a^{\alpha}\sigma_{\alpha}} 
\end{align}
where $\langle\ldots \rangle_{\text{axions}}$ implies setting the axions to their VEVs.
This exact result determines the Hessian for the $b^{a}$-axions,
thereby generalising the
approximate expression in \cite{Hristov:2008if}.
%Eq.~(5.12) 
Notice that we have not made any assumptions about the minimum of the $t^{\alpha}$ or $s$ which typically need to be determined numerically.

\subsubsection*{Simplest LVS set-up with arbitrary $h^{1,1}_{-}$}

To extract more information from \eqref{eq:NonSUSYHessT},
we look at the simplest set-up with $h^{1,1}_{+}=2$ and arbitrary $h^{1,1}_{-}$ and restrict to the leading order contributions in a volume expansion.
For the volume at the minimum $\hat{k}_{\alpha a}=0$ for the $b^{a}$ axions,
we can write
\begin{equation}
\cV=\dfrac{1}{6}\left (\kappa_{111}(t^{1})^{3}+\kappa_{222}(t^{2})^{3}\right )=d_{1}\tau_{1}^{3/2}-d_{2}\tau_{2}^{3/2}
\end{equation}
in terms of
\begin{equation}
d_{1}=\dfrac{\sqrt{2}}{3\sqrt{\kappa_{111}}}\kom d_{2}=\dfrac{\sqrt{2}}{3\sqrt{\kappa_{222}}}\, .
\end{equation}
One then obtains the following relations
\begin{equation}\label{eq:IDLVSExHess} 
\tau_{2}=\dfrac{\p \cV}{\p t^{2}}=\dfrac{1}{2}k_{2}\kom t^{2}=-\sqrt{\dfrac{2\tau_{2}}{\kappa_{222}}}\kom k_{22}=\kappa_{222}t^{2}=-\dfrac{2\sqrt{\tau_{2}}}{3d_{2}}\, .
\end{equation}

Recall from \eqref{eq:CondWZeroSN} that the leading order SUSY conditions for $S$ imply
\begin{equation}
4\, \langle s \rangle^2 \, |w_{H}|^{2}  = |W_0|^2=4|w_{F}|^{2}\, .
\end{equation}
Clearly, $\langle s \rangle$ is corrected by terms suppressed in the volume which we ignore for the moment.
In the limit $\epsilon_{2} = \frac{1}{4a_{2}\tau_2}\ll 1$ and $\cV\gg 1$, the minimum for the Kähler moduli is determined by \cite{Balasubramanian:2005zx}
\begin{align}
\label{Tstab1}
\cV &= \frac{3d_{2} \, \sqrt{\tau_2}\,(1-4\epsilon_{2})}{4a_{2}  (1 - \epsilon_{2})}\,\frac{|W_0|}{|A_{2}|}  \,e^{a_{2} \tau_2}
\simeq \frac{3d_{2} \, \sqrt{\tau_2}}{4a_{2}}\,\frac{|W_0|}{|A_{2}|}  \,e^{a_{2}\tau_{2}} \,, \\
\dfrac{\hat{\xi}}{2}  &= \frac{d_{2}(1-4\epsilon_{2})}{(1-\epsilon_{2})^2}\,\tau_{2}^{3/2}
\simeq d_{2}\tau_{2}^{3/2}\,.
\label{Tstab2}
\end{align}
Altogether,
\eqref{eq:NonSUSYHessT} becomes, after ignoring terms $\sim \gamma_{2}$ as well as taking care of the last term in the last line,\footnote{Similar results for the Hessian for LVS models with $h^{1,1}_{\pm}=2$ were previously obtained in \cite{Ben-Dayan:2014lca}. Notice though that in the convention of \cite{Ben-Dayan:2014lca} an additional minus sign is introduced in \eqref{eq:TachyonHessianNonSUSYEd3}.}
\begin{align}\label{eq:TachyonHessianNonSUSYEd3} 
\p_{b^{d}}\p_{b^{c}}V\bigl |_{\hat{k}_{\alpha a}=0}&=\hat{k}_{2 c d}\; \dfrac{3d_{2}\sqrt{\tau_{2}} |W_0|^{2}\ee^{\mathbb{K}}}{4\cV^{3}}\begin{cases}
2\bigl [ 1+a_{2}\tau_{2} \bigl ] +\,\cO(\cV^{-1})&\hat{k}_{1ab}=0\;\forall a,b \\
\bigl [ 1+2a_{2}\tau_{2} \bigl ]+\,\cO(\cV^{-1/3}) &\hat{k}_{1ab}\neq 0 \, .
\end{cases}
\end{align}
To identify potential tachyons at the vacuum,
we recall that
\begin{equation}
\cG_{cd}=- \frac{\hat{k}_{1cd}t^{1}+\hat{k}_{2cd}t^{2}}{4\, {\cal Y}}
\end{equation}
must be positive definite as a proper metric where $t^{1}>0$ and $t^{2}<0$.
In general,
we thus expect:
\begin{itemize}
\item if $\hat{k}_{1cd}=0$ for all $c,d$, then $\hat{k}_{2 cd}>0$ is positive definite and hence no tachyons.
\item if $\hat{k}_{1 cd}\neq 0$ for some $c,d$, then $\hat{k}_{2 cd}>0$ and $\hat{k}_{2 cd}<0$ are allowed since there is no restriction coming from demanding $\cG_{cd}>0$.
In this case, tachyons can appear depending on the model's intersection structure with their number being determined by the number of negative eigenvalues of $\hat{k}_{2 c d}$.
\end{itemize}

\noindent
As a final comment, we stress that the statements and results derived in this section are applicable specifically to $\mathrm{AdS}_{4}$ vacua considering only $F$-term contributions.
Thus far, we neglected both uplifting contributions to $\mathrm{dS}_{4}$ minima as well as $D$-term scalar potentials.
Generically,
at least the former are independent of odd axions $\hat{c}^{a}$ and $b^{a}$ when written in terms of volume moduli which is why we expect the above analysis to easily extend to scenarios with de Sitter uplifts.
%be directly implementable to the models equipped with the de Sitter uplifting proposals.
In the presence of chiral matter on e.g. D7/D3-branes,
the associated $D$-term contributions \cite{Blumenhagen:2007sm, Blumenhagen:2008zz} can have an induced dependence on both $T_{\alpha}$ and $G^{a}$, see e.g. \cite{Cicoli:2012vw}.
Given that such effects are highly model dependent,
we refrained from adding them to our present considerations, albeit it should be straight forward once a particular background configuration has been established.
%Further sources ignored in our analysis are NS5/D5-brane potentials which play a prominent role in axion monodromy \cite{Silverstein:2008sg,McAllister:2008hb}.

%%%%%%%%%%%%%%%%%%%%%%%%%%%%%%%%%%%%%%%%%%%%%%%%%%%%%%%%%%%%%%%%%%%%%%%%%%%%%%%%%%%%%%%%%%%%%%%%
%%%%%%%%%%%%%%%%%%%%%%%%%%%%%%%%%%%%%%%%%%%%%%%%%%%%%%%%%%%%%%%%%%%%%%%%%%%%%%%%%%%%%%%%%%%%%%%%
%%%%%%%%%%%%%%%%%%%%%%%%%%%%%%%%%%%%%%%%%%%%%%%%%%%%%%%%%%%%%%%%%%%%%%%%%%%%%%%%%%%%%%%%%%%%%%%%
%%%%%%%%%%%%%%%%%%%%%%%%%%%%%%%%%%%%%%%%%%%%%%%%%%%%%%%%%%%%%%%%%%%%%%%%%%%%%%%%%%%%%%%%%%%%%%%%

\section{Analysing the scalar potential for $W= W(U^i, S, T_\alpha, G^a)$}
\label{sec:GDepSup}

In our previous considerations, the R-R axions $c^a$ are never stabilised as no scalar potential is generated when the superpotential is independent of $G^{a}$. In such cases, the shift symmetry of the $c^a$ axions protects the flatness. In this section, we discuss superpotentials  $W \equiv W(G^a)$ which are suitable for stabilising all moduli simultaneously.
We present a form of the scalar potential which can be applied to specific models by merely giving the inputs of a superpotential depend on all moduli.

We define the superpotential
\begin{equation}
W=W_{0}(U^{i},S)+W_{1}(U^{i},S,T_{\alpha},G^{a})
\end{equation}
where $W_{0}$ might be associated with the 3-form flux background,
whereas $W_{1}$ can arise from D5-instantons or fluxed E3-instantons.
Depending on the microscopic details,
the $T_{\alpha}$ and $G^{a}$ dependence in $W_{1}$ might decouple,
at least at leading order in some instanton expansion.
From our master formula \eqref{eq:master2},
we have already derived the most general expression for the $F$-term scalar potential \eqref{eq:ScalarPotGModGen}.
In the remainder of this section, we apply this result to explicit superpotentials.

\subsection{Geometric flux superpotential} %: $W(G) = W_0 + \mho_a\, G^a$}

At the tree level, there can be several possibilities to induce the K\"ahler moduli dependence along with the odd moduli in the scalar potential, especially after including (non-)geometric fluxes \cite{Aldazabal:2006up,Benmachiche:2006df, Grana:2006hr,Robbins:2007yv,Blumenhagen:2015kja,Shukla:2016xdy,Shukla:2016hyy}. We initially consider the following ansatz for the superpotential with an explicit dependence on the odd moduli $G^a$,
\bea
\label{eq:WG-Lin}
&& W_1(G^a) = \mho_a\, G^a \quad \implies \quad \frac{\partial W_1}{\partial G^a} = \mho_a.
\eea
where $\mho_a$ is quantity which depends on the metric flux\footnote{Not all the components of geometric fluxes are allowed as the same are constrained by a set of quadratic flux constraints coming from the NS-NS Bianchi identities \cite{Grana:2006hr, Robbins:2007yv}.}, and can generically depend on the complex structure moduli as well. However,
we take $\mho_a$ to be some complex number for now. Subsequently the scalar potential becomes,
\bea
& & V(b^a, c^a) = \ee^{{\cal K}} \,\biggl( \dfrac{4l-3\tilde{\xi}}{l-3\tilde{\xi}}\, |W|^{2}+ \mho_a \left(s\, {\cal G}^{ab} + \gamma_1\, b^a\, b^b \right) \ov{\mho_b} - \, 4\, s\, b^a \text{Im}\bigl [W\ov{\mho}_{a}\bigl ]\biggl ) \, .
\eea
Therefore, the simplified scalar potential is a quadratic function of $b^a$ and $c^a$ moduli taking the following form,
\bea
& & V(b^a, \hat{c}^a) = 4\, \ee^{{\cal K}} \,\biggl[\hat{c}^a {\cal P}_{ab}\,\hat{c}^b + \hat{c}^a \, {\cal P}_a +  \frac{1}{4} \, \gamma_1 \, b^a {\cal P}_{ab}\,b^b - b^a \, {\cal Q}_a + {\cal P}_0 \biggr],
\eea
where we assume that complex structure is fixed (i.e., $W_{0}=\text{const}.$).
We further use the redefined combination of the $c^a$ axion $\hat{c}^a = c^a + c_0\, b^a$ (recall \eqref{eq:ConventionModAx}), along with (setting $\tilde{\xi}=0$ for convenience)
\bea
& & {\cal P}_{ab} = \text{Re}(\mho_a)\,\text{Re}(\mho_b) + \text{Im}(\mho_a)\,\text{Im}(\mho_b), \\
& & {\cal P}_a = 2\,\text{Re}(W_0) \text{Re}(\mho_a) + 2\, \text{Im}(W_0) \text{Im}(\mho_a), \nonumber\\
& & {\cal Q}_a = s\,\text{Re}(W_0) \text{Im}(\mho_a) - s\, \text{Im}(W_0) \text{Re}(\mho_a), \nonumber\\
& & {\cal P}_0 = \text{Re}({W_0})^2 + \text{Im}({W_0})^2 + \frac{s}{4}\, \text{Re}(\mho_a)\,{\cal G}^{ab}\,\text{Re}(\mho_b) + \frac{s}{4}\, \text{Im}(\mho_a)\,{\cal G}^{ab} \, \text{Im}(\mho_b). \nonumber
\eea
All the quantities ${\cal P}_{ab}, {\cal P}_a, {\cal Q}_a$ and ${\cal P}_0$ are independent of the odd axions $\{b^a, c^a\}$.
The exremisation of the potential with respect to the $\{b^a, \hat{c}^a\}$ axions results in the following conditions,
\bea
& &  {\cal P}_a +2\,  {\cal P}_{ab} \, \hat{c}^b = 0, \qquad {\cal Q}_a - \frac{1}{2} \,\gamma_1 {\cal P}_{ab}\,b^b = 0,
\eea
which leads to
\bea
& & \big\langle V(b^a, \hat{c}^a) \big\rangle = - 4\, \ee^{{\cal K}} \,\biggl[-\dfrac{1}{2} \langle \hat{c}^a \rangle \, {\cal P}_a +  \frac{1}{2} \, \langle b^a \rangle \, {\cal Q}_a - {\cal P}_0 \biggr]\, .
\eea
With the choice of variables $\{b^a, \hat{c}^a\}$, the Hessian turns out to be block diagonal, and $V_{ij}$ evaluated at the extremum reads
\bea
\label{eq:Hessian-geomeG}
& & \left\langle \frac{\partial^2 V}{\partial \hat{c}^a \partial{\hat{c}}^b} \right\rangle = 8\, \ee^{{\cal K}}{\cal P}_{ab}, \qquad \left\langle \frac{\partial^2 V}{\partial {b}^a \partial{{b}}^b} \right\rangle = 2\, \gamma_1\, \ee^{{\cal K}}{\cal P}_{ab}.
\eea
For the current ``two-field" analysis using $\{b^a, c^a\}$ let us have some leading order estimates for the axion decay constants so that we could have some estimates for the axionic masses as well. For that purpose we consider,
\bea
& & K_{G^a \ov{G^b}} \partial_\mu G^a \, \ov\partial^\mu \ov{G^b} = \frac{1}{2} f_{ab} \left(\partial_\mu\hat{b}^a \partial^\mu\hat{b}^b + \partial_\mu\hat{c}^a \partial^\mu\hat{c}^b \right)
\eea 
where 
\bea
& & \hat{b}^a = s\, b^a, \quad \hat{c}^a = c^a + c_0\, b^a, \quad f_{ab} = 2 K_{G^a \ov{G^b}} = \frac{2{\cal G}_{ab}}{s} + \frac{9 {\cal G}^{\alpha\beta} \hat{k}_{\alpha a}\hat{k}_{\beta b}}{2{\cal Y}^2} \simeq \frac{2{\cal G}_{ab}}{s} .
\eea
This results in a leading order decay constant matrix of the following form,
\bea
& & f_{b^ab^b} = 2\, s\, {\cal G}_{ab}, \qquad  f_{c^ac^b} = \frac{2\,{\cal G}_{ab}}{s}
\eea
Using (\ref{eq:Hessian-geomeG}) and recall that for the tree level K\"ahler potential $\gamma_1 = 4\, s^2$, and subsequently the $b^a$ axions turn out to be the massive as compared to the $\hat{c}^a$ axions,
\bea
& & \frac{m_{b^a}^2}{m_{\hat{c}^a}^2}  = \langle s \rangle ^2 = \frac{1}{g_s^2} \, 
\eea
However, if one works with a field $\hat{b}^a= s \, b^a$, and then this two field dynamics shows that the masses of $\hat{c^a}$ and $\hat{b}^a$ remain the same. This is well anticipated given that the source of the shift symmetry breaking for both the axions $\{b^a, \hat{c}^a\}$
is the same coupling in the superpotential, $W_1 = \mho_a\, G^a$. This observation about the ``no mass-splitting" for the $\hat{c}^a$ and $\hat{b}^a = s \, b^a$ fields at tree level is on the same footing in the canonical normalisation, and needs to be checked in generic scenarios when more moduli are part of the dynamics!

Similar to the two-field analysis of the axio-dilaton $\{c_0, s\}$ in Sect.~\ref{sec:MassSplittingBBHL}, the BBHL correction can induce some mass-splitting among the odd axions which reads
\bea
& & \frac{m_{{b}^a}^2}{m_{\hat{c}^a}^2} = \frac{1}{4} \langle \gamma_1 \rangle =\dfrac{1}{g_{s}^{2}}\left (1 + \frac{3\, \hat\xi}{4({\cal V}-\hat\xi)}\right )\, .
\eea
Given that the deviation is suppressed by powers of the volume,
this mechanism does not allow for mass hierarchies between the $\hat{c^a}$ and $\hat{b}^a$ axions. Nevertheless, let us mention that the $b^a$ axions are always heavier than the $c^a$ axions by the appearance of the string coupling.

\subsection{D5-brane gaugino condensates} %: $W(G) = W_0 + A\, \ee^{i\, n_a\, G^a}$}

A second source of $G^{a}$-dependent contributions to the superpotential is generated by gaugino condensation on spacetime filling D5-branes or ED1-brane instantons wrapping internal 2-cycles \cite{Grimm:2007xm,Grimm:2007hs, Ben-Dayan:2014zsa, Ben-Dayan:2014lca}.
%Similar effects can in principle also come from as analysed e.g. in \cite{Grimm:2007hs,Ben-Dayan:2014lca}.
However, contrary to the case of D5/ED1-contributions to the Kähler potential,
it remains unclear whether such effects can actually be generated in the superpotential since they survive the limits $\cV\rightarrow \infty$ and $g_{s}\rightarrow 0$ contrary to general expectations \cite{Witten:1996bn}.
Nonetheless, let us assume that such effects from D5-branes are generated in which case the superpotential is given by
\begin{equation}
W_1(G^a) =  B\, \ee^{i\, n_a\, G^a}
\end{equation}
where we treat $B$ as well as $W_{0}$ as some complex numbers.
Then, the scalar potential becomes (setting $\tilde{\xi}=0$)
\begin{align}
V(b^a, c^a)& = \ee^{{\cal K}} \,\biggl(4|W|^{2}  + n_a\, n_b \left(s\, {\cal G}^{ab}  + \gamma_1\, b^a\, b^b \right) |W_1|^2+ \, 4\, s\, b^a n_{a}\text{Re}(W\ov W_{1})\biggr)\, .
\end{align}
Assuming that $B=|B|\ee^{i\lambda}$ and $W_0 = |W_0| \ee^{i \theta}$ we have the following form of the scalar potential,
\begin{align}
V(b^a, \hat{c}^a) &= 4\, \ee^{{\cal K}} \,\biggl[|W_0|^2 +  |B|\, |W_0|\, \ee^{-s\, n_a b^a} (2 + s n_a\, b^a) \, \cos[n_a \hat{c}^a+\lambda - \theta]  \nonumber\\
&\quad+ |B|^2 \, \ee^{-2\, s\, n_a b^a} (1 + s n_a\, b^a) + \frac{|B|^2}{4} \left(s\, {\cal G}^{ab} + \gamma_1 \, b^a \, b^b \right)\, n_a\, n_b \, \ee^{- 2\, s\,n_a b^a} \biggr]\, ,
\end{align}
where the $\hat{c}^a$ axions appears as an oscillatory function, while the $b^a$ axions also have exponentially suppressed contributions. Assuming $|B|\neq 0$, the two extremising conditions are given as,
\bea
& (i). & \quad n_a\,(2 + s n_b\, b^b) \sin[n_b \hat{c}^b+\lambda - \theta] = 0, \\
& (ii). & \quad \frac{n_a\, |W_0|}{|B| \, \ee^{-\, s\, n_b b^b}}\, (1 + s \, n_b\, b^b) \, \cos[n_c \hat{c}^c+\lambda - \theta] + \, n_a (1 + 2\, s n_c\, b^c)  \nonumber\\
& & \quad\;+ \, \frac{n_a}{2} \,\left(s\, ({\cal G}^{bc}n_b n_c)  + \frac{\gamma_1}{s} \, (n_b b^b)\, (s\, n_c b^c -1)\, \right) = 0\, . \nonumber
\eea
For $n_{a}\neq 0$, we find the solutions\footnote{We can also have e.g. $s n_b\, b^b=-2$ which however leads to $\langle\p_{\hat{c}^{c}}\p_{\hat{c}^{d}}V\rangle=0$ which is not a proper minimum of the potential.}
\begin{equation}
n_b \langle\hat{c}^b\rangle+\lambda - \theta=n\pi
\end{equation}
so that
\begin{equation}\label{eq:SolD5BAx} 
|B| \, \ee^{-\, s\, n_b b^b}=\dfrac{-4|W_{0}|\, (1 + s \, n_b\, b^b) (-1)^{n}}{2(1 + 2\, s n_c\, b^c)+\left(s\, {\cal G}^{ab} + \gamma_1 \, b^a \, b^b \right)\, n_a\, n_b-\dfrac{\gamma_{1}}{s}(n_{b}b^{b})}\, .
\end{equation}

%|B| \, \ee^{-\, s\, n_b b^b}=\dfrac{-4|W_{0}|\, (1 + s \, n_b\, b^b) (-1)^{n}}{2+\left(s\, {\cal G}^{ab} + 4s^{2} \, b^a \, b^b \right)\, n_a\, n_b}\, .

The Hessian is given by
\begin{align}
\p_{\hat{c}^{c}}\p_{\hat{c}^{d}}V&=-4n_{c}n_{d}\, \ee^{{\cal K}} \,   |B|\, |W_0|\, \ee^{-s\, n_a b^a} (2 + s n_a\, b^a) \, \cos[n_a \hat{c}^a+\lambda - \theta]\, ,\nn \\
\p_{b^{c}}\p_{b^{d}}V&=-\dfrac{s^{3} n_a\, b^a\, \p_{\hat{c}^{c}}\p_{\hat{c}^{d}}V}{2 + s n_a\, b^a} +4s^{2}n_{c}n_{d}\, \ee^{{\cal K}} |B|^2\biggl [2-4s  \, b^b \, n_b\nn\\
&\quad+ \left(s\, {\cal G}^{ab} +4s^{2} \, b^a \, b^b \right)\, n_a\, n_b\biggl ] \, \ee^{- 2\, s\,n_a b^a}\, .
\end{align}
%\p_{b^{c}}\p_{b^{d}}V&=4s^{2}n_{c}n_{d}\, \ee^{{\cal K}} \,\biggl[  |B|\, |W_0|\, \ee^{-s\, n_a b^a} ( s n_a\, b^a) \, \cos[n_a \hat{c}^a+\lambda - \theta]  \nonumber\\
%& \quad +|B|^2\left [2-4s  \, b^b \, n_b+ \left(s\, {\cal G}^{ab} +4s^{2} \, b^a \, b^b \right)\, n_a\, n_b\right ] \, \ee^{- 2\, s\,n_a b^a} \biggr]\, .
At leading order ignoring $\hat{\xi}$, we set $\gamma_{1}=4s^{2}$ in \eqref{eq:SolD5BAx} so that
\begin{align}
\langle\p_{\hat{c}^{c}}\p_{\hat{c}^{d}}V\rangle&=n_{c}n_{d}\;\ee^{{\cal K}} \,  \dfrac{8|W_{0}|^{2}\, (1 + s \, n_b\, b^b)  (2 + s n_a\, b^a)}{2+\left(s\, {\cal G}^{ab} + 4s^{2} \, b^a \, b^b \right)\, n_a\, n_b}\, ,\nn\\
\langle\p_{b^{c}}\p_{b^{d}}V\rangle&=  \dfrac{s^{2}\, \langle\p_{\hat{c}^{c}}\p_{\hat{c}^{d}}V\rangle}{\left [2+\left(s\, {\cal G}^{ab} + 4s^{2} \, b^a \, b^b \right)\, n_a\, n_b\right ] (2 + s n_a\, b^a)}\,\biggl[  4+2s\cG^{ab}n_{a}n_{b}\nn\\
&\quad -6sb^{a}n_{a}+sb^{a}n_{a}\left (4s^{2}b^{b}b^{c}+s\cG^{bc}\right )n_{b}n_{c}  \biggr]
\end{align}
%\langle\p_{b^{c}}\p_{b^{d}}V\rangle&=s^{2}n_{c}n_{d}\; \ee^{{\cal K}} \dfrac{8|W_{0}|^{2}\, (1 + s \, n_b\, b^b)}{\left [2+\left(s\, {\cal G}^{ab} + 4s^{2} \, b^a \, b^b \right)\, n_a\, n_b\right ]^{2}}\,\biggl[  4+2s\cG^{ab}n_{a}n_{b}\nn\\
%&\quad -6sb^{a}n_{a}+sb^{a}n_{a}\left (4s^{2}b^{b}b^{c}+s\cG^{bc}\right )n_{b}n_{c}  \biggr]
where in both expressions $\langle\ldots\rangle$ for the VEVs is implicitly used for $s$ and $b^{a}$.
Hence
\begin{align}
\dfrac{\langle\p_{b^{c}}\p_{b^{d}}V\rangle}{\langle\p_{\hat{c}^{c}}\p_{\hat{c}^{d}}V\rangle}&= s^{2}\,\biggl[  1-\dfrac{8s  \, b^b \, n_b(1+sb^{a}n_{a})}{\left [2+\left(s\, {\cal G}^{ab} + 4s^{2} \, b^a \, b^b \right)\, n_a\, n_b\right ](2+sb^{a}n_{a})} \biggr]\, .
\end{align}
At $b^{a}=0$, the RHS reduces to $s^{2}$ as expected, though $b^{a}=0$ generically does not solve the condition \eqref{eq:SolD5BAx}.
Away from $b^{a}=0$, the mass splitting is at most polynomial in $b^{a}$.
Including the BBHL correction systematically as in Sect.~\ref{sec:MassSplittingBBHL} leads to additional corrections which are, however, suppressed by additional powers of the volume.
For convenience, we neglected any $T_{\alpha}$-dependence in the superpotential which contributes further terms to $\p_{b^{c}}\p_{b^{d}}V$ as previously determined in \eqref{eq:NonSUSYHessT}.
Even though this certainly induces a mass splitting between the $b^{a}$ and $\hat{c}^{a}$ axions,
it is unclear whether this is always sufficient to engineer mass hierarchies,
though this has been observed in many instances via suitable arrangements of the $n_{a}$ \cite{Berg:2009tg,Ben-Dayan:2014zsa,Ben-Dayan:2014lca}.

Before we continue,
we collect a final formula including an even moduli dependence as in \eqref{eq:SuperPotentialE3ETC}.
Indeed,
a general superpotential will include non-perturbative sources from D7-, D5- and D3-branes in which case
\begin{equation}\label{eq:SuperpotentialSTGD5} 
W=w_{F}+Sw_{H}+\sum_{\alpha=1}^{h_{+}^{1,1}}\, A_{\alpha}\ee^{-ia^{\alpha}T_{\alpha}}+\sum_{a=1}^{h_{-}^{1,1}}\, B_{a}\ee^{i n_{a}G^{a}}
\end{equation}
As discussed in Sect.~\ref{sec:NonPertTSup},
we ignore any $U^{i}$ or $S$ dependence in both prefactors $A_{\alpha}$ and $B_{a}$.
Plugging \eqref{eq:SuperpotentialSTGD5} into \eqref{eq:master2},
we find 
\begin{equation}
V=V^{\text{pert}}+V^{\text{np1}}+V^{\text{np2}}+V^{G}
\end{equation}
where $V^{\text{pert}},V^{\text{np1}},V^{\text{np2}}$ are defined in \eqref{eq:ScalarPotED3} and where
\begin{align}
V^{G}
&= \ee^{{\cal K}} \,\sum_{a=1}^{h^{1,1}_{-}}\, \biggl \{2\left (1+2sb^{a}n_{a}\right )\text{Re}\bigl [(w_{F}+c_{0}w_{H}) \bar{B}_{a}\ee^{-in_{a}\overline{G}^{a}}\bigl ]\nn\\[0.8em]
&\quad-2\left [s+b^{a}n_{a}\left (2s^{2}+ \gamma_1 \right )\right ]\text{Im}\bigl [ w_H \bar{B}_{a}\ee^{-in_{a}\overline{G}^{a}} \bigl ] \\[0.8em]
&\quad+\sum_{b=1}^{h^{1,1}_{-}}\, B_{a}\ov{B}_{b}\ee^{in_{a}G^{a}-in_{b}\ov{G}^{b}}\left (1+2s(b^{a}n_{a}+b^{b}n_{b})+n_{a}n_{b}\left [s\, {\cal G}^{ab} + \gamma_1\, b^a b^b\right ]\right ) \nn\\[0.8em]
&\quad+\sum_{\alpha=1}^{h^{1,1}_{+}}\biggl (2(1+a^{\alpha}(k_{\alpha}-s\hat{k}_{\alpha}))+4sb^{a}n_{a}\nn\\
&\quad-a^{\alpha}n_{a}\left [ 2\,s \, {\cal G}^{ab} \, \hat{{k}}_{\alpha b} + (\gamma_1 \, \hat{{k}}_{\alpha}\,+ \gamma_2\, {k}_\alpha)\, b^a \right ]\biggl ) \text{Re}\bigl [ A_{\alpha}\ov{B}_{a}\ee^{-ia^{\alpha}T_{\alpha}-in_{a}\overline{G}^{a}} \bigl ] \biggr \}\nn\, .
\end{align}
Again,
the scalar potential is known exactly without resorting to any small coupling or large volume expansion.
Given the way the $\hat{c}^{a}$ enter the potential,
we notice that \eqref{eq:LVSAxionMinima} still provides a solution of the stationary point conditions for $c_{0}$ and $\tilde{\rho}_{\alpha}$ provided
\begin{equation}
B_{a}=|B_{a}|\ee^{i\theta_{a}}\kom n_{a}\langle \hat{c}^{a}\rangle+\theta_{a}+\lambda_{F}=\pi\, .
\end{equation}
However,
solving the stationary point conditions for the NS-NS axions $b^{a}$ analytically is challenging as can already be seen from the simplified situation \eqref{eq:SolD5BAx} above.
This is similar to determining the VEVs for the $4$-cycle volumes in LVS for which analytic solutions can only be obtained in the large volume limit.

\subsection{Fluxed E3-instanton superpotential}\label{sec:FluxedE3Inst} 
 %: $W= W_0 + \sum_{\Sigma} \Theta_\Sigma({S}, G^a) \ee^{- i\, a_\Sigma^\alpha\, T_\alpha}$}

Another possibility to induce an explicit $G^{a}$ dependence in the scalar potential is through non-perturbative effects arising from the fluxed $E3$-instantons or via gaugino-condensation effects with magnetised-branes \cite{Grimm:2007xm,Grimm:2011dj,Gao:2013rra}. By modularity arguments, the non-perturbative superpotential from fluxed D3/D7-instantons wrapping divisors $D_{E}$ needs to have the following form \cite{Grimm:2007xm}
\begin{equation}
W_{np} ({S}, G^a, T_\alpha) =  \sum_{E} A_E\, \Theta_{E}({S}, G^a)\, \ee^{-i \, n_E{}^\alpha\, T_{\alpha}}\, .
\end{equation}
Here, $A_{E}$ denotes the 1-loop determinant for fluctuations
around the instanton, which only depends  on the complex structure moduli
and the $D7$-brane deformations.
Further,
we can write the modular function as
\begin{equation}\label{eq:ModFuncDef} 
\Theta_{E}(S,G^{a})=\sum_{{\cal F}_E} \, \ee^{i\beta_{E}S}\;\ee^{iq_{Ea}\, G^{a}}
\end{equation}
in terms of the flux dependent quantities
\begin{equation}
{q}_{Ea} =   \hat{k}_{\alpha a b}\, a_E^\alpha\,  {\cal F}_E^b\kom \beta_{E}={\textstyle\frac{1}{2}}
\Big[q_{Ec} \, {\cal F}_E^c  +2\pi
\int_{D_E}  {{\cal F}}_{E}^v \wedge {{\cal F}}_{E}^v\Big]\, .
\end{equation}
The gauge flux $\cF_{E}$ is separated into the components ${\cal F}_E^a$ from pulling back bulk 2-forms onto the brane
and variable flux ${{\cal F}}_{E}^v$ supported only on 2-cycles inside $D_{E}$ \cite{Grimm:2011dj}.
The sum over admissible gauge flux in \eqref{eq:ModFuncDef} is such that the functions $\Theta_{E}(S,G^{a})$ become appropriate holomorphic Jacobi forms which can be computed in particular limits in moduli space \cite{Grimm:2007xm}.

It is straight forward to plug the superpotential into the scalar potential \eqref{eq:ScalarPotGModGen} which will be analysed in an upcoming publication \cite{Cicoli:2021phenoOdd}.
Here, we simply point out that the form of $\Theta_{E}$ in \eqref{eq:ModFuncDef} is such that contributions to the mass of R-R axions $\hat{c}^{a}$ are exponentially suppressed by $\ee^{-\beta_{E}s}$.
At small string coupling $g_{s}=\langle s\rangle^{-1}$,
one thus expects to obtain mass hierarchies between the $\hat{c}^{a}$ and $b^{a}$ axions since the later receive further mass contributions from the inverse Kähler metric as computed in \eqref{eq:NonSUSYHessT}; a fact that has already been appreciated in \cite{Grimm:2007hs}.
This observation motivates further exploring the phenomenological implications of such scenarios which potentially open up new avenues towards inflation in string theory.

\section{Phenomenological implications}\label{sec:PhenoImplications}

Axions are ubiquitous in string compactifications borrowing their shift symmetries from the gauge redundancies of $p$-form fields in the higher-dimensional theories.
Generic models are expected to contain of $\cO(10-100)$ axionic fields \cite{Demirtas:2018akl,Carta:2020ohw,Mehta:2021pwf,Broeckel:2021dpz} making them highly attractive for model building purposes.
In fact, the rich cosmology of axions \cite{Marsh:2015xka} makes them the perfect target for the study of phenomenological implications of string theory, see e.g. \cite{Kim:2004rp,Svrcek:2006yi,Grimm:2007hs,Arvanitaki:2009fg,Cicoli:2012sz,Pajer:2013fsa,Baumann:2014nda}.

The results of the previous sections provide a systematic approach to computing the exact $F$-term scalar potentials for general $\cN=1$ CY orientifold compactifications and arbitrary superpotentials.
From a phenomenological point of view,
this is desirable because it enables us to derive simple conditions for stabilising moduli and for avoiding tachyonic directions
that, whenever analytic methods cease to work, can easily be implemented on a computer.
Further,
the precise notion of sub-leading terms in the volume and string coupling is necessary to have proper control over inflationary potentials.
This is of particular relevance for models of fibre inflation \cite{Cicoli:2008gp} or poly-instanton inflation \cite{Cicoli:2011ct,Blumenhagen:2012ue} where the scalar potential pieces determining the minimum are suppressed by $\cV^{-\kappa}$ with $\kappa>3$.

In our analysis,
we were mostly concerned with the general dependence of scalar potentials on odd axions $b^{a}$ and $\hat{c}^{a}$, especially due to their outstanding role in cosmological model building such as in axion monodromy \cite{Silverstein:2008sg,McAllister:2008hb}.
The basic idea in axion monodromy is the breaking of the discrete shift symmetry of axions by branes or fluxes in order to obtain a sequence of non-periodic branches.
Despite initial efforts in describing moduli stabilisation in these set-ups \cite{McAllister:2008hb},
a systematic understanding still remains a key challenge where
%Given that moduli stabilisation in these set-ups remains a key challenge,
our results provide the golden opportunity to make significant progress in future endeavours \cite{Cicoli:2021phenoOdd}.
More generally, a variety of inflationary models could benefit from our novel insights such as those based on alignment and hierarchical mixing of odd axions \cite{Berg:2009tg,Ben-Dayan:2014zsa,Gao:2014uha,Ben-Dayan:2014lca} or more recent set-ups like harmonic hybrid inflation \cite{Carta:2020oci}. On top of that, the plethora of potentially ultra-light axion-like particles in our set-ups may have direct applications to stringy realisations of
Dark Matter \cite{Hui:2016ltb},
Dark Radiation \cite{Cicoli:2012aq,Higaki:2012ar,Hebecker:2014gka,Cicoli:2015bpq},
Dark Energy \cite{Kaloper:2005aj,
Kaloper:2008qs,
Panda:2010uq,
Cicoli:2012tz,
Blaback:2013fca,
DAmico:2018mnx},
as well as astrophysics \cite{Cicoli:2014bfa,Cicoli:2017zbx}.
We hope to come back to these questions in the near future.

\section{Conclusions and future directions}
\label{sec_conclusions}

The quest for fully reliable string constructions with all moduli stabilised in well-controlled de Sitter minima in synergy with realistic particle phenomenology and cosmology remains to large extent unfulfilled.
A key obstacle is the systematic derivation of vacuum structures from $\cN=1$ scalar potentials including (non-)perturbative corrections as well as additional discrete parameters such as fluxes and triple intersection numbers.
In this article, we made substantial progress in this direction by computing explicit and exact expressions for $\alpha^{\prime}$- and $g_{s}$-corrected $F$-term scalar potentials without having to utilise any additional $4$D approximation.
Further, we revisited several issues pertaining to the stabilisation of odd moduli for which our observations provide a much sought after unifying framework.

The main results of this paper are the three master formulae for $\cN=1$ $F$-term scalar potentials derived in Sect.~\ref{sec:MasterFormulas} which are perfectly suited for stabilising closed string moduli in general type IIB CY orientifold compactifications.
These were obtained from exact identities for derivatives of the Kähler potential and, in particular, for the inverse Kähler metric at higher order in the $\alpha^{\prime}$ expansion using the tree level $(\alpha^{\prime})^{3}$ effects of \cite{Becker:2002nn}.
Further,
we touched upon higher order corrections in the closed string loop and non-perturbative D-instanton expansion.
Both are dictated by $\mathrm{SL}(2,\mathbb{Z})$ invariance of the 10D Einstein frame action through appropriate modular forms where in orientifold backgrounds a subgroup $\Gamma_{S}\subset \mathrm{SL}(2,\mathbb{Z})$ is expected to survive as a symmetry of the $4$D effective action \cite{Grimm:2007xm}.
We showed that even in this case, closed expressions for the inverse Kähler metric can be obtained giving rise to a compact formula for the $F$-term scalar potential.

The remainder of this paper was concerned with studying the nature of the scalar potential for a variety of different superpotentials.
In the simplest case of the GVW superpotential $W(U^i, S)$ depending only on complex structure moduli $U^{i}$ and the axio-dilaton $S$,
we showed explicitly that the scalar potential is independent of the $b^{a}$ axions.
This is of course expected in the absence of non-perturbative effects where the axionic shift symmetries are left untouched.
In addition, we briefly elaborated on splitting the masses of the universal axion $c_{0}$ and the dilaton $s$ through the BBHL correction.

In the presence of an explicit $T_{\alpha}$ dependence in the superpotential $W(U^i, S, T_\alpha)$,
the NS-NS axions $b^{a}$ are explicitly featured in the scalar potential.
We derived a general form of the $F$-term scalar potential for both non-perturbative effects from D3-/D7-branes generalising the results of \cite{AbdusSalam:2020ywo}
and non-geometric fluxes.
We illustrated the effectiveness of our results by deriving closed expressions for the Hessian for $b^{a}$ axions.
For SUSY minima,
we reproduced the results of \cite{Conlon:2006tq} where each unfixed R-R axion $\hat{c}^{a}$ has an associated tachyonic superpartner $b^{a}$.
In the non-SUSY case,
we formulated conditions on the intersection structure determining the presence and number of potential tachyonic directions.

Subsequently, we provided all the necessary tools to extract the $F$-term scalar potential for arbitrary superpotentials $W(U^i, S, T_\alpha, G^a)$.
This is of paramount importance for advancing our understanding of moduli stabilisation in generic set-ups with $h^{1,1}_{-}\neq 0$.
In this context, our results have direct applications to inflationary models based on odd axions, in particular axion monodromy \cite{Silverstein:2008sg,McAllister:2008hb}.
We derived explicit expressions for geometric flux and non-perturbative D5-gaugino superpotentials.
A particularly promising class of superpotentials from fluxed D3/D7-branes capable of inducing mass hierarchies between the NS-NS and R-R axions will be discussed in a forthcoming publication \cite{Cicoli:2021phenoOdd}.

In the present work, we restricted mostly to perturbative corrections to the K\"ahler potential arising from tree level $(\alpha^\prime)^3$-effects in $10$D.
Hence,
the stabilisation of the $C_2$-axions has been sourced by superpotential contributions only.
However, there is another possibility of stabilising odd moduli via worldsheet-instanton corrections to the K\"ahler potential \cite{RoblesLlana:2006is}.
The modular completion of such corrections results in a modified expression for the volume ${\cal Y}$ featured in the K\"ahler potential \eqref{eq:KgenKGS} given by \cite{Grimm:2007xm}
\begin{equation}
\label{eq:YwithWS}
{\cal Y} = {\cal V} + \frac{\zeta}{4} \, f_{0}({S}, \ov{S}) - 4 \, g({S}, \ov{S}, G^a, \ov G^a) \, .
\end{equation}
Here $f_{0}({S}, \ov{S})$ is defined in \eqref{eq:ModFunctionDef} and
\begin{equation}
g({S}, \ov{S}, G^a, \ov G^a) =\sum_{\substack{\beta\in H_{2}^{-}(X,\mathbb{Z}) \\[0.2em] (n, m)\neq(0,0)}} \, n_\beta^0\; \frac{s^{3/2}}{|n+ m {S}|^3}  \cos\biggl[(n+ {S} \, m) \frac{k_a^\beta (G^a - \ov G^a)}{{S} -\ov{S}} - m k_a^\beta G^a\biggr]
\end{equation}

\vspace*{-0.25cm}

\noindent in terms of integer genus zero Gopakumar-Vafa invariants $n_\beta^0$ \cite{Gopakumar:1998ii,Gopakumar:1998jq} and $k_{a}^\beta=\int_{\beta}\, \nu^{a}$ for a basis $\nu^{a}\in H^{2}_{-}(X,\mathbb{Z})$.
Thus far,
we limited ourselves to considering the $f_{0}({S}, \ov{S})$ piece without including the (modular completed version of the) worldsheet and D1-instanton effects encoded in $g({S}, \ov{S}, G^a, \ov G^a)$.
Similarly, given that worldsheet instantons such as the one in \eqref{eq:InstCorrections} have received a lot of attention recently,
see in particular \cite{Demirtas:2019sip,Demirtas:2021nlu},
it would be desirable to add these corrections in our general formalism.
As in the case of $f_{0}$, there exist closed expressions for the derivatives of polylogarithms which should allow for a straight forward generalisation of our framework.

Further, we focussed on analysing large complex structure pre-potentials which are commonly studied in the literature due their relevance in the context of mirror symmetry.
However,
the precise structure of pre-potentials highly depends on the monodromy symmetries and the additional data of the asymptotic regime in moduli space around which the periods are being expanded, see in particular \cite{Bastian:2021eom}.
Recently,
it was also suggested in \cite{Bastian:2021hpc} that other types of boundaries in moduli space can lead to small flux superpotentials with large mass hierarchies.
It would thus be instructive to extend our general results to other classes of pre-potentials.

Finally,
let us stress again that we studied mostly AdS$_{4}$ vacua ignoring additional uplifting sources, $D$-terms as well as open string moduli.
Clearly,
these effects are critical for the construction of fully explicit models including Standard Model sectors \cite{Cicoli:2021dhg} and de Sitter vacua from e.g. T-brane backgrounds \cite{Cicoli:2015ylx}.
Along the lines of \cite{Balasubramanian:2004uy,Westphal:2006tn,AbdusSalam:2007pm,Blaback:2013qza}, it would also be interesting to explore the option of getting de Sitter from the sources already discussed throughout the paper.
Treating such contributions notoriously remains challenging,
but our results build a solid foundation towards finding de Sitter minima with fully stabilised moduli in string theory.

\acknowledgments
We would like to thank Nicole E. Bollan and Fernando Quevedo for initial collaboration on this project.
We also thank Veronica Guidetti and Francisco Pedro for useful discussions.
AS acknowledges support by the German Academic Scholarship Foundation and by DAMTP through an STFC studentship. PS is grateful to Paolo Creminelli, Atish Dabholkar and Fernando Quevedo for their support, and also would like to thank the INFN-Bologna for hospitality during the initial stage of the work.

\appendix

\section{Useful relations for intermediate computations}
\label{app:UsefulIdentities}

In section~\ref{sec_scalarpotential},
we use the following identities
\begin{align}\label{eq:identitiesVariables} 
\partial_S t^\alpha &= \dfrac{i}{4} \hat{k}_\beta  k^{\alpha \beta}\kom\partial_{G^a} t^\alpha = - \dfrac{i}{2} k^{\alpha \beta} \hat{k}_{\beta a}\kom\partial_{T_\beta} t^\alpha = \dfrac{i}{2} k^{\alpha \beta}
\end{align}
to compute the derivatives of the Kähler potential.
Using the explicit expressions of the K\"ahler derivatives in Eq.~(\ref{eq:derK}) and of the inverse K\"ahler metric in Eqs.~(\ref{eq:InvK})--(\ref{eq:gamma123}), we find the following useful relations,
\bea
\label{eq:identities0}
& & K_{S} \, K^{{S} \ov{S}} = \frac{i \, s \, \left(4 \, {\cal V}-\hat{\xi }\right) \left(2 \, {\cal V}\, + \, 4 \, \hat{\xi }-\hat{k}_0 \,\, s\right)}{2 \left(\, {\cal V}-\hat{\xi }\right) \left(2 \, {\cal V} + \hat{\xi }\right)} = -\, K_{\ov{S}} \, K^{\ov{S} {S}}\,,\nonumber\\
& & K_{G^a} \, K^{G^a \ov{S}} = \frac{i \, \hat{k}_0 \, s^2 \, \left(4 \, {\cal V}-\hat{\xi }\right)}{\left(\, {\cal V}-\hat{\xi }\right) \left(2 \, {\cal V} + \hat{\xi }\right)} = -\, K_{\ov{G}^a} \, K^{\ov{G}^a {S}}\, ,\nonumber\\
& & K_{T_\alpha} \, K^{T_\alpha \ov{S}} = \frac{i \, s \, \left(\hat{k}_0 \, s \, \left(\hat{\xi }-4 \, {\cal V}\right)-18 \, \hat{\xi } \, {\cal V}\right)}{2 \left(\, {\cal V}-\hat{\xi}\right) \left(2 \, {\cal V}+\hat{\xi }\right)} = -\, K_{\ov{T}_\alpha} \, K^{\ov{T}_\alpha {S}}\, ,\nonumber\\
& & K_{S} \, K^{{S} \ov{G}^b} = -\frac{i \, s \, b^b \left(4 \,{\cal V}-\hat{\xi }\right) \left(\hat{k}_0 s-2 \left(2 \hat{\xi}+\,{\cal V}\right)\right)}{2 \left(\,{\cal V}-\hat{\xi }\right) \left(2 \,{\cal V}+\, \hat{\xi }\right)} = -\, K_{\ov{S}} \, K^{\ov{S} G^b} \,, \nonumber\\
& & K_{G^a} \, K^{G^a \ov{G}^b} = \frac{i \, s\, b^b  \left(\hat{k}_0 \, s \, \left(4 \,{\cal V}-\hat{\xi}\right)\right)}{({\,{\cal V}-\hat{\xi }})\, (2 \,{\cal V}+\hat{\xi })} -2\, i \, s\, b^b= -\, K_{\ov{G}^a} \, K^{\ov{G}^a G^b} \, ,\nonumber\\
& & K_{T_\alpha} \, K^{T_\alpha \ov{G}^b} = -\frac{i \, s \, b^b \left(4 \hat{\xi }^2+\hat{k}_0 s \left(4\,{\cal V}-\hat{\xi }\right)-8 \,{\cal V}^2+22 \hat{\xi } \,{\cal V}\right)}{2 \left(\,{\cal V}-\hat{\xi }\right) \left(2 \,{\cal V}+\, \hat{\xi }\right)} = -\, K_{\ov{T}_\alpha} \, K^{\ov{T}_\alpha G^b} \, ,\nonumber\\
& & K_{S} \, K^{{S} \ov{T}_\beta} = \frac{i \left(4 \hat{\xi }-\hat{k}_0 s+2 \,{\cal V}\right) \left(\hat{\xi } \left(3 {k} _{\beta }-s \hat{{k}}_{\beta }\right)+4 s \,{\cal V} \hat{{k} }_{\beta }\right)}{4 \left(\,{\cal V}-\hat{\xi }\right) \left(2 \,{\cal V}+\, \hat{\xi }\right)} = -\, K_{\ov{S}} \, K^{\ov{S} T_\beta}\, , \nonumber\\
& & K_{G^a} \, K^{G^a \ov{T}_\beta} = \frac{i \, s \,\hat{k}_0 \left(\hat{\xi } \left(3 {k} _{\beta }-s \hat{{k} }_{\beta}\right)+4 s \,{\cal V} \hat{{k} }_{\beta }\right)}{2\, ({\cal V}-\hat{\xi }) \left(2 \,{\cal V}+\, \hat{\xi }\right)} -2\, i \, s \, \hat{{k} }_{\beta } = -\, K_{\ov{G}^a} \, K^{\ov{G}^a  T_\beta}\, , \nonumber\\
& & \hskip-0.0cm K_{T_\alpha} \, K^{T_\alpha \ov{T}_\beta} = -\frac{i}{4 \left(\,{\cal V}-\hat{\xi }\right) \left(2 \,{\cal V}+\, \hat{\xi }\right)} \biggl[{k} _{\beta } \left(8 \hat{\xi}^2+\hat{\xi } \left(3 \hat{k}_0 s+2 \,{\cal V}\right)+8 \,{\cal V}^2\right) \nonumber\\
& & \hskip1.5cm +s \hat{{k} }_{\beta } \left(8 \hat{\xi}^2+\hat{k}_0 s \left(4 \,{\cal V}-\hat{\xi }\right)-16 \,{\cal V}^2 + 26 \, \hat{\xi } \,{\cal V}\right)\biggr] = -\, K_{\ov{T}_\alpha} \, K^{\ov{T}_\alpha  T_\beta}\,.
\eea
In addition, one verifies that
\bea
\label{eq:identities00}
& & K_{S} \, K^{{S} \ov{S}}\, K_{\ov{S}} = \frac{\left(4 \,{\cal V}-\hat{\xi }\right) \left(4 \hat{\xi }-\hat{k}_0 s+2 \,{\cal V}\right)^2}{4 \left(\,{\cal V}-\hat{\xi }\right) \left(\hat{\xi }+2 \,{\cal V}\right)^2} \,, \\
& & K_{S} \, K^{{S} \ov{G}^a}\, K_{\ov{G}^a} = \frac{\hat{k}_0 s \left(4 \,{\cal V}-\hat{\xi}\right) \left(4 \hat{\xi }-\hat{k}_0 s+2 \,{\cal V}\right)}{2 \left(\,{\cal V}-\hat{\xi }\right) \left(\hat{\xi }+2\,{\cal V}\right)^2} = K_{G^a} \, K^{G^a \ov{S}}\, K_{\ov{S}} \,, \nonumber\\
& & K_{S} \, K^{{S} \ov{T}_\alpha}\, K_{\ov{T}_\alpha} = -\frac{\left(4 \hat{\xi }-\hat{k}_0 s+2 \,{\cal V}\right) \left(\hat{k}_0 s \left(4 \,{\cal V}-\hat{\xi }\right)+18 \hat{\xi } \,{\cal V}\right)}{4 \left(\,{\cal V}-\hat{\xi }\right) \left(\hat{\xi }+2 \,{\cal V}\right)^2} = K_{T_\alpha} \,K^{{T}_\alpha \ov{S}}\, K_{\ov{S}} \, ,\nonumber\\
& & K_{G^a} \, K^{G^a \ov{G}^b}\, K_{\ov{G}^b} =  \frac{\hat{k}_0 s \left(\hat{k}_0 \, s \, \left(4 \,{\cal V}-\hat{\xi }\right)-4 \,{\cal V}^2+2 \hat{\xi } \left(\hat{\xi}+\,{\cal V}\right)\right)}{\left(\,{\cal V}-\hat{\xi }\right) \left(\hat{\xi }+2 \,{\cal V}\right)^2}\, , \nonumber\\
& & K_{G^a} \, K^{G^a \ov{T}_\alpha}\, K_{\ov{T}_\alpha} =\frac{\hat{k}_0 s
   \left(-4 \hat{\xi }^2+\hat{k}_0 s \left(\hat{\xi }-4 \,{\cal V}\right)+8 \,{\cal V}^2-22 \hat{\xi } \,{\cal V}\right)}{2
   \left(\,{\cal V}-\hat{\xi }\right) \left(\hat{\xi }+2 \,{\cal V}\right)^2} =  K_{T_\alpha} \, K^{T_\alpha \ov{G}^a}\, K_{\ov{G}^a}\, ,\nonumber\\
& & K_{T_\alpha} \, K^{T_\alpha \ov{T}_\beta}\, K_{\ov{T}_\beta} = \frac{\hat{k}_0^2 s^2 \left(4 \,{\cal V}-\hat{\xi }\right)+4 \hat{k}_0 s \left(2 \hat{\xi }^2-4
   \,{\cal V}^2+11 \hat{\xi } \,{\cal V}\right)+12 \,{\cal V} \left(4 \hat{\xi }^2+4 \,{\cal V}^2+\hat{\xi }
   \,{\cal V}\right)}{4 \left(\,{\cal V}-\hat{\xi }\right) \left(\hat{\xi }+2 \,{\cal V}\right)^2} \,. \nonumber
\eea
Note that we utilised the shorthand notations given in Eq.~(\ref{eq:shorthands}). Also we used $k_0 = {{k}}_\alpha \, t^\alpha =  6 {\cal V}$ and $\hat{{k}}_\alpha \, t^\alpha = \hat{k}_0 = - 4\, {\cal Y}\, {\cal G}_{ab}\, b^a\, b^b$ following from the definitions of moduli space metrics in Eq.~(\ref{eq:genMetrices}).

In the context of two-step moduli stabilisation schemes, like KKLT and LVS, in which the complex structure moduli ($U^i$) and the axio-dilaton (${S}$) are stabilised at the leading order, while the K\"ahler and odd moduli ($T_\alpha$ and $G^a$) are stabilised at subleading order,
the following identities are commonly used
\bea
\label{eq:identities000}
& & K_{A'} \, K^{A' \ov{B'}}\, K_{\ov{B'}} = \frac{3\,{\cal V} \left(4 \, \hat{\xi}^2+4 \, {\cal V}^2 + \hat{\xi } \, {\cal V}\right)}{\left({\cal V}-\hat{\xi }\right) \left(\hat{\xi }+2 \, {\cal V}\right)^2} + \frac{\hat{k}_0^2 \, s^2 \, \left(4 \, {\cal V}-\hat{\xi }\right)-36 \, \hat{k}_0 \, \hat{\xi } \, s \, {\cal V}}{4 \left(\, {\cal V}-\hat{\xi }\right) \left(\hat{\xi }+2 \, {\cal V}\right)^2}\, , \nonumber\\
& & K_{A'} \, K^{A' \ov{B'}}\, K_{\ov{B'}} -3 =\frac{3\, \hat{\xi }\left({\cal V}^2+7 \hat{\xi } \, {\cal V} + \, \hat{\xi }^2\right)}{\left(\, {\cal V}-\hat{\xi }\right) \left(\hat{\xi }+2 \, {\cal V}\right)^2} + \frac{\hat{k}_0^2 \,s^2 \,\left(4 \, {\cal V}-\hat{\xi }\right)-36 \,\hat{k}_0 \,\hat{\xi } \,s \, {\cal V}}{4 \left(\, {\cal V}-\hat{\xi }\right) \left(\hat{\xi }+2 \, {\cal V}\right)^2}\, .
\eea
Here, the sum over indices $A'$ and $B'$ runs over the $\{T_\alpha, G^a\}$ chiral variables without the axio-dilaton (${S}$).

%%%%%%%%%%%%%%%%%%%%%%%%%%%%%%%%%%%%%%%%%%%%%%%%%%%%
%
\newpage
\bibliographystyle{JHEP}
\bibliography{reference}
\end{document}